\documentclass{article}

\usepackage[left = 20 mm, right = 20 mm, top = 30 mm, bottom = 20 mm]{geometry}
\usepackage{amsmath, amssymb, verbatim, graphicx, subfig, caption, slashed, makecell, cite, jheppub}
\usepackage[toc,page]{appendix}

\newcommand{\minus}{\scalebox{0.75}[1.0]{$-$}}

\allowdisplaybreaks

\begin{document}

\title{Construction of lepton mass matrices and TeV-scale phenomenology in the minimal left-right symmetric model}

\author{Chang-Hun Lee}
\affiliation{Maryland Center for Fundamental Physics and Department of Physics, \\
University of Maryland, College Park, \\
Maryland 20742, U.S.A.}
\emailAdd{changhun@terpmail.umd.edu}

\begin{abstract}
{We develop a systematic procedure of constructing lepton mass matrices that satisfy all the experimental constraints in the light lepton sector of the minimal left-right symmetric model with type-I seesaw dominance. This method is unique since it is applicable to the most general cases of type-I seesaw  with complex electroweak vacuum expectation values in the model. With this method, we investigate the TeV-scale phenomenology in the normal hierarchy without fine-tuning of model parameters, focusing on the charged lepton flavour violation, neutrinoless double beta decay, and electric dipole moments of charged leptons. We examine the predictions for typical ranges of associated observables such as branching ratios of rare lepton decays, and study how those experimental constraints affect the model parameter space. The most notable result is that the regions of parameter space that allow small light neutrino masses have been constrained by the present experimental bounds from charged lepton flavour violation. Furthermore, we also find that the mass of the lightest heavy neutrino should be relatively small in order to satisfy those experimental constraints.}
\end{abstract}

\maketitle

\section{Introduction}
The Standard Model (SM) of particle physics is a chiral theory with a broken parity symmetry, and the left-right symmetric model is an extension of the SM with the parity symmetry restored at high energies \cite{LNC, NLRS, LRSVP}. Its extended particle content (e.g.~right-handed (RH) neutrinos and gauge bosons) allows us not only to find the solution to the parity problem of the SM but also to solve the problem of understanding the neutrino masses via the seesaw mechanism~\cite{mutoeg9, UTBN, SUGRA, numassSPV}. If the scale of parity restoration is in the few TeV range, we can expect new physics signals that are not present in the SM in planned future experiments. For example, since the lepton number for each flavour in the left-right symmetric model is no longer an exact symmetry of nature as in the SM, it is possible to observe charged lepton flavour violation (CLFV) processes such as $\mu \to e \gamma$ or lepton number violation effects through neutrinoless double beta decay ($0 \nu \beta \beta$). Furthermore, since the left-right symmetric model not only has more particles but also has more sources of CP violation not present in the SM such as complex Yukawa couplings and vacuum expectation values (VEV), we can also expect large CP violating effects such as the electric dipole moment (EDM) of a charged lepton. In this paper, these aspects of the left-right symmetric model will be discussed.

In the lepton sector of the minimal left-right symmetric model (MLRSM), of which a brief review is provided in section \ref{sec:MLRSM}, we have four mass matrices: the charged lepton mass matrix $M_\ell$, the Dirac neutrino mass matrix $M_D$, and the left-handed and RH Majorana neutrino mass matrices $M_L$ and $M_R$. The light neutrino mass matrix $M_\nu$ is determined by $M_D$, $M_L$, and $M_R$ through the seesaw mechanism $M_\nu \approx M_L - M_D M_R^{-1} M_D^\mathsf{T}$. Since we have experimental data on the masses of charged leptons and the squared-mass differences of neutrinos as well as their mixing angles, $M_\ell$ is completely known in the charged lepton mass basis and $M_\nu$ is also partially determined in its own mass basis and in the charged lepton mass basis. The neutrino mass matrices $M_D$, $M_L$, and $M_R$ are nonetheless completely unknown, and constructing those matrices compatible with experimental data is a nontrivial problem, not only because $M_\ell$ and $M_D$ in the MLRSM are determined from common Yukawa couplings and electroweak VEV's, but also because those Yukawa coupling matrices have a specific structure (i.e.~Hermitian or symmetric) in a specific basis (i.e.~symmetry basis) due to the discrete symmetry (i.e.~parity or charge conjugation symmetry) of the model that realizes the manifest left-right symmetry at high energies.

For simplicity, we may assume that the electroweak VEV's are all real, in which case $M_\ell$ and $M_D$ have the same structure (i.e.~Hermitian or symmetric) as the Yukawa coupling matrices. Since they maintain that structure in any basis, we can work in the charged lepton mass basis where $M_\ell$ is completely determined so that we can practically forget about it while keeping the structure of mass matrices. Now using that structure itself, we can find $M_R$ from known $M_D$ \cite{LRSMD} or alternatively find $M_D$ from known $M_R$ \cite{LRSMDM}. Without loss of generality, however, we can make only one of two electroweak VEV's real by gauge transformation. Furthermore, for the TeV-scale MLRSM, $M_D$ assumed or constructed in such ways usually requires fine-tuning of Yukawa couplings and VEV's, and it would be rather difficult to make natural predictions for the TeV-scale phenomenology of the MLRSM using those mass matrices.

In this paper, we develop a different approach appropriate for the case of type-I dominance (i.e.~$M_L = 0$) with complex electroweak VEV's: (i) the Yukawa coupling matrices with a desired structure are constructed from $M_\ell$ in the symmetry basis; (ii) $M_D$ is determined from those Yukawa couplings as well as the electroweak VEV's, and $M_R$ is calculated from $M_D$ we have found. Since Yukawa couplings are explicitly constructed and $M_D$ is calculated from them, fine-tuned $M_D$ can only appear rarely. With this method, we collect a huge amount of data points that satisfy all the major experimental constraints, and conduct a comprehensive study of the TeV-scale phenomenology of the model, focusing on the CLFV, $0 \nu \beta \beta$, and EDM's of charged leptons.

There are several works which studied CLFV and $0 \nu \beta \beta$ in the MLRSM: in reference \cite{LRSMLFV}, those effects were discussed in the type-I or type-II seesaw dominance, and several processes of $0 \nu \beta \beta$ were examined in detail; in reference \cite{LRSMST}, CLFV and $0 \nu \beta \beta$ processes were investigated also in type-I or type-II dominance with emphasis on the allowed masses of doubly charged scalar fields; in reference \cite{LRSMCLFV}, the type-I+II seesaw contributions were simultaneously considered as in references \cite{LRSMD} and \cite{LRSMDM}, but with richer results on the phenomenology; in reference \cite{LRSMPLFV}, the CLFV effects were studied in detail also in the type-I+II seesaw cases by a slightly different method from the one originally proposed by reference \cite{LRSMD}. However, the common features of those works are: (i) real electroweak VEV's were explicitly or implicitly assumed, and (ii) $M_D$ or $M_R$ was chosen for numerical analysis without considering the issue of fine-tuning. Even though we can still obtain meaningful results focusing on specific regions of parameter space with rich phenomenologies, it is important to investigate the predictions of the model in a more natural situation. Furthermore, some works assumed that the tree-level contribution to $\mu \to eee$ is always dominant over the type-I contribution in their analyses. We will also see that this is an inadequate assumption.

This paper is organized as follows: in section \ref{sec:MLRSM}, a brief review on the MLRSM is provided; in section \ref{sec:Mmass}, we systematically construct lepton mass matrices that satisfy the experimental constraints in the light lepton sector; in section \ref{sec:LRSMTeV}, the conditions for the TeV-scale MLRSM is investigated, and its phenomenology is studied; all the expressions of observables of CLFV, $0 \nu \beta \beta$, and EDM's of charged leptons used in this paper are summarized in appendix \ref{sec:expObs}; the benchmark model parameters and their predictions are provided in appendix \ref{sec:BM}.

\section{Minimal left-right symmetric model} \label{sec:MLRSM}
In this section, we briefly review the MLRSM. The gauge group of the MLRSM is
\begin{align}
	\text{SU}(2)_L \otimes \text{SU}(2)_R \otimes \text{U}(1)_{B - L},
\end{align}
and the representations of the leptons are
\begin{align}
	L_{Li}' = \left( \begin{array}{c} \nu_{Li}' \\ \ell_{Li}' \end{array} \right) \sim (\textbf{2}, \textbf{1}, -1), &\qquad
	L_{Ri}' = \left( \begin{array}{c} \nu_{Ri}' \\ \ell_{Ri}' \end{array} \right) \sim (\textbf{1}, \textbf{2}, -1)
\end{align}
where $i$ is the flavour index. The bi-doublet scalar field is given by
\begin{align}
	\Phi = \left( \begin{array}{cc} \phi_1^0 & \phi_2^+ \\ \phi_1^- & \phi_2^0 \end{array} \right) \sim (\textbf{2}, \textbf{2}, 0),
\end{align}
and the triplet scalar fields are
\begin{align}
	\Delta_L = \left( \begin{array}{cc} \delta_L^+ / \sqrt{2} & \delta_L^{++} \\ \delta_L^0 & -\delta_L^+ / \sqrt{2} \end{array} \right) \sim (\textbf{3}, \textbf{1}, 2), \qquad
	\Delta_R = \left( \begin{array}{cc} \delta_R^+ / \sqrt{2} & \delta_R^{++} \\ \delta_R^0 & -\delta_R^+ / \sqrt{2} \end{array} \right) \sim (\textbf{1}, \textbf{3}, 2).
\end{align}
The Lagrangian terms of Yukawa interactions are written as
\begin{align}
	\mathcal{L}_Y^\ell &= -\overline{L_{Li}'} (f_{ij} \Phi + \tilde{f}_{ij} \tilde{\Phi}) L_{Rj}'
	- h_{Lij} \overline{L_{Li}'^c} i \sigma_2 \Delta_L L_{Lj}' - h_{Rij} \overline{L_{Ri}'^c} i \sigma_2 \Delta_R L_{Rj}'
	+ \text{H.c.}
\end{align}
where
\begin{align}
	\tilde{\Phi} \equiv \sigma_2 \Phi^* \sigma_2
	= \left( \begin{array}{cc} \phi_2^{0*} & -\phi_1^+ \\ -\phi_2^- & \phi_1^{0*} \end{array} \right).
\end{align}
Here, $\psi^c \equiv \mathsf{C} \psi^*$, and thus $\overline{\psi^c} = -\psi^\mathsf{T} \mathsf{C}$ where $\mathsf{C} = i \gamma^2 \gamma^0$ is the charge conjugation operator in the Dirac-Pauli representation. Note that $h_L$ and $h_R$ are symmetric matrices. Without loss of generality, we can write the VEV's of scalar fields as
\begin{align}
	\Phi = \left( \begin{array}{cc} \kappa_1 / \sqrt{2} & 0 \\ 0 & \kappa_2 e^{i \alpha} / \sqrt{2} \end{array} \right), \qquad
	\Delta_L = \left( \begin{array}{cc} 0 & 0 \\ v_L e^{i \theta_L} / \sqrt{2} & 0 \end{array} \right), \qquad
	\Delta_R = \left( \begin{array}{cc} 0 & 0 \\ v_R / \sqrt{2} & 0 \end{array} \right).
	\label{eq:VEV}
\end{align}
After spontaneous symmetry breaking, the mass matrix of charged leptons is written as
\begin{align}
	M_\ell = \frac{1}{\sqrt{2}} (f \kappa_2 e^{i \alpha} + \tilde{f} \kappa_1),
\end{align}
and the neutrino mass term is given by
\begin{align}
	\mathcal{L}_\nu^\text{mass} = -\frac{1}{2} (\overline{\nu_L'} \ \overline{\nu_R'^c})
		\left( \begin{array}{cc} M_L & M_D \\ M_D^\mathsf{T} & M_R \end{array} \right)
		\left( \begin{array}{c} \nu_L'^c \\ \nu_R' \end{array} \right) + \text{H.c.}
\end{align}
where
\begin{align}
	M_D = \frac{1}{\sqrt{2}} (f \kappa_1 + \tilde{f} \kappa_2 e^{-i \alpha}), \qquad
	M_L = \sqrt{2} h_L^* v_L e^{-i \theta_L}, \qquad
 	M_R = \sqrt{2} h_R v_R.
\end{align}
When $v_L \ll \kappa_1, \kappa_2 \ll v_R$, the light neutrino mass matrix is given by the seesaw mechanism
\begin{align}
	M_\nu \approx M_L - M_D M_R^{-1} M_D^\mathsf{T}.
\end{align}
In this paper, we only consider the case of type-I dominance by assuming $v_L = 0$, and the light neutrino mass matrix is given by the type-I seesaw formula
\begin{align}
	M_\nu \approx -M_D M_R^{-1} M_D^\mathsf{T}.
	\label{eq:seesaw}
\end{align}
We denote the mass eigenstates of the light and heavy neutrinos as $\nu_i$ and $N_i$ $(i = 1, 2, 3)$, respectively. The charged gauge bosons $W_L^-$, $W_R^-$ in the gauge basis can be written in terms of the mass eigenstates $W_1^-$, $W_2^-$ as
\begin{align}
	\left( \begin{array}{c} W_L^- \\ W_R^- \end{array} \right)
	= \left( \begin{array}{cc}
		\cos{\xi} & \sin{\xi} e^{i \alpha} \\
		-\sin{\xi} e^{-i \alpha} & \cos{\xi}
	\end{array} \right)
	\left( \begin{array}{c} W_1^- \\ W_2^- \end{array} \right)
\end{align}
where $\xi$ is the $W_L$-$W_R$ mixing parameter given by
\begin{align}
	\xi \approx -\frac{\kappa_1 \kappa_2}{v_R^2}.
\end{align}
The masses of charged gauge bosons are
\begin{align}
	m_{W_1}^2 \approx \frac{1}{4} g^2 v_\text{EW}^2, \qquad
	m_{W_2}^2 \approx \frac{1}{2} g^2 v_R^2
\end{align}
where $v_\text{EW} = \sqrt{\kappa_1^2 + \kappa_2^2} = 246$ GeV is the VEV of the SM. In addition, the masses of neutral gauge bosons $Z_1$, $Z_2$, $A$ are given by
\begin{align}
	m_{Z_1}^2 \approx \frac{g^2 v_\text{EW}^2}{4 \cos^2{\theta_W}}, \qquad
	m_{Z_2}^2 \approx \frac{g^2 \cos^2{\theta_W} v_R^2}{\cos{2\theta_W}}, \qquad
	m_A^2 = 0
\end{align}
where $\theta_W$ is the Weinberg angle. We can identify $W_1$, $Z_1$, $A$ as $W$, $Z$, the photon of the SM, respectively. The neutral gauge bosons $W_L^3$, $W_R^3$, $B$ in the gauge basis are expressed in terms of the mass eigenstates as
\begin{align}
	\left( \begin{array}{c} W_L^3 \\ W_R^3 \\ B \end{array} \right)
	&= \left( \begin{array}{ccc}
		1 & 0 & 0 \\
		0 & \cos{\zeta_1} & \sin{\zeta_1} \\
		0 & -\sin{\zeta_1} & \cos{\zeta_1}
	\end{array} \right)
	\left( \begin{array}{ccc}
		\cos{\zeta_2} & 0 & \sin{\zeta_2} \\
		0 & 1 & 0 \\
		-\sin{\zeta_2} & 0 & \cos{\zeta_2}
	\end{array} \right)
	\left( \begin{array}{ccc}
		\cos{\zeta_3} & \sin{\zeta_3} & 0 \\
		-\sin{\zeta_3} & \cos{\zeta_3} & 0 \\
		0 & 0 & 1
	\end{array} \right)
	\left( \begin{array}{c} Z_1 \\ Z_2 \\ A \end{array} \right)
\end{align}
where
\begin{align}
	\zeta_1 = \sin^{-1}{(\tan{\theta_W})}, \qquad
	\zeta_2 \approx \theta_W, \qquad
	\zeta_3 \approx -\frac{g^2 \sqrt{\cos{2\theta_W}} v_\text{EW}^2}{4 \cos^2{\theta_W} m_{Z_2}^2}.
	\label{eq:NGmixing}
\end{align}
For the MLRSM with a manifest left-right symmetry before spontaneous symmetry breaking, we need a discrete symmetry which could be either the parity symmetry or the charge conjugation symmetry. In case of the parity symmetry, we have the relationships of fields and Yukawa couplings given by
\begin{align}
	L_{Li}' \leftrightarrow L_{Ri}', \qquad
	\Delta_L \leftrightarrow \Delta_R, \qquad
	\Phi \leftrightarrow \Phi^\dagger, \qquad
	f = f^\dagger, \qquad
	\tilde{f} = \tilde{f}^\dagger, \qquad
	h_L = h_R,
\end{align}
and in case of the charge conjugation symmetry
\begin{align}
	L_{Li}' \leftrightarrow L_{Ri}'^c, \qquad
	\Delta_L \leftrightarrow \Delta_R^*, \qquad
	\Phi \leftrightarrow \Phi^\mathsf{T}, \qquad
	f = f^\mathsf{T}, \qquad
	\tilde{f} = \tilde{f}^\mathsf{T}, \qquad
	h_L = h_R^*.
\end{align}
We consider only the parity symmetry here. This symmetry is manifest in a specific basis in the flavour space, which we call the symmetry basis. The scalar potential invariant under the parity symmetry is written as
\footnotesize
\begin{align}
	V &= -\mu_1^2 \text{Tr} \big[ \Phi^\dagger \Phi \big]
		- \mu_2^2 \left( \text{Tr} \big[ \Phi^\dagger \tilde{\Phi} \big] + \text{Tr} \big[ \tilde{\Phi}^\dagger \Phi \big] \right)
		- \mu_3^2 \left( \text{Tr} \big[ \Delta_L^\dagger \Delta_L \big] + \text{Tr} \big[ \Delta_R^\dagger \Delta_R \big] \right) \nonumber \\
	&\qquad + \lambda_1 \text{Tr} \big[ \Phi^\dagger \Phi \big]^2
		+ \lambda_2 \left( \text{Tr} \big[ \Phi^\dagger \tilde{\Phi} \big]^2 + \text{Tr} \big[ \tilde{\Phi}^\dagger \Phi \big]^2 \right)
		+ \lambda_3 \text{Tr} \big[ \Phi^\dagger \tilde{\Phi} \big] \text{Tr} \big[ \tilde{\Phi}^\dagger \Phi \big]
		+ \lambda_4 \text{Tr} \big[ \Phi^\dagger \Phi \big] \left( \text{Tr} \big[ \Phi^\dagger \tilde{\Phi} \big] + \text{Tr} \big[ \tilde{\Phi}^\dagger \Phi \big] \right) \nonumber \\
	&\qquad + \rho_1 \left( \text{Tr} \big[ \Delta_L^\dagger \Delta_L \big]^2 + \text{Tr} \big[ \Delta_R^\dagger \Delta_R \big]^2 \right)
		+ \rho_2 \left( \text{Tr} \big[ \Delta_L^\dagger \Delta_L^\dagger \big] \text{Tr} \big[ \Delta_L \Delta_L \big] + \text{Tr} \big[ \Delta_R^\dagger \Delta_R^\dagger \big] \text{Tr} \big[ \Delta_R \Delta_R \big] \right) \nonumber \\
	&\qquad + \rho_3 \text{Tr} \big[ \Delta_L^\dagger \Delta_L \big] \text{Tr} \big[ \Delta_R^\dagger \Delta_R \big]
		+ \rho_4 \left( \text{Tr} \big[ \Delta_L^\dagger \Delta_L^\dagger \big] \text{Tr} \big[ \Delta_R \Delta_R \big] + \text{Tr} \big[ \Delta_L \Delta_L \big] \text{Tr} \big[ \Delta_R^\dagger \Delta_R^\dagger \big] \right) \nonumber \\
	&\qquad + \alpha_1 \text{Tr} \big[ \Phi^\dagger \Phi \big] \left( \text{Tr} \big[ \Delta_L^\dagger \Delta_L \big] + \text{Tr} \big[ \Delta_R^\dagger \Delta_R \big] \right)
	+ \left\{ \alpha_2 e^{i \delta_2} \left( \text{Tr} \big[ \Phi^\dagger \tilde{\Phi} \big] \text{Tr} \big[ \Delta_L^\dagger \Delta_L \big] + \text{Tr} \big[ \tilde{\Phi}^\dagger \Phi \big] \text{Tr} \big[ \Delta_R^\dagger \Delta_R \big] \right) + \text{H.c.} \right\} \nonumber \\
	&\qquad + \alpha_3 \left( \text{Tr} \big[ \Phi \Phi^\dagger \Delta_L \Delta_L^\dagger \big] + \text{Tr} \big[ \Phi^\dagger \Phi \Delta_R \Delta_R^\dagger \big] \right)
	+ \beta_1 \left( \text{Tr} \big[ \Phi^\dagger \Delta_L^\dagger \Phi \Delta_R \big] + \text{Tr} \big[ \Phi^\dagger \Delta_L \Phi \Delta_R^\dagger \big] \right) \nonumber \\
	&\qquad + \beta_2 \left( \text{Tr} \big[ \Phi^\dagger \Delta_L^\dagger \tilde{\Phi} \Delta_R \big] + \text{Tr} \big[ \tilde{\Phi}^\dagger \Delta_L \Phi \Delta_R^\dagger \big] \right)
	+ \beta_3 \left( \text{Tr} \big[ \tilde{\Phi}^\dagger \Delta_L^\dagger \Phi \Delta_R \big] + \text{Tr} \big[ \Phi^\dagger \Delta_L \tilde{\Phi} \Delta_R^\dagger \big] \right).
\end{align}
\normalsize
In this paper, we study the TeV-scale MLRSM without fine-tuning, for which $\kappa_1 \gg \kappa_2$ is one of the sufficient conditions, as we will see in section \ref{sec:LRSMTeV}. The physical scalar fields and their masses when $v_L = 0$ and $v_R \gg \kappa_1 \gg \kappa_2$ are summarized in table \ref{tab:scalar} \cite{LRSMCP}.
\begin{table}[ht]
	\center
	\begin{tabular}{|l||l|}
		\hline
		Physical scalar fields & Mass-squared \\ \hline
		$h^0 = \sqrt{2} \text{Re}[\phi_1^{0*} + \epsilon_2 e^{-i \alpha} \phi_2^0]$ & $\frac{1}{2} (4\lambda_1 - \alpha_1^2 / \rho_1) \kappa_1^2 + \frac{1}{2} \alpha_3 v_R^2 \epsilon_2^2$ \\ \hline
		$H_1^0 = \sqrt{2} \text{Re}[-\epsilon_2 e^{i \alpha} \phi_1^{0*} + \phi_2^0]$ & $\frac{1}{2} \alpha_3 v_R^2$ \\ \hline
		$H_2^0 = \sqrt{2} \text{Re}[\delta_R^0]$ & $2 \rho_1 v_R^2$ \\ \hline
		$H_3^0 = \sqrt{2} \text{Re}[\delta_L^0]$ & $\frac{1}{2} (\rho_3 - 2 \rho_1) v_R^2$ \\ \hline
		$A_1^0 = \sqrt{2} \text{Im}[-\epsilon_2 e^{i \alpha} \phi_1^{0*} + \phi_2^0]$ & $\frac{1}{2} \alpha_3 v_R^2$ \\ \hline
		$A_2^0 = \sqrt{2} \text{Im}[\delta_L^0]$ & $\frac{1}{2} (\rho_3 - 2 \rho_1) v_R^2$ \\ \hline
		$H_1^+ = \delta_L^+$ & $\frac{1}{2} (\rho_3 - 2 \rho_1) v_R^2 + \frac{1}{4} \alpha_3 \kappa_1^2$ \\ \hline
		$H_2^+ = \phi_2^+ + \epsilon_2 e^{i \alpha} \phi_1^+ + \frac{1}{\sqrt{2}} \epsilon_1 \delta_R^+$ & $\frac{1}{2} \alpha_3 \big( v_R^2 + \frac{1}{2} \kappa_1^2 \big)$ \\ \hline
		$\delta_R^{++}$ & $2 \rho_2 v_R^2 + \frac{1}{2} \alpha_3 \kappa_1^2$ \\ \hline
		$\delta_L^{++}$ & $\frac{1}{2} (\rho_3 - 2 \rho_1) v_R^2 + \frac{1}{2} \alpha_3 \kappa_1^2$ \\ \hline
	\end{tabular}
	\caption{Physical scalar fields and their masses in the MLRSM when $v_L = 0$ and $v_R \gg \kappa_1 \gg \kappa_2$. Here, $\epsilon_1 \equiv \kappa_1 / v_R$ and $\epsilon_2 \equiv \kappa_2 / \kappa_1$. The SM Higgs field is identified as $h^0$. Note that $m_{H_1^+} \approx m_{\delta_L^{++}}$ for $v_R \gg v_\text{EW}$. The mixing between $\delta_L^{++}$ and $\delta_R^{++}$ is assumed to be small, although it could be large in principle for relatively small values of $\rho_3 - 2 \rho_1$ and $v_R$ \cite{LRSMPLFV}. It is, however, a good assumption even for such cases if we introduce an additional assumption $\beta_1, \beta_3 \lesssim \mathcal{O}(10^{-1})$.}
	\label{tab:scalar}
\end{table}

\section{Construction of lepton mass matrices} \label{sec:Mmass}
In this section, we discuss the procedure to construct lepton mass matrices that satisfy the experimental constraints in the light lepton sector (i.e.~light neutrino masses and mixing angles) in case of type-I dominance. The Yukawa coupling matrices $f$, $\tilde{f}$ in the symmetry basis are Hermitian due to the parity symmetry before spontaneous symmetry breaking. However, the mass matrices $M_\ell$ and $M_D$ in the same basis do not have such structures when the electroweak VEV's are complex, and it is therefore a non-trivial problem to construct mass matrices that would give Yukawa couplings with the right structure in the symmetry basis and simultaneously satisfy all the constraints in the light lepton sector.

The procedure to construct such lepton mass matrices is as follows: (i) first, we find $M_\ell$ in the symmetry basis that gives the right masses of charged leptons, and build up $f$, $\tilde{f}$, and VEV's out of it. The solutions are not unique; (ii) $M_D$ is constructed in the straightforward way from the Yukawa couplings and VEV's we have obtained, and $M_R$ can also be easily calculated from this $M_D$ and the type-I seesaw formula of equation \ref{eq:seesaw}.

Since the masses of charged leptons are already known, $M_\ell$ in the symmetry basis can be easily constructed from
\begin{align}
	M_\ell = V^\ell_L M_\ell^c V^{\ell \dagger}_R
\end{align}
where $V^\ell_L$ and $V^\ell_R$ are arbitrary unitary matrices and $M_\ell^c$ is the diagonal matrix which has charged lepton masses as its entries. The superscript $c$ denotes mass matrices in the charged lepton mass basis, and we always assume that matrices without any superscript are in the symmetry basis. Note that $V^\ell_L$ and $V^\ell_R$ are totally different matrices in general even with a manifest discrete symmetry when the electroweak VEV's are complex. With the parity symmetry, we have $M_\ell = A e^{i \alpha} + B$ ($A \equiv f \kappa_2 / \sqrt{2}$, $B \equiv \tilde{f} \kappa_1 / \sqrt{2}$) where $A$, $B$ are Hermitian matrices. Therefore, for the rest of step (i), we claim that, for an arbitrary matrix $M$, it is always possible to find Hermitian matrices $A$, $B$ such that $M = A e^{i \alpha} + B$.

In order to prove it, we explicitly construct Hermitian matrices $A$, $B$ that satisfy $M = A e^{i \alpha} + B$. First, we write $A_{ij} = |A_{ij}| e^{i \theta_{ij}}$ and $B_{ij} = |B_{ij}| e^{i \phi_{ij}}$ where $\theta_{ji} = -\theta_{ij}$ and $\phi_{ji} = -\phi_{ij}$. Then, we have $M_{ij} = |A_{ij}| e^{i (\alpha + \theta_{ij})} + |B_{ij}| e^{i \phi_{ij}}$ and $M_{ji} = |A_{ij}| e^{i (\alpha - \theta_{ij})} + |B_{ij}| e^{-i \phi_{ij}}$. From these expressions, it is straightforward to derive
\begin{align}
	2 |A_{ij}| \sin{\alpha} = \pm \sqrt{\text{Re}[M_{ji} - M_{ij}]^2 + \text{Im}[M_{ji} + M_{ij}]^2}
\end{align}
and
\begin{align}
	\tan{\theta_{ij}} &= \frac{\text{Re}[M_{ji} - M_{ij}]}{\text{Im}[M_{ji} + M_{ij}]}.
	\label{eq:Aphs}
\end{align}
Note that two different values of $\theta_{ij}$ are allowed in the range $-\pi < \theta_{ij} < \pi$ for each pair of $i, j$. In addition, since $|\sin{\alpha}| \leq 1$, we must have
\begin{align}
	|A_{ij}| \geq \frac{1}{2} \sqrt{\text{Re}[M_{ji} - M_{ij}]^2 + \text{Im}[M_{ji} + M_{ij}]^2}
\end{align}
which sets the lower bound of $|A_{ij}|$ for given $M$. If $|A_{ij}| \neq 0$, we can write
\begin{align}
	\sin{\alpha} = \pm \frac{1}{2 |A_{ij}|} \sqrt{\text{Re}[M_{ji} - M_{ij}]^2 + \text{Im}[M_{ji} + M_{ij}]^2}.
\end{align}
Now we choose an arbitrary real number $|A_{11}|$ that satisfies
\begin{align}
	|A_{11}| > \big| \text{Im}[M_{11}] \big|,
\end{align}
and determine $\alpha$ from
\begin{align}
	\sin{\alpha} = \pm \frac{\big| \text{Im}[M_{11}] \big|}{|A_{11}|}.
\end{align}
Note that four different values of $\alpha$ are allowed in the range $-\pi < \alpha < \pi$. We can find all the other $|A_{ij}|$ from
\begin{align}
	|A_{ij}| &= \frac{1}{2 |\sin{\alpha}|} \sqrt{\text{Re}[M_{ji} - M_{ij}]^2 + \text{Im}[M_{ji} + M_{ij}]^2} \\
		&= \frac{|A_{11}|}{2 \big| \text{Im}[M_{11}] \big|} \sqrt{\text{Re}[M_{ji} - M_{ij}]^2 + \text{Im}[M_{ji} + M_{ij}]^2}.
	\label{eq:Aabs}
\end{align}
By equations \ref{eq:Aabs} and \ref{eq:Aphs}, $A$ is completely determined. Alternatively we can write 
\begin{align}
	A_{ij} &= \pm \frac{1}{2 |\sin{\alpha}|} \big( \text{Im}[M_{ji} + M_{ij}] + i \text{Re}[M_{ji} - M_{ij}] \big) \\
	&= \pm \frac{|A_{11}|}{2 \big| \text{Im}[M_{11}] \big|} \big( \text{Im}[M_{ji} + M_{ij}] + i \text{Re}[M_{ji} - M_{ij}] \big).
\end{align}
It is now trivial to find $B$ from $B = M - A e^{i \alpha}$, and explicitly
\begin{align}
	\text{Re}[B_{ij}] = \frac{1}{2} \text{Re}[M_{ji} + M_{ij}] - \text{Re}[A_{ij}] \cos{\alpha}, \qquad
	\text{Im}[B_{ij}] = -\frac{1}{2} \text{Im}[M_{ji} - M_{ij}] - \text{Im}[A_{ij}] \cos{\alpha},
\end{align}
or
\begin{align}
	B_{ij} = \frac{1}{2} \big( \text{Re}[M_{ji} + M_{ij}] - i \text{Im}[M_{ji} - M_{ij}] \big) - A_{ij} \cos{\alpha}.
\end{align}
Note that $A$ and $B$ are indeed Hermitian matrices. Since we have two choices of $A_{ij}$ for each pair of $i, j$ as well as each choice of $\alpha$ and $|A_{11}|$, there are $2^6$ choices of $A$ for each $\alpha$ and $|A_{11}|$ as we have three diagonal and three off-diagonal independent components in $A$. Moreover, since we have four choices of $\alpha$ for each $|A_{11}|$, there are total $2^6 \cdot 4 = 256$ different choices of $A$, $B$, and $\alpha$ for each choice of $|A_{11}|$. We use this method to construct lepton mass matrices in the TeV-scale MLRSM.

\section{TeV-scale phenomenology of the minimal left-right symmetric model} \label{sec:LRSMTeV}

\subsection{Conditions for the TeV-scale minimal left-right symmetric model}
In the MLRSM, $M_\ell$ and $M_D$ are determined from common Yukawa couplings and VEV's: $f$, $\tilde{f}$, $\kappa_1$, and $\kappa_2 e^{i \alpha}$. Hence, it would be natural if the largest component of $M_D$ is $\mathcal{O}(1)$ GeV, since the largest component of $M_\ell$ should be comparable to $m_\tau \sim \mathcal{O}(1)$ GeV. However, this implies that the smallest heavy neutrino mass should be larger than $\mathcal{O}(10^{10})$ GeV, since $M_\nu$ is determined from the seesaw formula of equation \ref{eq:seesaw} and the present upper bound of the light neutrino mass is $m_\nu \lesssim \mathcal{O}(0.1)$ eV \cite{Planck}.

For the TeV-scale MLRSM, i.e.~0.1 TeV $\lesssim m_N \lesssim$ 100 TeV, we need $|M_{Dij}| \lesssim 10^{-3}$ GeV. Since $M_D = (f \kappa_1 + \tilde{f} \kappa_2 e^{-i \alpha}) / \sqrt{2}$ in the MLRSM, its largest component could be as small as $10^{-3}$ GeV when the corresponding components of $f \kappa_1$ and $\tilde{f} \kappa_2 e^{-i \alpha}$ almost cancel each other, which is however unnatural. One solution to avoid such cancellation is that either $f \kappa_2$ or $\tilde{f} \kappa_1$ is dominant in $M_\ell$ while $\tilde{f} \kappa_2$ and $f \kappa_1$ are both small and comparable to each other in $M_D$. Note that we need hierarchies in both Yukawa couplings and VEV's to satisfy this condition. Even though it is good enough if only a few components of either $f \kappa_2$ or $\tilde{f} \kappa_1$ that correspond to $m_\tau$ and $m_\mu$ are dominant in $M_\ell$, we assume that all the components of either $f \kappa_2$ or $\tilde{f} \kappa_1$ are dominant over the others for simplicity.

Now we write $A \equiv f \kappa_2 / \sqrt{2}$ and $B \equiv \tilde{f} \kappa_1 / \sqrt{2}$, and thus $M_\ell = A e^{i \alpha} + B$, as before. When $|A_{ij}| \ll |B_{ij}|$, $M_\ell$ must be close to a Hermitian matrix, which is equivalent to $V_L^{\ell \dagger} V_R^\ell \approx 1$. When $|A_{ij}| \gg |B_{ij}|$, we have $M_\ell \approx A e^{i \alpha}$, which implies that $M_\ell e^{-i \alpha}$ is approximately Hermitian, i.e.~$V_L^{\ell \dagger} V_R^\ell \approx e^{i \alpha}$. Note that we need the condition on mixing matrices in addition to the conditions on the Yukawa couplings and VEV's since constructing $M_\ell$ from mixing matrices is one of the first steps to construct all the mass matrices.

In this paper, we only consider the first case, i.e.~$|A_{ij}| \ll |B_{ij}|$. For simplicity, we could assume $A = 0$, for which we need either $f = 0$ or $\kappa_2 = 0$. In these cases, the mass matrices are rather simple: $M_\ell = \tilde{f} \kappa_1 / \sqrt{2}$, $M_D = \tilde{f} \kappa_2 e^{-i \alpha} / \sqrt{2}$ if $f = 0$, and $M_\ell = \tilde{f} \kappa_1 / \sqrt{2}$, $M_D = f \kappa_1 / \sqrt{2}$ if $\kappa_2 = 0$. However, $f = 0$ is the limiting case of an extreme hierarchy between two Yukawa coupling matrices $f$ and $\tilde{f}$, which is rather unnatural. Furthermore, we must have $M_\ell \propto M_D \propto \tilde{f}$, and thus $M_D$ is diagonal in the mass basis of charged leptons, which means that we have to resort to only restrictive structures of mass matrices. On the other hand, with the condition $\kappa_2 = 0$, the $W_L$-$W_R$ mixing parameter $\xi \approx -\kappa_1 \kappa_2 / v_R^2$ vanishes, and we have to lose the rich phenomenology dependent upon $\xi$, especially the EDM's of charged leptons. Therefore, we do not introduce these extreme conditions.

In summary, for the TeV-scale MLRSM without fine-tuning in $M_D$, we can assume the conditions either that (i) $f_{ij} \ll \tilde{f}_{ij}$ and $\kappa_1 \gg \kappa_2$, when $M_\ell$ is approximately Hermitian, i.e. $V_L^\ell \approx V_R^\ell$, or that (ii) $f_{ij} \gg \tilde{f}_{ij}$ and $\kappa_1 \ll \kappa_2$, when $M_\ell e^{-i \alpha}$ is approximately Hermitian, i.e. $V_L^\ell \approx V_R^\ell e^{-i \alpha}$. We study the first case here.

\subsection{Numerical procedure}
In this paper, we only consider the normal hierarchy in light neutrino masses. The procedure to calculate all the model parameters that determine the phenomenology of the MLRSM in type-I dominance is as follows:
\begin{enumerate}
	\item{Randomly generate the lightest light neutrino mass $m_{\nu_1}$, and calculate $m_{\nu_2} = \sqrt{m_{\nu_1}^2 + \Delta m_{21}^2}$ and $m_{\nu_3} = \sqrt{m_{\nu_1}^2 + \Delta m_{31}^2}$.}
	
	\item{Calculate $M_\nu^c$ from $M_\nu^c = U_\text{PMNS} M_\nu^\text{diag} U_\text{PMNS}^\mathsf{T}$ where $M_\nu^c$ and $M_\nu^\text{diag}$ are the light neutrino mass matrices in the charged lepton and light neutrino mass bases, respectively. The mixing matrix $U_\text{PMNS}$ is the Pontecorvo-Maki-Nakagawa-Sakata (PMNS) matrix whose CP phases are also randomly generated.}
	
	\item{Randomly generate $V^\ell_L$, $V^\ell_R$, and calculate $M_\ell = V^\ell_L M_\ell^c V^{\ell \dagger}_R$ where $M_\ell$ and $M_\ell^c$ are charged lepton mass matrices in the symmetry and charged lepton mass bases, repectively.} 
	
	\item{Find $A \equiv f \kappa_2 / \sqrt{2}$, $B \equiv \tilde{f} \kappa_1 / \sqrt{2}$ from $M_\ell = A e^{i \alpha} + B$ using the method discussed in section \ref{sec:Mmass}. Randomly generate $\kappa_2$, and calculate $f$, $\tilde{f}$ from $A$, $B$.}
	
	\item{Calculate $M_D = (f \kappa_1 + \tilde{f} \kappa_2 e^{-i \alpha}) / \sqrt{2}$ from $f$, $\tilde{f}$, $\alpha$, $\kappa_2$, $\kappa_1 = \sqrt{v_\text{EW}^2 - \kappa_2^2}$, and find $M_D^c = V^{\ell \dagger}_L M_D V^\ell_R$ where $M_D^c$ is the Dirac neutrino mass matrix in the charged lepton mass basis.}
	
	\item{Calculate $M_R^c$ from the type-I seesaw formula $M_\nu^c = -M_D^c M_R^{c -1} M_D^{c \mathsf{T}}$ where $M_R^c$ is the RH neutrino mass matrix in the charged lepton mass basis.}
	
	\item{Construct the $6 \times 6$ neutrino mass matrix $M_{\nu N}^c$ from $M_D^c$ and $M_R^c$, and find the $6 \times 6$ mixing matrix $V_{\nu N}$ that diagonalizes $M_{\nu N}^c$.}
\end{enumerate}
Here, the $6 \times 6$ neutrino mass matrix $M_{\nu N}^c$ in the charged lepton mass basis is written as
\begin{align}
	M_{\nu N}^c = \left( \begin{array}{cc} 0 & M_D^c \\ M_D^{c \mathsf{T}} & M_R^c \end{array} \right),
\end{align}
and this matrix is diagonalized by the $6 \times 6$ unitary matrix $V_{\nu N}$:
\begin{align}
	M_{\nu N}^\text{diag} = V_{\nu N}^\mathsf{T} M_{\nu N}^c V_{\nu N}
\end{align}
where $M_{\nu N}^\text{diag}$ is the diagonal matrix with positive entries. Following the convention of reference \cite{LRSMLFV}, we write
\begin{align}
	V_{\nu N}^* = \left( \begin{array}{cc} U & S \\ T & V \end{array} \right)
\end{align}
where $U$, $S$, $T$, and $V$ are $3 \times 3$ mixing matrices. Note that $U = U_\text{PMNS}$. The straightforward numerical diagonalization might not work appropriately because of the hierarchy in the components of $M_{\nu N}^c$. Instead, $V_{\nu N}$ is calculated in two steps: 
\begin{align}
	V_{\nu N} = V_{\nu N 1} V_{\nu N 2}
\end{align}
where
\begin{align}
	V_{\nu N 1} = \left( \begin{array}{cc} 1 & -M_D^c M_R^{c -1} \\ -M_R^{c -1} M_D^{c \mathsf{T}} & -1 \end{array} \right), \qquad
	V_{\nu N 2} = \left( \begin{array}{cc} U^* & 0 \\ 0 & -V^* \end{array} \right).
\end{align}
Here, $V_{\nu N 1}$ transforms $M_{\nu N}$ into the block-diagonal matrix
\begin{align}
	M_{\nu N}^\text{BD} = \left( \begin{array}{cc} M_\nu^c & 0 \\ 0 & M_R^c + M_R^{c -1} M_D^{c \mathsf{T}} M_D^c + M_D^{c \mathsf{T}} M_D^c M_R^{c -1} \end{array} \right),
\end{align}
and $V_{\nu N 2}$ is the matrix that diagonalizes $M_{\nu N}^\text{BD}$. In addition, we use the standard parametrization of the PMNS matrix:
\begin{align}
	U_\text{PMNS} &= \left( \begin{array}{ccc}
		1 & 0 & 0 \\
		0 & \cos{\theta_{23}} & \sin{\theta_{23}} \\
		0 & -\sin{\theta_{23}} & \cos{\theta_{23}} \\
	 \end{array} \right)
	 \left( \begin{array}{ccc}
		\cos{\theta_{13}} & 0 & \sin{\theta_{13}} e^{-i \delta_D} \\
		0 & 1 & 0 \\
		-\sin{\theta_{13}} e^{i \delta_D} & 0 & \cos{\theta_{13}} \\
	 \end{array} \right)
	 \left( \begin{array}{ccc}
		\cos{\theta_{12}} & \sin{\theta_{12}} & 0 \\
		-\sin{\theta_{12}} & \cos{\theta_{12}} & 0 \\
		0 & 0 & 1 \\
	 \end{array} \right) \nonumber \\
	 &\qquad \qquad \times \left( \begin{array}{ccc}
		1 & 0 & 0 \\
		0 & e^{-i \delta_{M1}} & 0 \\
		0 & 0 & e^{-i \delta_{M2}} \\
	 \end{array} \right)
\end{align}
where $\delta_D$ and $\delta_{Mi}$ are Dirac and Majorana CP phases, respectively. On the other hand, we parametrize $V_L^\ell$ and $V_R^\ell$ as
\begin{align}
	V = V_1 V_2 V_3
\end{align}
where
\begin{align}
	V_1 &= \left( \begin{array}{ccc}
		1 & 0 & 0 \\
		0 & e^{-i \delta_2} & 0 \\
		0 & 0 & e^{-i \delta_3} \\
	 \end{array} \right), \\
	V_2 &= \left( \begin{array}{ccc}
		1 & 0 & 0 \\
		0 & \cos{\theta_{23}} & \sin{\theta_{23}} \\
		0 & -\sin{\theta_{23}} & \cos{\theta_{23}} \\
	 \end{array} \right)
	 \left( \begin{array}{ccc}
		\cos{\theta_{13}} & 0 & \sin{\theta_{13}} e^{-i \delta_1} \\
		0 & 1 & 0 \\
		-\sin{\theta_{13}} e^{i \delta_1} & 0 & \cos{\theta_{13}} \\
	 \end{array} \right)
	 \left( \begin{array}{ccc}
		\cos{\theta_{12}} & \sin{\theta_{12}} & 0 \\
		-\sin{\theta_{12}} & \cos{\theta_{12}} & 0 \\
		0 & 0 & 1 \\
	 \end{array} \right), \\
	V_3 &= \left( \begin{array}{ccc}
		e^{-i \delta_4} & 0 & 0 \\
		0 & e^{-i \delta_5} & 0 \\
		0 & 0 & e^{-i \delta_6} \\
	 \end{array} \right).
\end{align}
Note that it is always possible to absorb $V_{R3}^\ell$ into $V_{L3}^\ell$ since $M_\ell = V^\ell_L M_\ell^c V^{\ell \dagger}_R$ where $M_\ell^c$ is a diagonal matrix. We can therefore write
\begin{align}
	V_L^\ell = V_{L1}^\ell V_{L2}^\ell V_{L3}^\ell, \qquad
	V_R^\ell = V_{R1}^\ell V_{R2}^\ell.
\end{align}
In addition, the Hermitian matrix $A~(\equiv f \kappa_2 / \sqrt{2})$ is parametrized as
\begin{align}
	A = \left( \begin{array}{ccc}
		A_{11} & |A_{12}| e^{i \theta_{A_{12}}} & |A_{13}| e^{i \theta_{A_{13}}} \\
		|A_{12}| e^{-i \theta_{A_{12}}} & A_{22} & |A_{23}| e^{i \theta_{A_{23}}} \\
		|A_{13}| e^{-i \theta_{A_{13}}} & |A_{23}| e^{-i \theta_{A_{23}}} & A_{33} \\
	 \end{array} \right)
\end{align}
where $A_{ii}$ are real numbers. The list of model parameters and the ranges where they are randomly generated are summarized in table \ref{tab:parameters}. Several appropriate constraints are imposed on some model parameters, and they are presented in table \ref{tab:constraints}. 
\begin{table}[ht]
	\center
	\begin{tabular}{|l||l|}
		\hline
		Parameter & Range \\ \hline
		$\log_{10}{(m_{\nu_1} / \text{eV})}$ & $\minus 4 - \log_{10}{2}$ \\ \hline
		$m_{W_R}$ & $2 - 35$ TeV \\ \hline
		$\log_{10}{(\kappa_2 / \text{GeV})}$ & $\minus 4 - 1$ \\ \hline
		\makecell[l]{$\delta_D$, $\delta_{M1}$, $\delta_{M2}$, \\
			$\theta_{L12}$, $\theta_{L13}$, $\theta_{L23}$, \\
			$\delta_{L1}$, $\delta_{L2}$, $\delta_{L3}$}
			& $\minus \pi - \pi$ rad \\ \hline
		$\delta_{L4}$ & ($\minus 1 - 1$)$ \cdot 10^{-3}$ rad \\ \hline
		$\log_{10}{(|A_{11}| / \text{GeV})}$ & $\log_{10}{\big| \text{Im}[M_{\ell 11}] \big|} - \log_{10}{\big( 5 \sqrt{2\pi} v_\text{EW} \big)}$ \\ \hline
		$\log_{10}{\alpha_3}$, $\log_{10}{\rho_2}$ & $\log_{10}{(1000~\text{GeV}^2 / v_R^2)} - \log_{10}{(5 \sqrt{4\pi})}$ \\ \hline
		$\log_{10}{(\rho_3 - 2 \rho_1)}$ & $\log_{10}{(1000~\text{GeV}^2 / v_R^2)} - \log_{10}{(15 \sqrt{4\pi})}$ \\ \hline
	\end{tabular}
	\caption{List of parameters and the ranges where those parameters are randomly generated. It is also assumed that $\delta_{L5} = \delta_{L6} = 0$, $\theta_{Rij} = \theta_{Lij}$, and $\delta_{Ri} = \delta_{Li}$ ($i,j = 1,2,3$). Here, $A$ is defined as $A \equiv f \kappa_2 / \sqrt{2}$, and $M_\ell =  V_L^\ell M_\ell^c V_R^{\ell \dagger}$ is the charged lepton mass matrix in the symmetry basis. The electroweak VEV is $v_\text{EW} = \sqrt{\kappa_1^2 + \kappa_2^2} = 246$ GeV, and $v_R = m_{W_R} \sqrt{2} / g$ $(g = 0.65)$ is the VEV of the SU(2)$_R$ triplet. Since Yukawa coupling matrices $f$, $\tilde{f}$ are constructed from given $M_\ell$ by the method presented in section \ref{sec:Mmass}, we explicitly consider only the condition $\kappa_1 \gg \kappa_2$ for the TeV-scale MLRSM. Any Yukawa couplings that do not satisfy $f_{ij} \ll \tilde{f}_{ij}$ can be excluded by filtering $M_R$ with large entries, which is one of the constraints given in table \ref{tab:constraints}. The ranges and values of $\delta_{L4}$, $\delta_{L5}$, $\delta_{L6}$, $\theta_{Rij}$, and $\delta_{Ri}$ are chosen to guarantee $V_R^\ell \approx V_L^\ell$ for TeV-scale $m_N$. In principle, we only need $\delta_{L4} \approx 0$, $\delta_{L5} \approx 0$, $\delta_{L6} \approx 0$, $\theta_{Rij} \approx \theta_{Lij}$, and $\delta_{Ri} \approx \delta_{Li}$ for $V_R^\ell \approx V_L^\ell$. However, for the parameters other than $\delta_{L4}$, it turned out that only extremely small deviations ($\lesssim 10^{-6}$) from the values assumed above are allowed to obtain TeV-scale $m_N$. Therefore, for convenience, only $\delta_{L4}$ is varied around 0 while all the other parameters are set to the fixed values mentioned above. The coupling constants $\alpha_3$, $\rho_2$, $\rho_3 - 2 \rho_1$ are assumed to be positive, which is a sufficient condition to have real masses of charged scalar fields. Note that slightly broader ranges than necessary are chosen for several parameters, in order to generate contour plots less distorted around the borders.}
	\label{tab:parameters}
	\center
	\begin{tabular}{|l||l|}
		\hline
		Parameter & Constraint \\ \hline
		$m_{H_1^+}$, $m_{H_2^+}$, $m_{\delta_L^{++}}$, $m_{\delta_R^{++}}$ & $>$ 500 GeV \\ \hline
		$|$Eigenvalues of $f$, $\tilde{f}, h|$, $\alpha_3$, $\rho_2$ & $< \sqrt{4\pi}$ \\ \hline
		$\rho_3 - 2 \rho_1$ & $< 3 \sqrt{4\pi}$ \\ \hline
		$|$Eigenvalues of $M_D|$ & $> 1$ keV \\ \hline
		$|$Eigenvalues of $M_R|$ & $100~\text{GeV} - \sqrt{8\pi} v_R$ \\ \hline
	\end{tabular}
	\caption{List of constraints imposed on several model parameters. The lower limits of scalar field masses are set to 500 GeV to safely neglect many loop diagrams by those charged scalar fields. Note that the upper limits of all the coupling constants are set to $\sqrt{4\pi}$. The lower limit of the eigenvalues of $M_D$ is appropriately chosen to avoid singularity in calculating $M_D^{-1}$. The constraint from the absence of the flavour changing neutral current in the quark sector requires $m_{H_1^0}, m_{H_2^+} \gtrsim 10$ TeV \cite{LRSMCP, LRSML}, which is not considered in this paper because the contribution of $H_2^+$ to CLFV is almost negligible, as shown in figures \ref{fig:mWR30vsmH2p}. The constraint from the SM Higgs mass $m_{h^0} = 125$ GeV is not explicitly considered as well, because we can always find $\lambda_1, \alpha_1$ that would give the correct Higgs mass for given $\rho_1, \alpha_3$ if $\epsilon_2 \lesssim 0.01$ and $m_{W_R} < 30$ TeV. The condition $\epsilon_2 \lesssim 0.01$ is found to be satisfied for all the data points due to the perturbativity constraint, as shown in figure \ref{fig:LPetnvseps2}.}
	\label{tab:constraints}
\end{table}

\subsection{Numerical results}
\begin{figure}[htp]
	\centering
	\subfloat[$\text{BR}_{\tau \to \mu \gamma}$ vs.~$\text{BR}_{\mu \to e \gamma}$]{
		\includegraphics[width = 0.3 \textwidth]{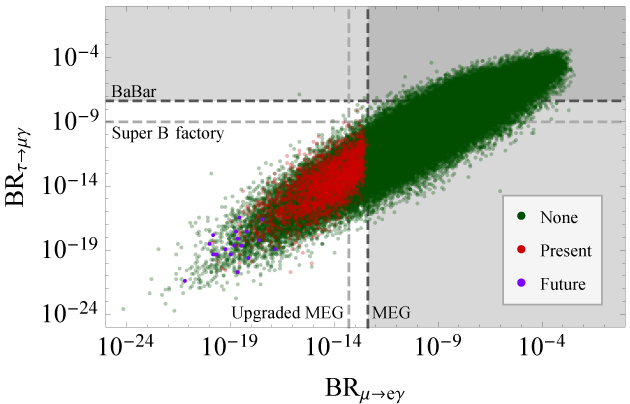}
		\label{fig:LPBRtmgvsBRmeg}
	}
	\subfloat[$\text{BR}_{\tau \to e \gamma}$ vs.~$\text{BR}_{\mu \to e \gamma}$]{
		\includegraphics[width = 0.3 \textwidth]{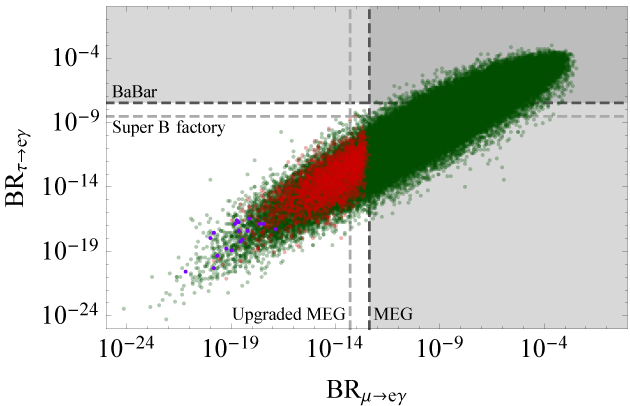}
		\label{fig:LPBRtegvsBRmeg}
	}
	\subfloat[$\text{BR}_{\mu \to eee}^\text{type-I}$ vs.~$\text{BR}_{\mu \to eee}^\text{tree}$]{
		\includegraphics[width = 0.3 \textwidth]{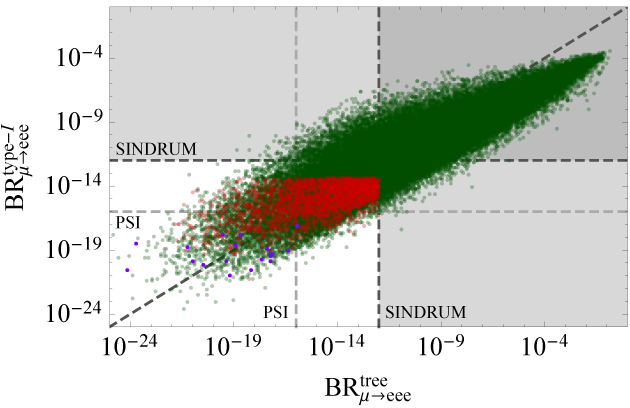}
		\label{fig:LPBRm3evsBRm3e}
	} \\
	\subfloat[$\text{BR}_{\mu \to eee}$ vs.~$\text{BR}_{\mu \to e \gamma}$]{
		\includegraphics[width = 0.3 \textwidth]{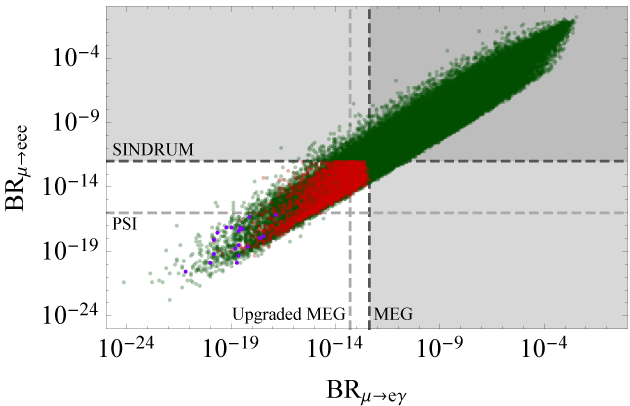}
		\label{fig:LPBRm3evsBRmeg}
	}
	\subfloat[$\text{BR}_{\mu \to eee}$ vs.~$\text{R}_{\mu \to e}^\text{Ti}$]{
		\includegraphics[width = 0.3 \textwidth]{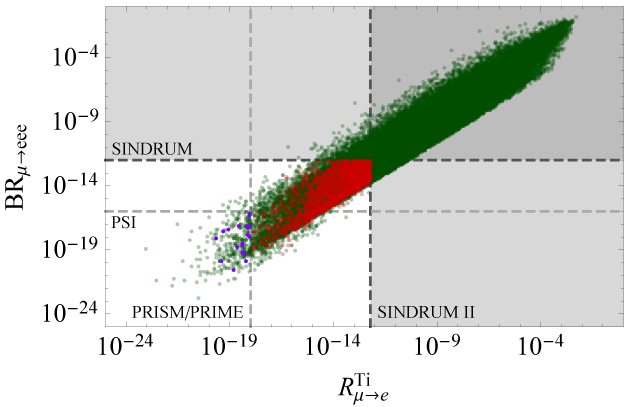}
		\label{fig:LPBRm3evsRTi}
	}
	\subfloat[$\text{R}_{\mu \to e}^\text{Ti}$ vs.~$\text{BR}_{\mu \to e \gamma}$]{
		\includegraphics[width = 0.3 \textwidth]{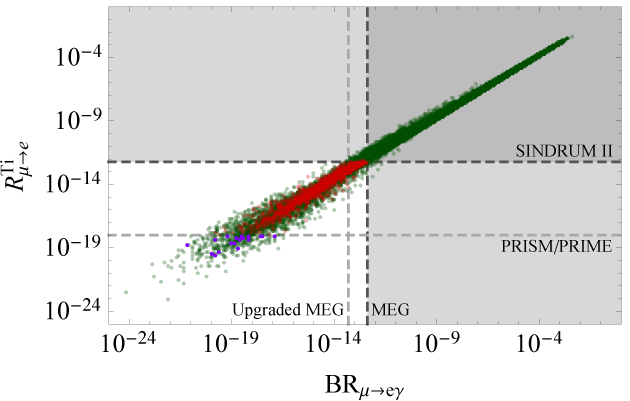}
		\label{fig:LPRTivsBRmeg}
	} \\
	\subfloat[$\text{R}_{\mu \to e}^\text{Al}$ vs.~$\text{R}_{\mu \to e}^\text{Ti}$]{
		\includegraphics[width = 0.3 \textwidth]{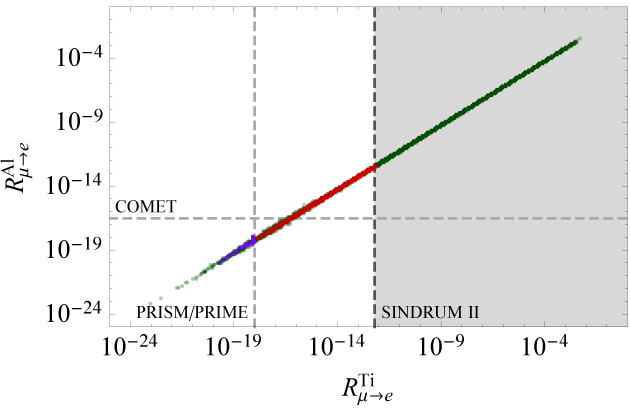}
		\label{fig:LPRAlvsRTi}
	}
	\subfloat[$\text{R}_{\mu \to e}^\text{Au}$ vs.~$\text{R}_{\mu \to e}^\text{Ti}$]{
		\includegraphics[width = 0.3 \textwidth]{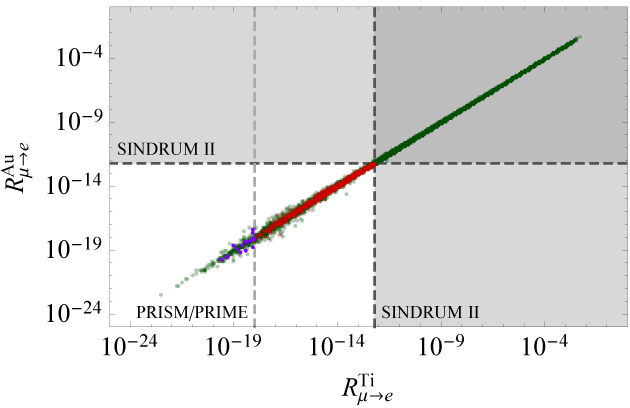}
		\label{fig:LPRAuvsRTi}
	}
	\subfloat[$\text{R}_{\mu \to e}^\text{Pb}$ vs.~$\text{R}_{\mu \to e}^\text{Ti}$]{
		\includegraphics[width = 0.3 \textwidth]{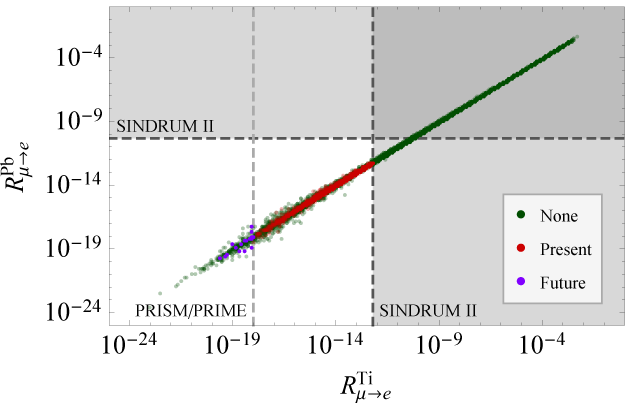}
		\label{fig:LPRPbvsRTi}
	}
	\caption{CLFV in the MLRSM for 2 TeV $< m_{W_R} <$ 30 TeV.
	The green dots are data points that satisfy only the experimental constraints from the light lepton masses and PMNS matrix. The red dots are data points that also satisfy present bounds from the CLFV, $0 \nu \beta \beta$, EDM's of charged leptons, and Planck observation. The purple dots are those that satisfy the strongest bounds from future experiments. The shaded regions are regions of parameter space excluded by present experimental bounds.
	Figures \ref{fig:LPBRtmgvsBRmeg} and \ref{fig:LPBRtegvsBRmeg} show that there exist only small chances that $\tau \to \mu \gamma$ or $\tau \to e \gamma$ could be detected in near-future experiments.
	In figure \ref{fig:LPBRm3evsBRm3e}, the tree-level and 1-loop contributions to $\mu \to eee$ are compared, and it shows that we should consider both when calculating $\text{BR}_{\mu \to eee}$.
	Figures \ref{fig:LPBRm3evsBRmeg}$-$\ref{fig:LPRTivsBRmeg} show the linear correlations among various CLFV effects.
	Note that the strongest future bounds on CLFV come from PRISM/PRIME and PSI, as clearly shown in figure \ref{fig:LPBRm3evsRTi}.
	Figures \ref{fig:LPRAlvsRTi}$-$\ref{fig:LPRPbvsRTi} show that the $\mu \to e$ conversion rates for various nuclei have very strong linear correlations with each other.
	The total number of data points is 83724 (total) = 81132 (green) + 2573 (red) + 19 (purple).}
	\label{fig:CLFV}
\end{figure}

\begin{figure}[htp]
	\centering
	\subfloat[$T_{1/2}^{0 \nu} \big|_\text{Ge}^\text{max}$ vs.~$|\eta_\nu|$]{
		\includegraphics[width = 0.3 \textwidth]{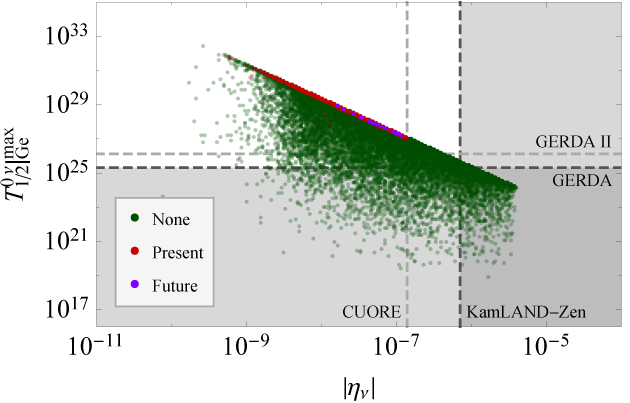}
		\label{fig:LPTGevsetn}
	}
	\subfloat[$T_{1/2}^{0 \nu} \big|_\text{Te}^\text{max}$ vs.~$|\eta_\nu|$]{
		\includegraphics[width = 0.3 \textwidth]{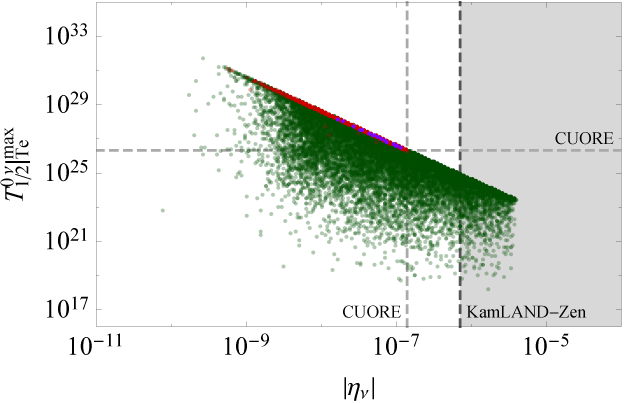}
		\label{fig:LPTTevsetn}
	}
	\subfloat[$T_{1/2}^{0 \nu} \big|_\text{Xe}^\text{max}$ vs.~$|\eta_\nu|$]{
		\includegraphics[width = 0.3 \textwidth]{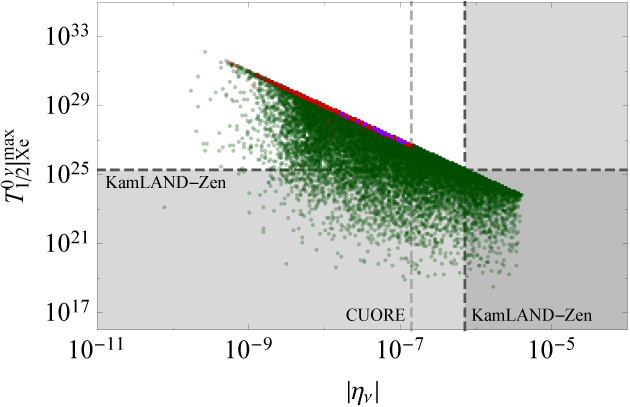}
		\label{fig:LPTXevsetn}
	} \\
	\subfloat[$|\eta_\nu|$ vs.~$|\eta^R_{N_R}|$]{
		\includegraphics[width = 0.3 \textwidth]{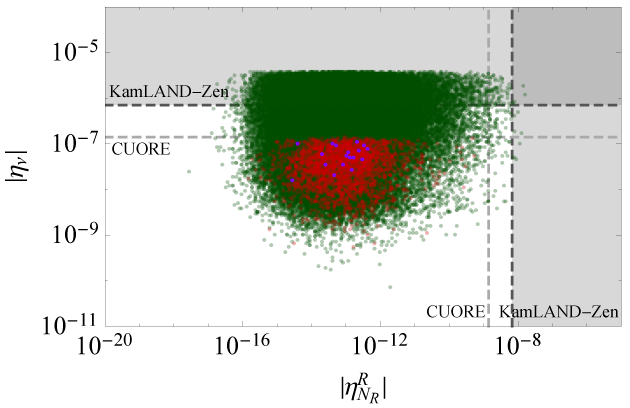}
		\label{fig:LPetnvsetRNR}
	}
	\subfloat[$|\eta_\nu|$ vs.~$|\eta_{\delta_R}|$]{
		\includegraphics[width = 0.3 \textwidth]{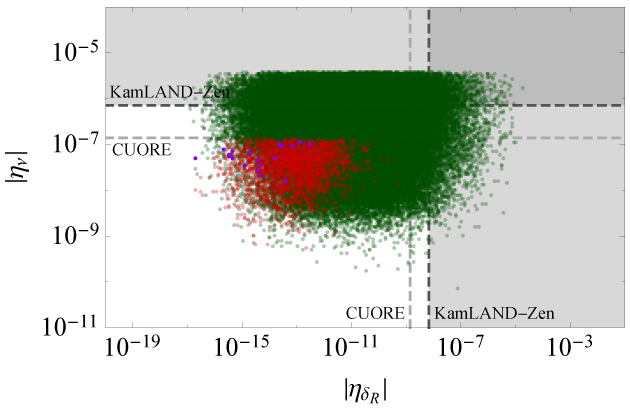}
		\label{fig:LPetnvsetdR}
	}
	\subfloat[$|\eta_{\delta_R}|$ vs.~$|\eta^R_{N_R}|$]{
		\includegraphics[width = 0.3 \textwidth]{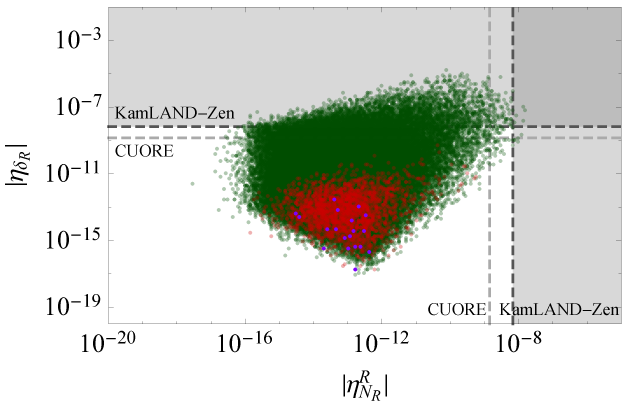}
		\label{fig:LPetdRvsetRNR}
	} \\
	\subfloat[$|\eta^L_{N_R}|$ vs.~$|\eta^R_{N_R}|$]{
		\includegraphics[width = 0.3 \textwidth]{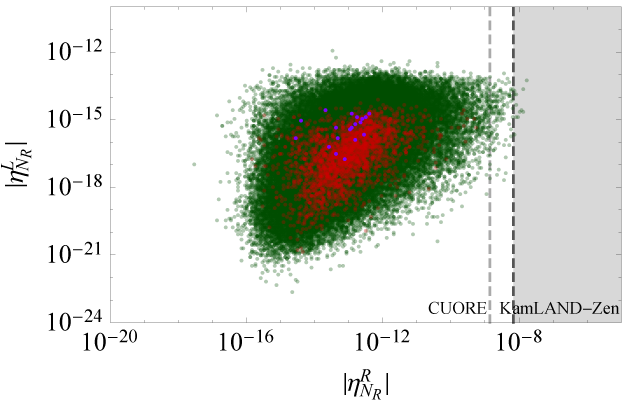}
		\label{fig:LPetLNRvsetRNR}
	}
	\subfloat[$|\eta_\eta|$ vs.~$|\eta_\lambda|$]{
		\includegraphics[width = 0.3 \textwidth]{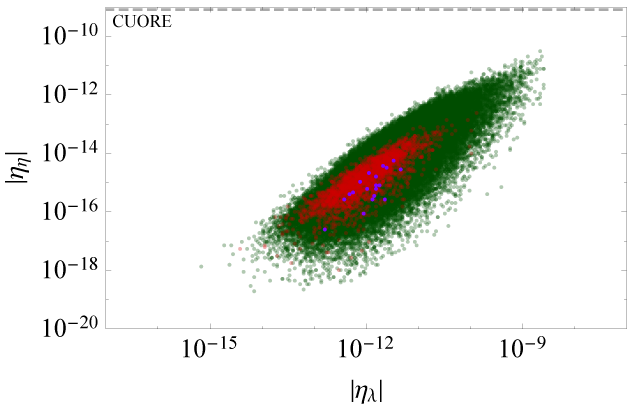}
		\label{fig:LPetetvsetl}
	}
	\subfloat[$|\eta_\nu|$ vs.~$m_{\nu_1}$]{
		\includegraphics[width = 0.3 \textwidth]{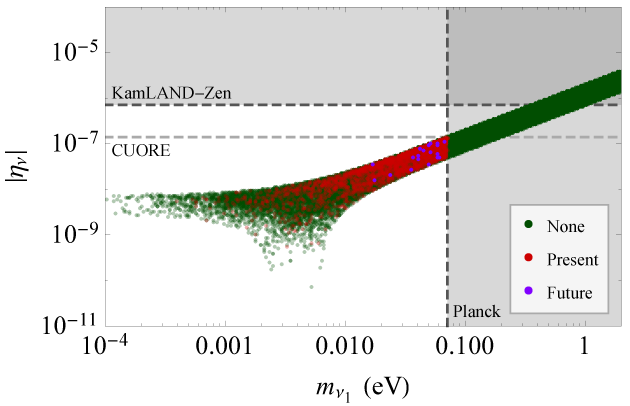}
		\label{fig:LPetnvsmnl}
	}
	\caption{Parameters of $0 \nu \beta \beta$ in the MLRSM for 2 TeV $< m_{W_R} <$ 30 TeV.
	Figures \ref{fig:LPTGevsetn}$-$\ref{fig:LPTXevsetn} show that only cases where $\eta_\nu$ dominantly determines $T_{1/2}^{0 \nu} \big|^\text{max}$ are allowed with a few exceptions by the present and future experimental bounds. Even though the contributions of $\eta^R_{N_R}$ and $\eta_{\delta_R}$ could be comparable to that of $\eta_\nu$ in principle, such cases have been actually almost excluded by the constraints from CLFV, as shown in figures \ref{fig:LPetnvsetRNR}$-$\ref{fig:LPetdRvsetRNR}.
	The contributions from $\eta_\eta$ or $\eta_\lambda$ are too small compared with experimental bounds, as shown in figure \ref{fig:LPetetvsetl}.
	Figure \ref{fig:LPetnvsmnl} shows that the present upper bound of the light Majorana neutrino mass from Planck is already below the bounds from KamLAND-Zen and CUORE, which means that $0 \nu \beta \beta$ processes are difficult to be detected in near-future experiments since the light neutrino exchange diagrams are dominant for most of the parameter space due to the CLFV constraints.}
	\label{fig:0nbb}
\end{figure}

\begin{figure}[ht]
	\centering
	\subfloat[$|d_\mu|$ vs.~$|d_e|$]{
		\includegraphics[width = 0.3 \textwidth]{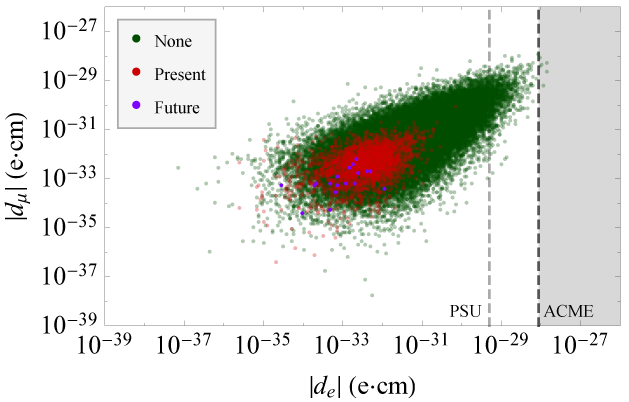}
		\label{fig:LPdmvsde}
	}
	\subfloat[$|d_\tau|$ vs.~$|d_e|$]{
		\includegraphics[width = 0.3 \textwidth]{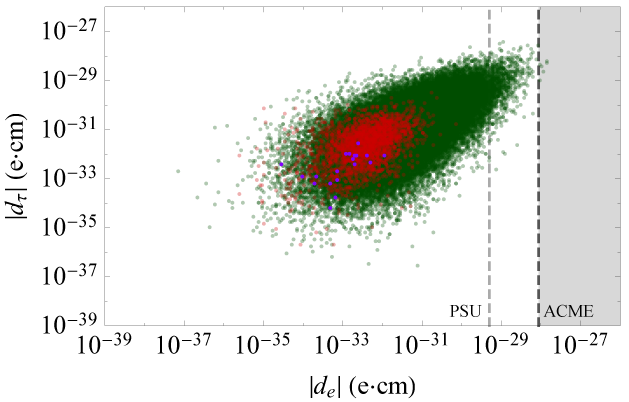}
		\label{fig:LPdtvsde}
	}
	\subfloat[$|d_e|$ vs.~$\text{R}_{\mu \to e}^\text{Ti}$]{
		\includegraphics[width = 0.3 \textwidth]{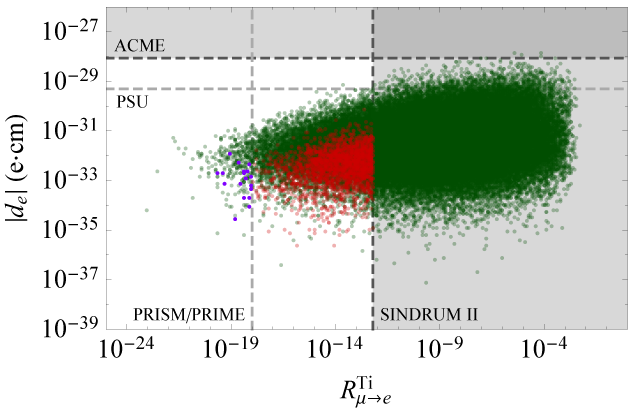}
		\label{fig:LPdevsRTi}
	}
	\caption{EDM's of charged leptons in the MLRSM for 2 TeV $< m_{W_R} <$ 30 TeV.
	The predicted values are found to be too small compared with the present and future bounds, since large EDM's require small $m_{W_R}$ whose regions of parameter space have been largely constrained as shown in figure \ref{fig:LPRTivsmWR}.
	Even though the correlations between EDM's and CLFV are rather weak, as shown in figure \ref{fig:LPdevsRTi}, the larger EDM's generally require the larger CLFV effects since $m_{W_R}$ affects both CLFV and EDM's.}
	\label{fig:EDM}
\end{figure}

\begin{figure}[ht]
	\centering
	\subfloat[$\text{R}_{\mu \to e}^\text{Ti}$ vs.~$m_{W_R}$]{
		\includegraphics[width = 0.3 \textwidth]{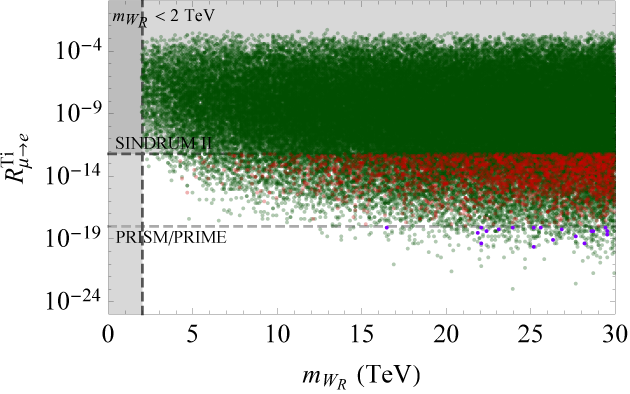}
		\label{fig:LPRTivsmWR}
	}
	\subfloat[$\text{R}_{\mu \to e}^\text{Ti}$ vs.~$m_{N_1}$]{
		\includegraphics[width = 0.3 \textwidth]{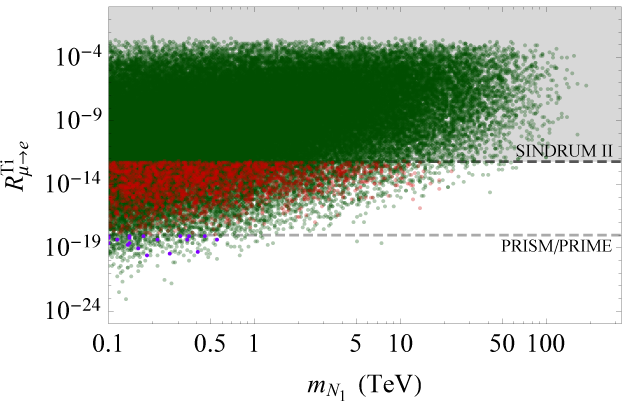}
		\label{fig:LPRTivsmN1}
	}
	\subfloat[$\text{R}_{\mu \to e}^\text{Ti}$ vs.~$m_{N_3}$]{
		\includegraphics[width = 0.3 \textwidth]{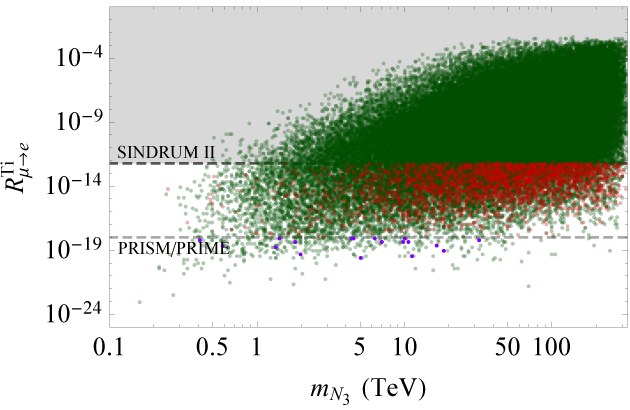}
		\label{fig:LPRTivsmN3}
	} \\
	\subfloat[$\text{R}_{\mu \to e}^\text{Ti}$ vs.~$m_{\nu_1}$]{
		\includegraphics[width = 0.3 \textwidth]{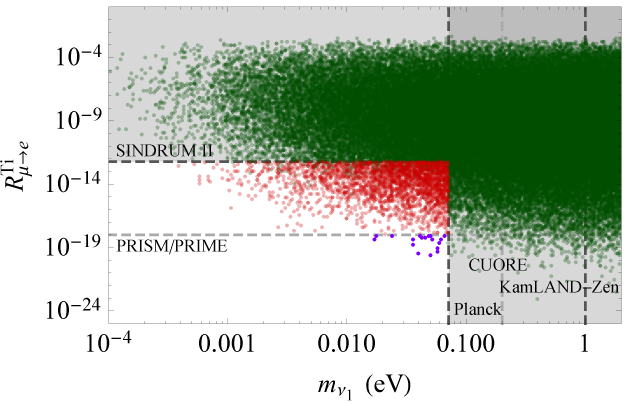}
		\label{fig:LPRTivsmnl}
	}
	\subfloat[$\text{BR}_{\mu \to e \gamma}$ vs.~$m_{\nu_1}$]{
		\includegraphics[width = 0.3 \textwidth]{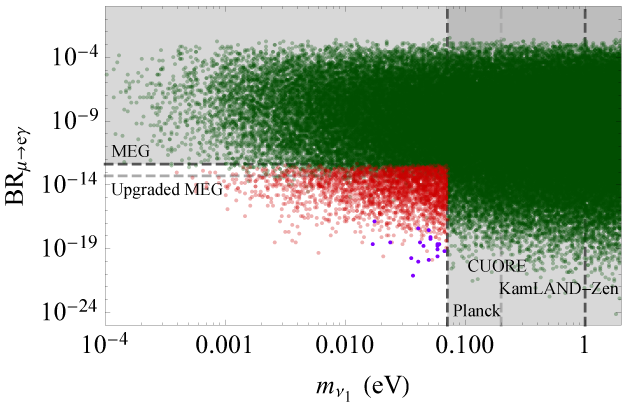}
		\label{fig:LPBRmegvsmnl}
	}
	\subfloat[$|\eta_\nu|$ vs.~$\epsilon_2$ $(\equiv \kappa_2 / \kappa_1)$]{
		\includegraphics[width = 0.3 \textwidth]{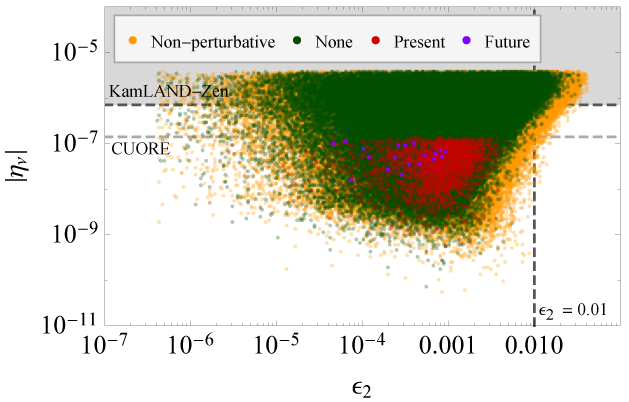}
		\label{fig:LPetnvseps2}
	}
	\caption{Figures \ref{fig:LPRTivsmWR}$-$\ref{fig:LPBRmegvsmnl} show the effect of CLFV constraints on the masses of neutrinos and the RH gauge boson. Here, $\text{R}_{\mu \to e}^\text{Ti}$ is chosen since it most clearly divides the colors of data points through its experimental bounds.
	The smaller values of the lightest light neutrino mass $m_{\nu_1}$ produce the larger CLFV effects, as in figures \ref{fig:LPRTivsmnl} and \ref{fig:LPBRmegvsmnl}, since they require the larger values of the heaviest heavy neutrino mass $m_{N_3}$ in most of the parameter space, as shown in figure \ref{fig:mN3vsmnlC}. As a result, the regions of parameter space of small light neutrino masses get constrained by the experimental bounds on CLFV.
	In figure \ref{fig:LPetnvseps2}, additional data points (yellow dots) are also presented in order to show the effects of the perturvativity constraints, and all the data points generated in the ranges of parameters given in table \ref{tab:parameters} are shown in this plot. For those yellow points, at least one of the coupling constants are larger than $\sqrt{4\pi}$ while the experimental constraints in the light neutrino sector are still satisfied. This figure shows that $\epsilon_2 \equiv \kappa_2 / \kappa_1 \lesssim 0.01$ is satisfied for all the data points due to the perturvativity constraints as well as the condition $\kappa_2 < 10$ GeV, and thus the Higgs mass constraint can be easily satisfied, as mentioned in table \ref{tab:constraints}.}
	\label{fig:LPvsM}
\end{figure}

\begin{figure}[ht]
	\centering
	\subfloat[$m_{W_R}$ vs.~$m_{N_1}$]{
		\includegraphics[width = 0.3 \textwidth]{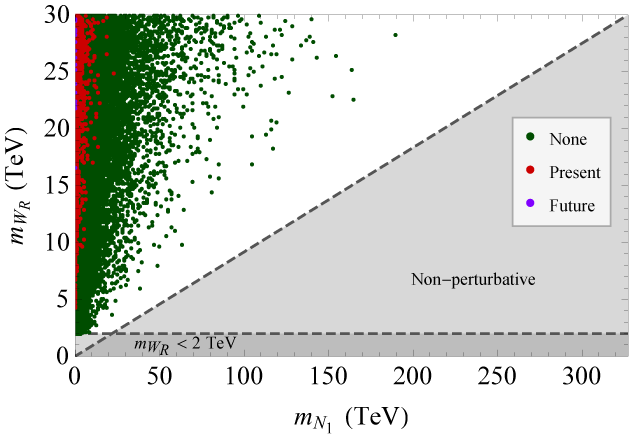}
		\label{fig:mWR30vsmN1CL}
	}
	\subfloat[$m_{W_R}$ vs.~$m_{N_1}$]{
		\includegraphics[width = 0.3 \textwidth]{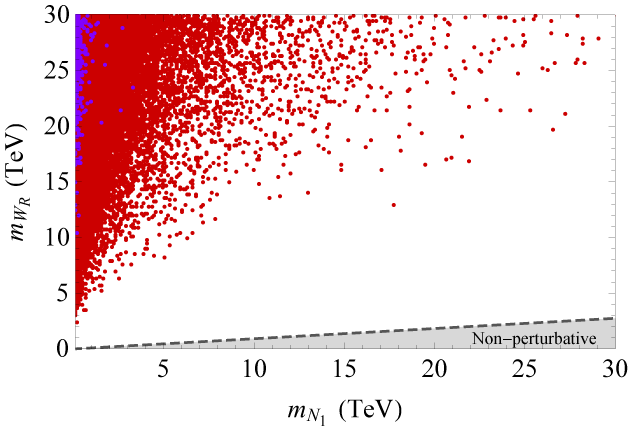}
		\label{fig:mWR30vsmN1L}
	}
	\subfloat[$m_{W_R}$ vs.~$m_{N_3}$]{
		\includegraphics[width = 0.3 \textwidth]{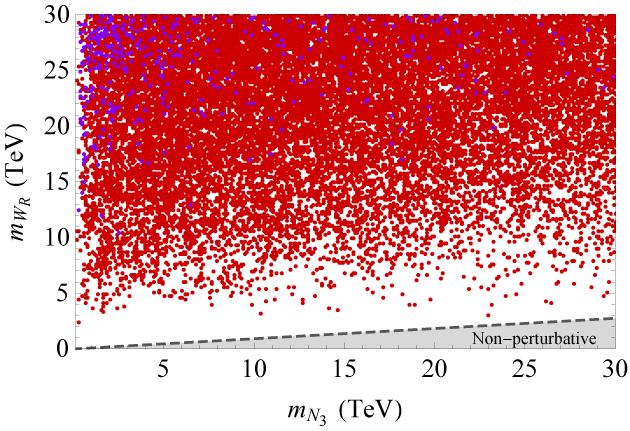}
		\label{fig:mWR30vsmN3L}
	}
	\caption{Masses of heavy neutrinos in the TeV-scale MLRSM for 2 TeV $< m_{W_R} <$ 30 TeV.
	For figure \ref{fig:mWR30vsmN1CL}, the same data set as in the previous plots are used to show the effect of the consraints from CLFV, $0 \nu \beta \beta$, EDM's, and Planck on the parameter space.
	The non-perturbative regions are where at least one coupling constant is larger than $\sqrt{4\pi}$. Note that green dots in figure \ref{fig:mWR30vsmN1CL} do not completely fill the available parameter space because of the constraints on masses and angles in the light lepton sector.
	For figures \ref{fig:mWR30vsmN1L} and \ref{fig:mWR30vsmN3L}, much more amount of data points was used to show how the present and future bounds constrain the parameter space.
	Figures \ref{fig:mWR30vsmN1CL} and \ref{fig:mWR30vsmN1L} show that the lightest heavy neutrino mass $m_{N_1}$ has been notably constrained by the experimental bounds, especially for smaller $m_{W_R}$.
	Figure \ref{fig:mWR30vsmN3L} is the plot on the heaviest heavy neutrino mass $m_{N_3}$, and it shows that only a small region of parameter space with small $m_{W_R}$ seems to have been excluded.
	Even though these plots in the linear scale are better in presenting the effect of experimental constraints on largest possible masses of heavy neutrinos, they do not correctly show the density distributions since the matrix $A~(\equiv f \kappa_2 / \sqrt{2})$ is generated in the logarithmic scale. Plots of $m_N$ in the logarithmic scale are presented in figure \ref{fig:mWR30}.
	For figures \ref{fig:mWR30vsmN1L} and \ref{fig:mWR30vsmN3L}, the data sets for figures \ref{fig:mWR30vsmN1} and \ref{fig:mWR30vsmN3} are used, respectively.}
	\label{fig:mWRvsmNL}
\end{figure}

\begin{figure}[ht]
	\centering
	\subfloat[$m_{W_R}$ vs.~$m_{N_1}$]{
		\includegraphics[width = 0.3 \textwidth]{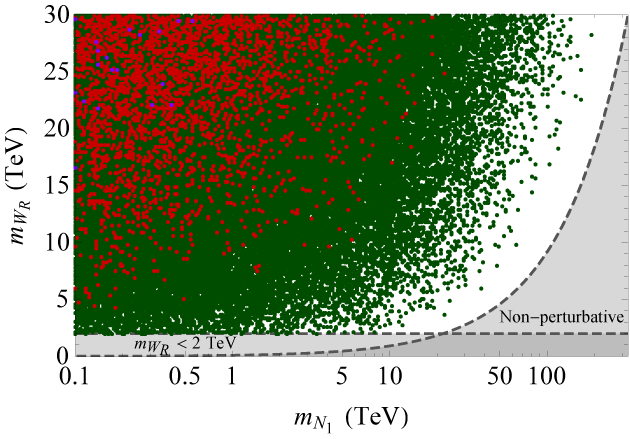}
		\label{fig:mWR30vsmN1C}
	}
	\subfloat[$m_{W_R}$ vs.~$m_{N_3}$]{
		\includegraphics[width = 0.3 \textwidth]{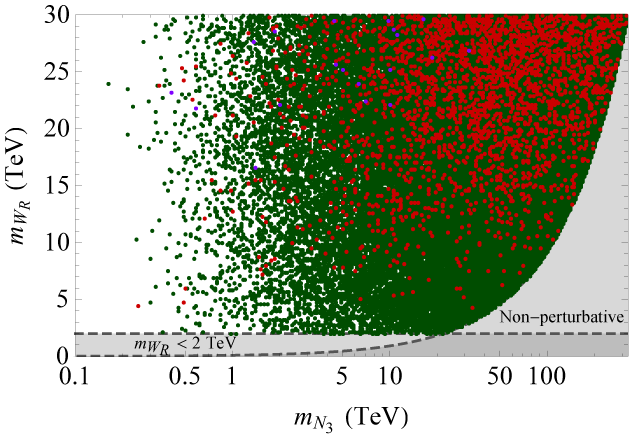}
		\label{fig:mWR30vsmN3C}
	}
	\subfloat[$m_{N_3}$ vs.~$m_{N_1}$]{
		\includegraphics[width = 0.3 \textwidth]{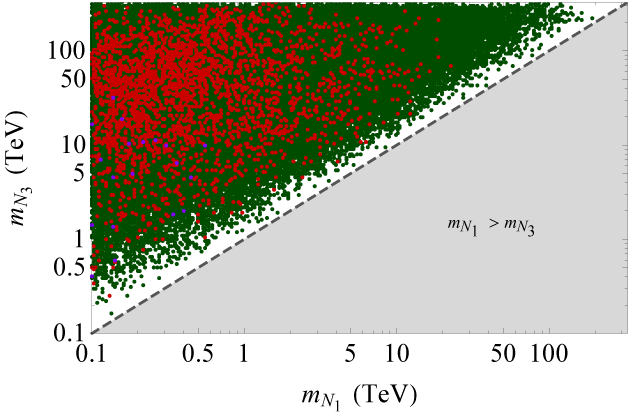}
		\label{fig:mN3vsmN1C}
	} \\
	\subfloat[$m_{W_R}$ vs.~$m_{\nu_1}$]{
		\includegraphics[width = 0.3 \textwidth]{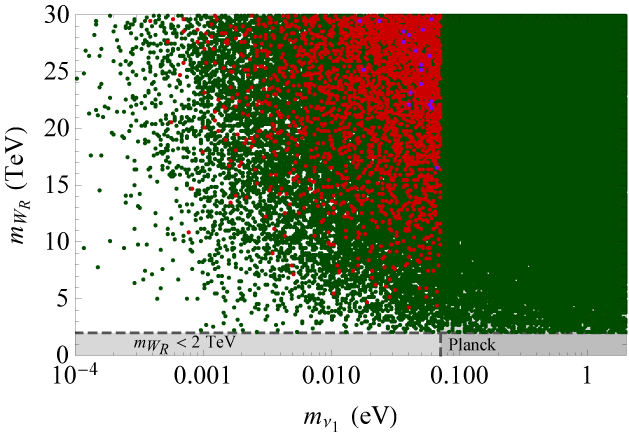}
		\label{fig:mWRvsmnlC}
	}
		\subfloat[$m_{N_1}$ vs.~$m_{\nu_1}$]{
		\includegraphics[width = 0.3 \textwidth]{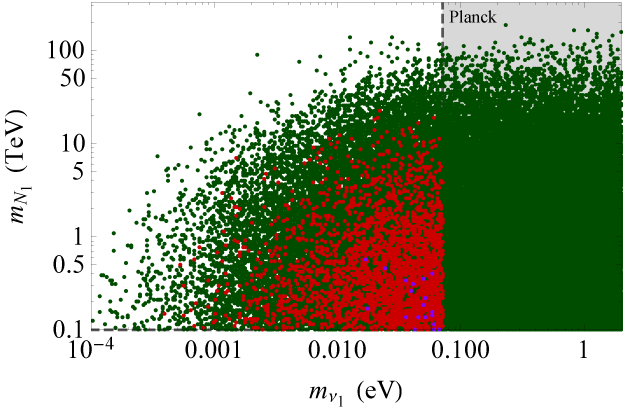}
		\label{fig:mN1vsmnlC}
	}
	\subfloat[$m_{N_3}$ vs.~$m_{\nu_1}$]{
		\includegraphics[width = 0.3 \textwidth]{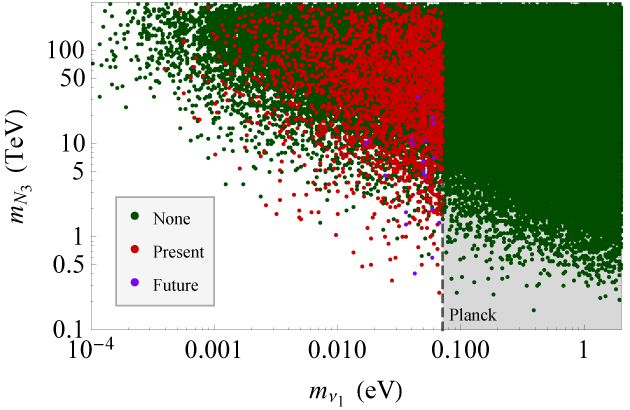}
		\label{fig:mN3vsmnlC}
	}
	\caption{Figures \ref{fig:mWR30vsmN1C}$-$\ref{fig:mWRvsmnlC} show the effect of experimental bounds on the masses of neutrinos and the RH gauge boson.
	Figures \ref{fig:mWR30vsmN1C} and \ref{fig:mWR30vsmN3C} show that the regions with smaller $m_{W_R}$ and larger $m_N$ are more affected by the present bounds on CLFV, $0 \nu \beta \beta$, and EDM's.
	Figures \ref{fig:mN1vsmnlC} and \ref{fig:mN3vsmnlC} show that, for smaller $m_{\nu_1}$, i.e.~for the light neutrino masses with a larger hierarchy, the heavy neutrino masses also generally need to have a larger hierarchy accordingly since $M_D$ itself does not have the structure that would give hierarchical light neutrino masses. Due to this effect, only larger $m_{W_R}$ is generally allowed for smaller $m_{\nu_1}$, as shown in figure \ref{fig:mWRvsmnl}, since large $m_{N_3}$ requires large $v_R$.}
	\label{fig:Mcorr}
\end{figure}

\begin{figure}[htp]
	\centering
	\subfloat[$m_{W_R}$ vs.~$m_{\nu_1}$]{
		\includegraphics[width = 0.3 \textwidth]{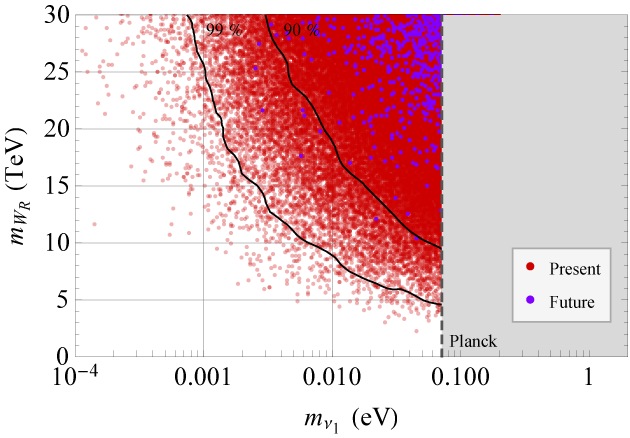}
		\label{fig:mWRvsmnl}
	}
	\subfloat[$m_{W_R}$ vs.~$m_{N_1}$]{
		\includegraphics[width = 0.3 \textwidth]{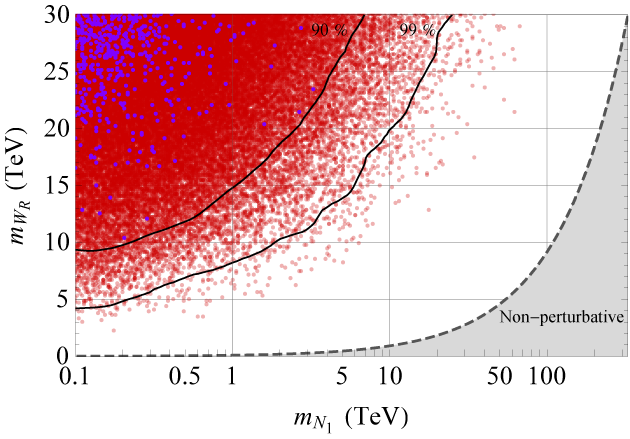}
		\label{fig:mWR30vsmN1}
	}
	\subfloat[$m_{W_R}$ vs.~$m_{N_3}$]{
		\includegraphics[width = 0.3 \textwidth]{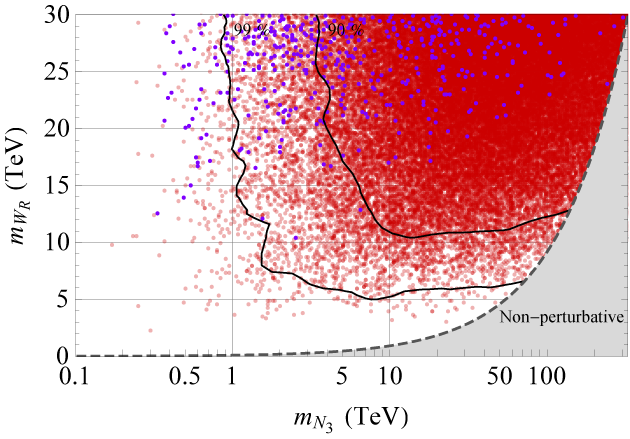}
		\label{fig:mWR30vsmN3}
	} \\
	\subfloat[$m_{W_R}$ vs.~$m_{H_1^+}$ $\big(\approx m_{\delta_L^{++}} \big)$]{
		\includegraphics[width = 0.3 \textwidth]{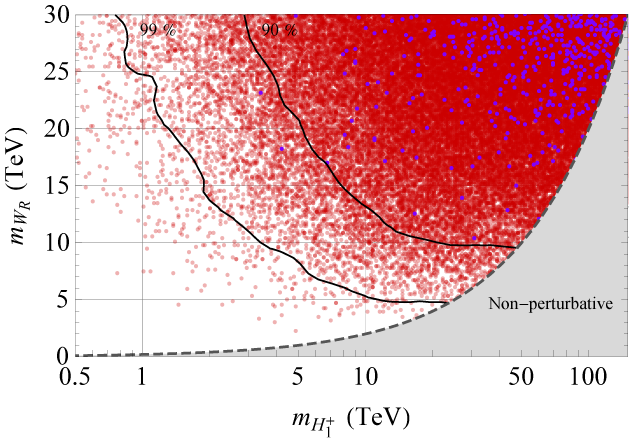}
		\label{fig:mWR30vsmH1p}
	}
	\subfloat[$m_{W_R}$ vs.~$m_{H_2^+}$]{
		\includegraphics[width = 0.3 \textwidth]{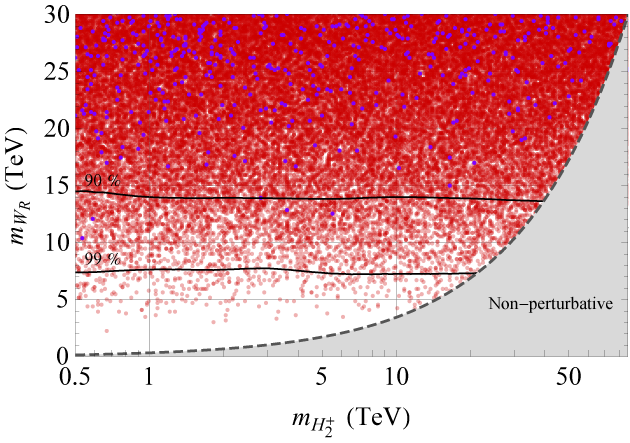}
		\label{fig:mWR30vsmH2p}
	}
	\subfloat[$m_{W_R}$ vs.~$m_{\delta_R^{++}}$]{
		\includegraphics[width = 0.3 \textwidth]{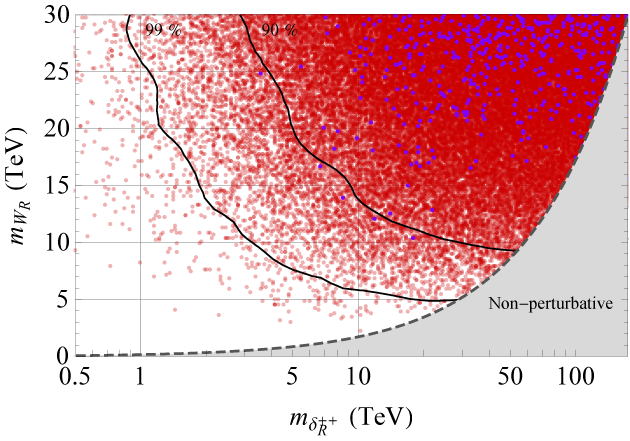}
		\label{fig:mWR30vsmdRpp}
	}
	\caption{Masses of neutrinos and charged scalar fields in the MLRSM for $m_{W_R} < 30$ TeV. The contours of 90 \% and 99 \% densities are also presented for illustration purposes. According to the 99 \% contour in figure \ref{fig:mWRvsmnl}, $m_{\nu_1} \sim 0.1$ eV for $m_{W_R} = 5$ TeV and $m_{\nu_1} \gtrsim 6 \cdot 10^{-3}$ eV for $m_{W_R} = 10$ TeV. In addition, the 99 \% contour in figure \ref{fig:mWR30vsmN1} shows that $m_{N_1} \lesssim 200$ GeV for $m_{W_R} < 5$ TeV and $m_{N_1} \lesssim 2$ TeV for $m_{W_R} < 10$ TeV. While the masses of $H_1^+$, $\delta_L^{++}$, and $\delta_R^{++}$ have been also constrained by the experimental bounds, the mass of $H_2^+$ which appears only in the $Z_1$-exchange diagrams of CLFV processes has been barely constrained, as shown in figure \ref{fig:mWR30vsmH2p}. Hence, the constraint of $m_{H_2^+} \gtrsim 10$ TeV from the absence of flavour changing neutral current in the quark sector is not considered in this paper. The total number of data points is 51971 = 51561 (red) + 410 (purple).
}
	\label{fig:mWR30}
\end{figure}

The present and future experimental bounds on CLFV, $0 \nu \beta \beta$, and EDM's of charged leptons are summarized in table \ref{tab:expbounds}. The upper bound of light neutrino masses from the Planck observation is also considered. The experimental bounds on the dimensionless parameters associated with the various processes of $0 \nu \beta \beta$ are given in table \ref{tab:exp0nbbDP}.
\begin{table}[ht]
	\center
	\begin{tabular}{|l||l|l|}
		\hline
		& Present bound & Future sensitivity \\ \hline
		BR$_{\mu \to e \gamma}$ & $< 4.2 \cdot 10^{-13}$ (MEG) \cite{MEG} & $< 5.0 \cdot 10^{-14}$ (Upgraded MEG) \cite{MEG (Sawada)} \\ \hline
		BR$_{\tau \to \mu \gamma}$ & $< 4.4 \cdot 10^{-8}$ (BaBar) \cite{BaBar} & $< 1.0 \cdot 10^{-9}$ (Super B factory) \cite{SBf} \\ \hline
		BR$_{\tau \to e \gamma}$ & $< 3.3 \cdot 10^{-8}$ (BaBar) \cite{BaBar} & $< 3.0 \cdot 10^{-9}$ (Super B factory) \cite{SBf} \\ \hline
		BR$_{\mu \to eee}$ & $< 1.0 \cdot 10^{-12}$ (SINDRUM) \cite{SINDRUM} & $< 1.0 \cdot 10^{-16}$ (PSI) \cite{PSI} \\ \hline
		R$_{\mu \to e}^\text{Al}$ & \makecell{$\cdot$} & $< 3.0 \cdot 10^{-17}$ (COMET) \cite{COMET} \\ \hline
		R$_{\mu \to e}^\text{Ti}$ & $< 6.1 \cdot 10^{-13}$ (SINDRUM II) \cite{SINDRUM II (RTi)} & $< 1.0 \cdot 10^{-18}$ (PRISM/PRIME) \cite{PRISM} \\ \hline
		R$_{\mu \to e}^\text{Au}$ & $< 6.0 \cdot 10^{-13}$ (SINDRUM II) \cite{COMET} & \makecell{$\cdot$} \\ \hline
		R$_{\mu \to e}^\text{Pb}$ & $< 4.6 \cdot 10^{-11}$ (SINDRUM II) \cite{SINDRUM II (RPb)} & \makecell{$\cdot$} \\ \hline
		$T_{1/2}^{0 \nu} \big|_\text{Ge}$ & $> 2.1 \cdot 10^{25}$ yrs. (GERDA) \cite{0nbb} & $> 1.35 \cdot 10^{26}$ yrs. (GERDA II) \cite{0nbb} \\ \hline
		$T_{1/2}^{0 \nu} \big|_\text{Te}$ & \makecell{$\cdot$} & $> 2.1 \cdot 10^{26}$ yrs. (CUORE) \cite{0nbb} \\ \hline
		$T_{1/2}^{0 \nu} \big|_\text{Xe}$ & $> 1.9 \cdot 10^{25}$ yrs. (KamLAND-Zen) \cite{0nbb} & \makecell{$\cdot$} \\ \hline
		$|d_e|$ & $< 8.7 \cdot 10^{-29}~e \cdot$cm (ACME) \cite{ACME} & $< 5.0 \cdot 10^{-30}~e \cdot$cm (PSU) \cite{PSU} \\ \hline
		$|d_\mu|$ & $< 1.9 \cdot 10^{-19}~e \cdot$cm (Muon $(g-2)$) \cite{Mu(g-2)} & \makecell{$\cdot$} \\ \hline
		$|d_\tau|$ & $\lesssim 5.0 \cdot 10^{-17}~e \cdot$cm (Belle) \cite{Belle} & \makecell{$\cdot$} \\ \hline
		$\sum_i^3 m_{\nu_i}$ & $< 0.23$ eV (Planck) \cite{Planck} & \makecell{$\cdot$} \\ \hline
	\end{tabular}
	\caption{Experimental bounds on CLFV, $0 \nu \beta \beta$, EDM's of charged leptons, and light neutrino masses. The actual present bounds on $d_\tau$ reported by Belle Collaboration are $-2.2 \cdot 10^{-17} e \cdot$cm $< \text{Re}[d_\tau] < 4.5 \cdot 10^{-17} e \cdot$cm and $-2.5 \cdot 10^{-17} e \cdot$cm $< \text{Im}[d_\tau] < 0.8 \cdot 10^{-17} e \cdot$cm. For the normal hierarchy, the constraint from the Planck observation corresponds to the upper bound of the lightest neutrino mass $m_{\nu_1} < 0.071$ eV.}
	\label{tab:expbounds}
	\center
	\begin{tabular}{|l||l|l|}
		\hline
		& Present bound (KamLAND-Zen) & Future sensitivity (CUORE) \\ \hline
		$|\eta_\nu|$ & $< 7.1 \cdot 10^{-7}$ & $< 1.4 \cdot 10^{-7}$ \\ \hline
		$|\eta^L_{N_R}|$ & $< 6.8 \cdot 10^{-9}$ & $< 1.4 \cdot 10^{-9}$ \\ \hline
		$|\eta^R_{N_R}|$ & $< 6.8 \cdot 10^{-9}$ & $< 1.4 \cdot 10^{-9}$ \\ \hline
		$|\eta_{\delta_R}|$ & $< 6.8 \cdot 10^{-9}$ & $< 1.4 \cdot 10^{-9}$  \\ \hline
		$|\eta_\lambda|$ & $< 5.7 \cdot 10^{-7}$ & $< 1.2 \cdot 10^{-7}$  \\ \hline
		$|\eta_\eta|$ & $< 3.0 \cdot 10^{-9}$ & $< 8.2 \cdot 10^{-10}$  \\ \hline
	\end{tabular}
	\caption{Experimental bounds on the dimensionless parameters associated with the various processes of $0 \nu \beta \beta$.
	The present bounds come from KamLAND-Zen, and the strongest future bounds are from CUORE \cite{0nbb}. To obtain each bound, the associated decay channel is assumed to be dominant over the others. Even though there exist regions of parameter space where contributions from $\eta_\nu$, $\eta^R_{N_R}$, and $\eta_{\delta_R}$ are comparable to each other, it does not invalidate the assumption at least for the data points of interest around the present and future bounds, since larger values of $|\eta^R_{N_R}|$ and $|\eta_{\delta_R}|$ are rarely allowed by the constraints from CLFV, as shown in figures \ref{fig:LPetnvsetRNR}$-$\ref{fig:LPetdRvsetRNR}.}
	\label{tab:exp0nbbDP}
\end{table}
The numerical results are presented in figures \ref{fig:CLFV}$-$\ref{fig:mWR30}. The plots on the various branching ratios and conversion rates of CLFV in the MLRSM for 2 TeV $< m_{W_R} <$ 30 TeV are given in figure \ref{fig:CLFV}. The results on the dimensionless parameters of $0 \nu \beta \beta$ for the same range of $m_{W_R}$ are presented in figure \ref{fig:0nbb}. The plots on the EDM's of charged leptons are presented in figure \ref{fig:EDM}. The effect of experimental constraints on the masses of the RH gauge boson, neutrinos, and scalar fields are shown in figures \ref{fig:LPvsM}$-$\ref{fig:mWR30}. The benchmark model parameters and their predicitons are given in appendix \ref{sec:BM}.

The most notable result is that the regions of parameter space that allow small light neutrino masses are largely constrained by the experimental bounds from CLFV as well as the constraints from the light neutrino mass and mixing angles. Since the type-I seesaw formula implies det($M_\nu$) $\approx$ det($M_D$)$^2$/det($M_R$), we need a hierarchy in the eigenvalues of $M_D$ or $M_R$ when light neutrino masses have a hierarchy. However, $M_D$ is determined from Yukawa couplings and VEV's, and it generally does not have the appropriate hierarchy in its eigenvalues to give hierachical light neutrino masses for most of the available parameter space. In other words, we generally need a hierarchy in the eigenvalues of $M_R$, i.e.~in the heavy neutrino masses as well, in order to obtain hierachical light neutrino masses. Since we are considering a range of $m_N$, i.e.~0.1 TeV $\lesssim m_N \lesssim$ 100 TeV, the cases of large hierarchies in light neutrino masses are supposed to get constrained accordingly. Furthermore, since the regions of parameter space with large $m_N$ are largely affected by the experimental constraints from CLFV, small light neutrino masses are disfavored by all those experimental constraints. These results are all clearly presented in several plots in figures \ref{fig:LPvsM}, \ref{fig:Mcorr}, and \ref{fig:mWR30}. For example, the 99 \% contour in figure \ref{fig:mWRvsmnl} shows that $m_{\nu_1} \sim 0.1$ eV for $m_{W_R} = 5$ TeV and $m_{\nu_1} \gtrsim 6 \cdot 10^{-3}$ eV for $m_{W_R} = 10$ TeV. Note that this does not necessarily mean that there exists a strict lower bound of the light neutrino mass for given $m_{W_R}$, since the results of this paper are based on the naturalness argument such as no fine-tuning in $M_D$. Note also that we can observe similar patterns in neutrino mass correlations in any type-I seesaw models, even in the simple extension of the SM only with gauge singlet neutrinos. The difference in the MLRSM, or in a more general class of the left-right symmetric model, is that we can have large CLFV effects and thus the experimental bounds on CLFV are constraining the light neutrino masses. Moreover, since the largest possible hierarchy in heavy neutrino masses is directly associated with $m_{W_R}$ and the regions of parameter space with smaller $m_{W_R}$ are more constrained by CLFV bounds, we can expect that the discovery of light $W_R$ as well as any improved experimental bounds on CLFV would largely constrain the regions of parameter space of the normal hierarchy.

Another interesting result is that the mass of the lightest heavy neutrino $m_{N_1}$ has been also notably constrained by the present experimental constraints, which is, of course, associated with the result on light neutrino masses just mentioned. This is shown in figures \ref{fig:mWR30vsmN1CL}, \ref{fig:mWR30vsmN1L}, \ref{fig:mWR30vsmN1C}, and \ref{fig:mWR30vsmN1}. For example, the 99 \% density contour of figure \ref{fig:mWR30vsmN1} shows that $m_{N_1} \lesssim 200$ GeV for $m_{W_R} = 5$ TeV and $m_{N_1} \lesssim 2$ TeV for $m_{W_R} = 10$ TeV. Due to the mass insertion in the Dirac propagators of heavy neutrinos in some CLFV processes, large heavy neutrino masses generally induce large CLFV effects. Figure \ref{fig:LPRTivsmN1} explicitly shows how the CLFV bound is constraining $m_{N_1}$. The heaviest heavy neutrino mass is also affected by the experimental bounds, although its effect is rather small, as shown in figures \ref{fig:mWR30vsmN3L}, \ref{fig:mWR30vsmN3C}, and \ref{fig:mWR30vsmN3}.

While the CLFV effects of muons could be large enough for the associated processes to be detected in near-future experiments, the branching ratios of tau decays are either too small or just around the sensitivities of future experiments, as shown figure \ref{fig:CLFV}. The experimental bounds of CLFV are also constraining small masses of charged scalar fields as well as the RH gauge boson, as shown in figure \ref{fig:mWR30}. As a result, the $0 \nu \beta \beta$ processes through the heavy neutrinos as well as RH gauge boson (denoted by $\eta^R_{N_R}$) and also processes through $\delta_R^{++}$ as well as the RH gauge boson (denoted by $\eta_{\delta_R}$) are both suppressed. Hence, for most data points that satisfy the present experimental constraints, the dominant contribution to $0 \nu \beta \beta$ comes from the process of the light neutrino exchange (denoted by $\eta_\nu$), as shown in figures \ref{fig:LPTGevsetn}$-$\ref{fig:LPTXevsetn}. However, since the upper bound of the light neutrino mass by Planck is already below the bounds of future experiments as shown in figure \ref{fig:LPetnvsmnl}, i.e.~the light neutrino exchange channel has been largely constrained by the Planck observation, the possibility to detect $0 \nu \beta \beta$ processes in near-future experiments is small. As for the EDM's of electrons, there seems to be also only small chances that they could be detected in near-future experiments as shown in figure \ref{fig:EDM}, since the largest possible EDM's of electrons are well below the future sensitivities of the planned experiement. In addition, the EDM's of muons and taus are too small compared with the present upper bounds. Note that the EDM's of charged leptons has been also constrained by the experimental bounds from CLFV, since large EDM's generally require small $m_{W_R}$ and large $m_N$ and such regions of parameter space are largely affected by those experimental constraints. Note also that, even with the relatively small values of the RH scale, i.e.~$v_R < 65$ TeV corresponding to $m_{W_R} < 30$ TeV, the observables of CLFV, $0 \nu \beta \beta$, and EDM's cover very wide ranges, e.g.~roughly $10^{-20} \lesssim \text{BR}_{\mu \to e \gamma} \lesssim 10^{-3}$ and $10^{-35}~e \cdot \text{cm} \lesssim |d_e| \lesssim 10^{-29}~e \cdot \text{cm}$. Hence, neither a success nor a failure in detecting one of these effects rules out even the TeV-scale MLRSM, unless any other experimental results are simultaneously considered.

\section{Conclusion}
In this paper, the procedure to construct lepton mass matrices is presented in the MLRSM of type-I dominance with the parity symmetry, and the conditions for the TeV-scale MLRSM without fine-tuning are also discussed, i.e.~either (i) $\kappa_1 \gg \kappa_2$ and $f_{ij} \ll \tilde{f}_{ij}$, which implies $V^\ell_L \approx V^\ell_R$, or (ii) $\kappa_1 \ll \kappa_2$ and $f_{ij} \gg \tilde{f}_{ij}$, which implies $V^\ell_L \approx V^\ell_R e^{-i \alpha}$. Based on these results, the phenomenology of the TeV-scale MLRSM is numerically investigated when the masses of light neutrinos are in the normal hierarchy, and the numerical results on how the present and future experimental bounds from the CLFV, $0 \nu \beta \beta$, EDM's of charged leptons, and Planck observation constrain the parameter space of the MLRSM are presented.

According to the numerical results, the regions of parameter space of small light neutrino masses have been constrained by the experimental bounds on CLFV effects, although it does not necessarily mean there exists a strict lower bound of light neutrino masses. The lightest heavy neutrino mass is also found to have been notably constrained by the present experimental bounds especially for small $m_{W_R}$. In addition, it has been shown that all the $0 \nu \beta \beta$ processes and the EDM's of charged leptons have been suppressed by the experimental constraints from CLFV, and we have at best only small chances to detect any of these effects in near-future experiments.

Note that the results of this paper are based on several nontrivial assumptions such as (i) type-I seesaw dominance, (ii) the parity symmetry, and (iii) the normal hierarchy in light neutrino masses. Furthermore, it should be emphasized that this paper is considering the TeV-scale phenomenology of the MLRSM without fine-tuning of model parameters. If fine-tuning is allowed, significantly different predictions could be made.


\section*{Acknowledgement}
The author would like to thank Dr.~R.N.~Mohapatra for valuable discussions and encouragement. This work is supported by the National Science Foundation grant NSF-PHY-1620074.

\appendix

\section{Expressions of observables} \label{sec:expObs}
In this paper, the expressions presented in reference \cite{LRSMLFV} are mostly used. The exceptions are the form factors $F^{Z_1}_R$ and $B_{RR}^{\mu eee}$: for $F^{Z_1}_R$, a mixed expression from references \cite{LRSMLFV} and \cite{LRSMQR} is used; for $B_{RR}^{\mu eee}$, the suppression factor $(m_{W_L} / m_{W_R})^2$ is multiplied to the whole expression. The normalized Yukawa couplings $\tilde{h}_L$ and $\tilde{h}_R$ are explicitly distinguished in this paper, since they are generally different even with the manifest left-right symmetry.

\subsection{Charged lepton flavour violation}
The normalized Yukawa couplings $\tilde{h}_L$, $\tilde{h}_R$ in the charged lepton mass basis are given by \cite{LFVnonSUSY}
\begin{align}
	\tilde{h}_L &\equiv \frac{2}{g} V_L^{\ell \mathsf{T}} h_L V_L^\ell
		= \frac{2}{g} V_L^{\ell \mathsf{T}} \frac{M_L^* e^{-i \theta_L}}{\sqrt{2} v_L} V_L^\ell, \\
	\tilde{h}_R &\equiv \frac{2}{g} V_R^{\ell \mathsf{T}} h_R V_R^\ell
		= \frac{2}{g} V_R^{\ell \mathsf{T}} \frac{M_R}{\sqrt{2} v_R} V_R^\ell
		= V_R^{\ell \mathsf{T}} \frac{M_R}{m_{W_R}} V_R^\ell.
\end{align}
Note that $\tilde{h}_L \neq \tilde{h}_R$ in general since $V_L^\ell \neq V_R^\ell$ for nonzero $\alpha$, although $h \equiv h_L = h_R$ with the parity symmetry. The loop functions of CLFV are given in appendix \ref{sec:loopf}.

\subsubsection{$\ell_a \to \ell_b \gamma$}
\begin{figure}[ht]
	\centering
	\subfloat[$G_L^\gamma$]{
		\includegraphics[width = 0.3 \textwidth]{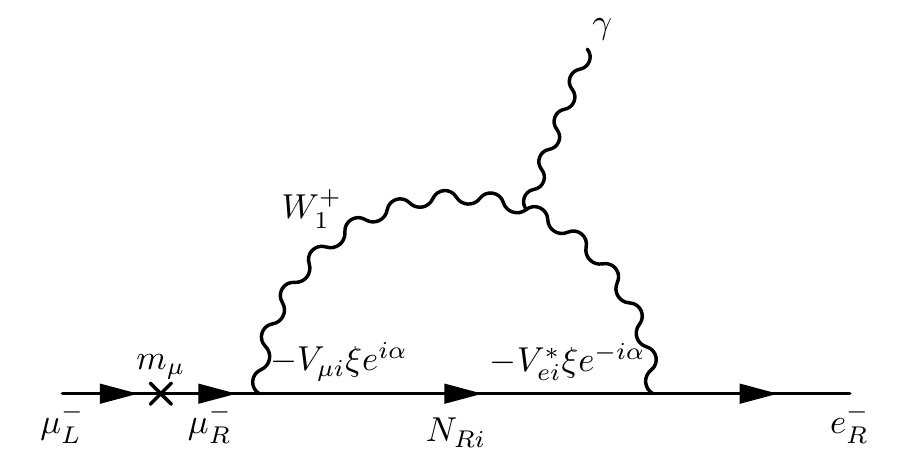}
		\label{fig:meg1}
	}
	\subfloat[$G_L^\gamma$]{
		\includegraphics[width = 0.3 \textwidth]{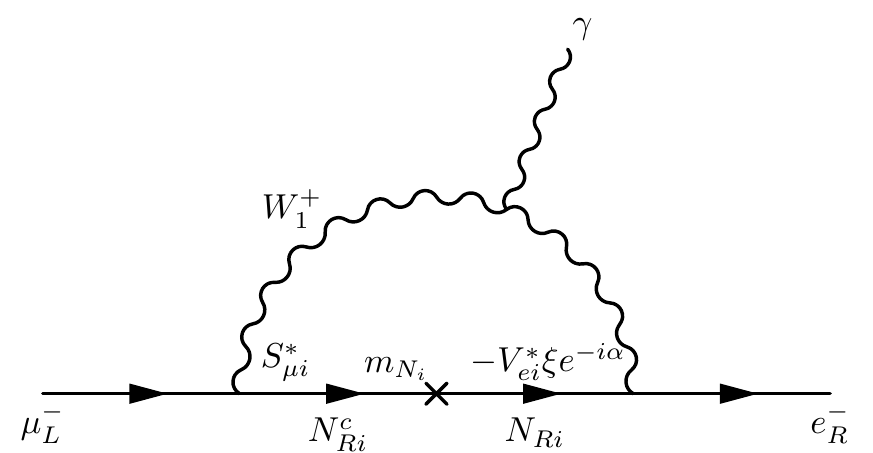}
		\label{fig:meg2}
	}
	\subfloat[$G_L^\gamma$]{
		\includegraphics[width = 0.3 \textwidth]{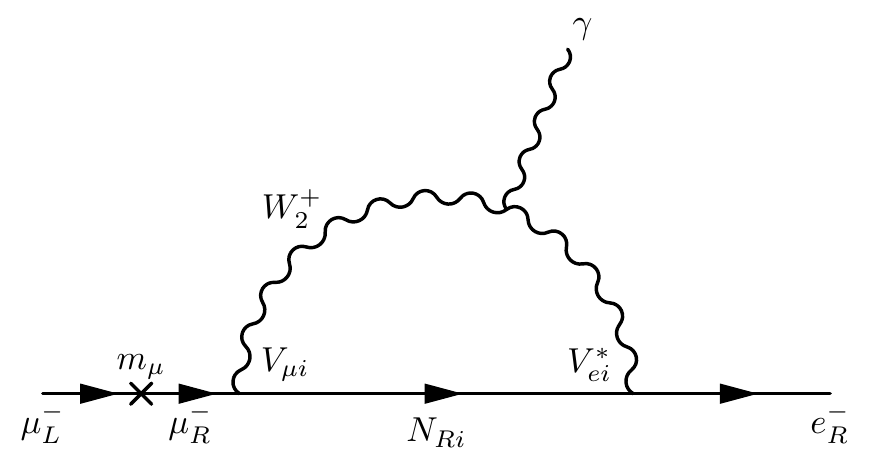}
		\label{fig:meg3}
	} \\
	\subfloat[$G_L^\gamma$]{
		\includegraphics[width = 0.3 \textwidth]{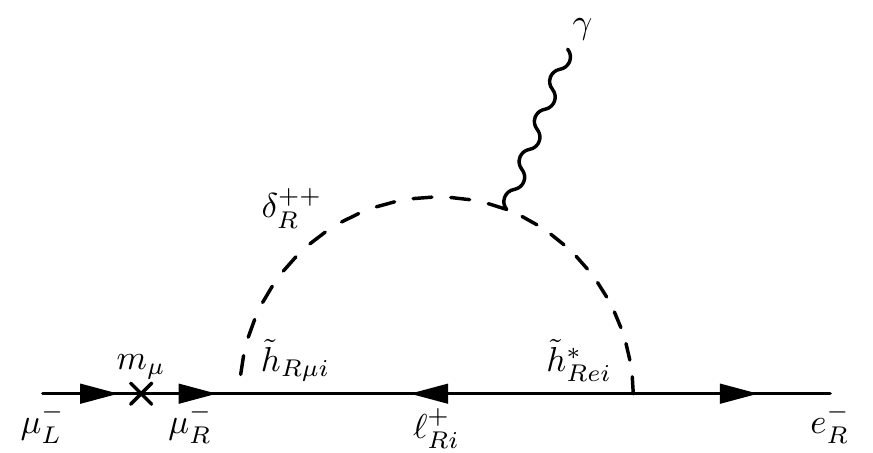}
		\label{fig:meg4}
	}
	\subfloat[$G_L^\gamma$]{
		\includegraphics[width = 0.3 \textwidth]{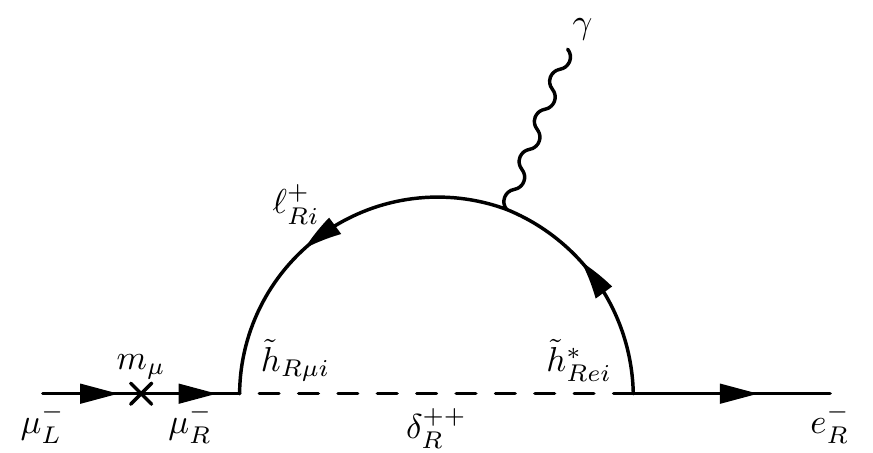}
		\label{fig:meg5}
	}
	\subfloat[$G_R^\gamma$]{
		\includegraphics[width = 0.3 \textwidth]{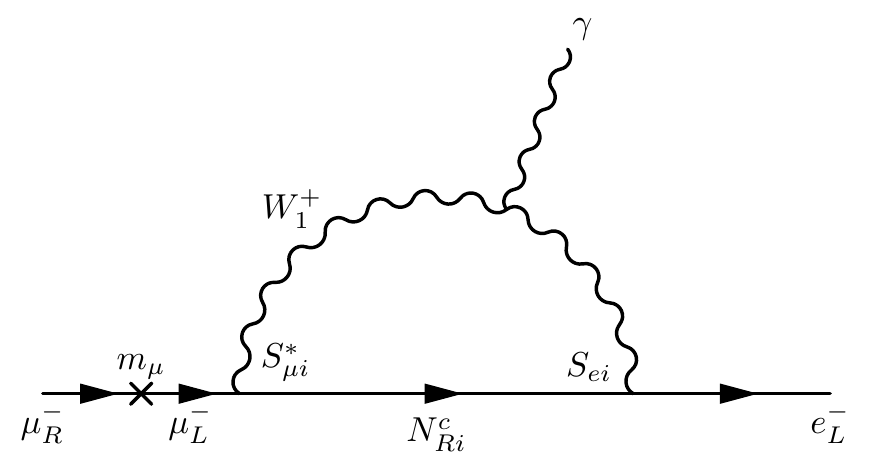}
		\label{fig:meg6}
	} \\
	\subfloat[$G_R^\gamma$]{
		\includegraphics[width = 0.3 \textwidth]{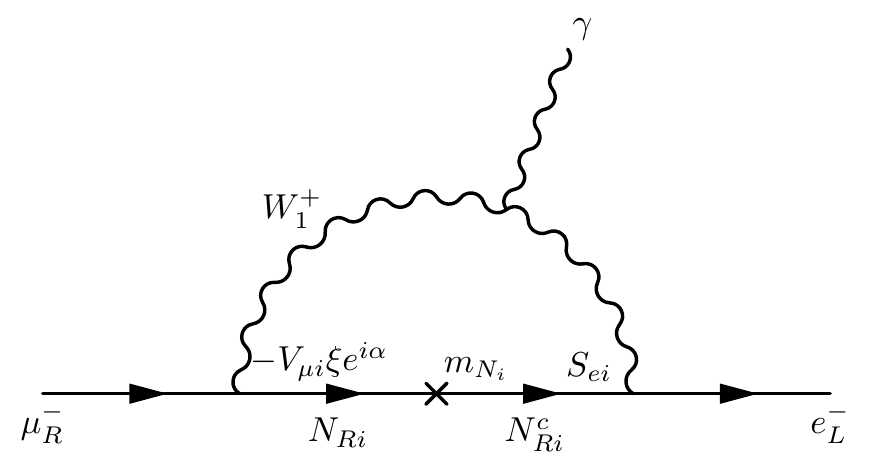}
		\label{fig:meg7}
	}
	\subfloat[$G_R^\gamma$]{
		\includegraphics[width = 0.3 \textwidth]{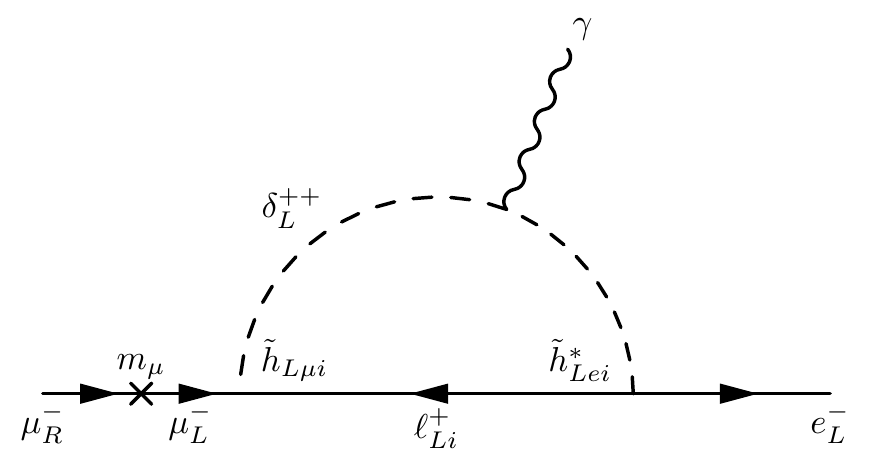}
		\label{fig:meg8}
	}
	\subfloat[$G_R^\gamma$]{
		\includegraphics[width = 0.3 \textwidth]{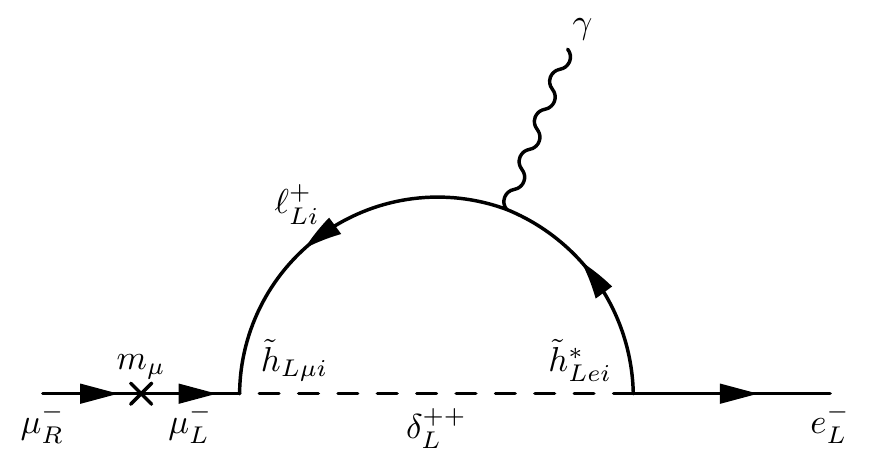}
		\label{fig:meg9}
	}
	\caption{Feynman diagrams of on-shell $\mu \to e \gamma$. Here, $W_L^+ \approx W_1^+ + \xi e^{-i \alpha} W_2^+$ and $W_R^+ \approx -\xi e^{i \alpha} W_1^+ + W_2^+$. Figures \ref{fig:meg1}$-$\ref{fig:meg5} contribute to $G_L^\gamma$, and figures \ref{fig:meg6}$-$\ref{fig:meg9} to $G_R^\gamma$. The arrows in neutrino propagators denote the directions of the propagation of $N_i = N_{Ri} + N_{Ri}^c$.}
	\label{fig:megFD}
\end{figure}

For on-shell decay $\ell_a \to \ell_b \gamma$, the branching ratio is given by
\begin{align}
	\text{BR}_{\ell_a \to \ell_b \gamma} = \frac{\alpha_W^3 s_W^2 m_{\ell_a}^5}{256\pi^2 m_{W_L}^4 \Gamma_{\ell_a}} \big( |G_L^\gamma|^2 + |G_R^\gamma|^2 \big)
\end{align}
where $\alpha_W \equiv g^2 / (4\pi)$, $s_W \equiv \sin{\theta_W}$, and $\Gamma_{\ell_a}$ is the decay rates of $\ell_a$: $\Gamma_\mu = 2.996 \cdot 10^{-19}$ GeV and $\Gamma_\tau = 2.267 \cdot 10^{-12}$ GeV \cite{PDG}. The form factors $G_L^\gamma$, $G_R^\gamma$ are given by
\begin{align}
	G_L^\gamma &= \sum_{i = 1}^3 \bigg[ V_{\mu i} V_{e i}^* \xi^2 G_1^\gamma(x_i) - S_{\mu i}^* V_{e i}^* \xi e^{-i \alpha} G_2^\gamma(x_i) \frac{m_{N_i}}{m_{\ell_a}}
		+ V_{\mu i} V_{e i}^* \frac{m_{W_L}^2}{m_{W_R}^2} G_1^\gamma (y_i) + \tilde{h}_{R \mu i} \tilde{h}_{Rei}^* \frac{2}{3} \frac{m_{W_L}^2}{m_{\delta_R^{++}}^2} \bigg], \\
	G_R^\gamma &= \sum_{i = 1}^3 \bigg[ S_{\mu i}^* S_{e i} G_1^\gamma(x_i) - V_{\mu i} S_{e i} \xi e^{i \alpha} G_2^\gamma(x_i) \frac{m_{N_i}}{m_{\ell_a}}
		+ \tilde{h}_{L \mu i} \tilde{h}_{Lei}^* \bigg( \frac{2}{3} \frac{m_{W_L}^2}{m_{\delta_L^{++}}^2} + \frac{1}{12} \frac{m_{W_L}^2}{m_{H_1^+}^2} \bigg) \bigg]
\end{align}
where $x_i = (m_{N_i} / m_{W_L})^2$ and $y_i = (m_{N_i} / m_{W_R})^2$. The initial and final charged leptons have opposite chiralities, and $L$ or $R$ in $G_{L,R}^\gamma$ denotes the chirality of the initial charged lepton. The Feynman diagrams of on-shell $\mu \to e \gamma$ are given in figure \ref{fig:megFD}.

\subsubsection{$\mu \to eee$}
\begin{figure}[ht]
	\centering
	\subfloat[]{
		\includegraphics[width = 0.3 \textwidth]{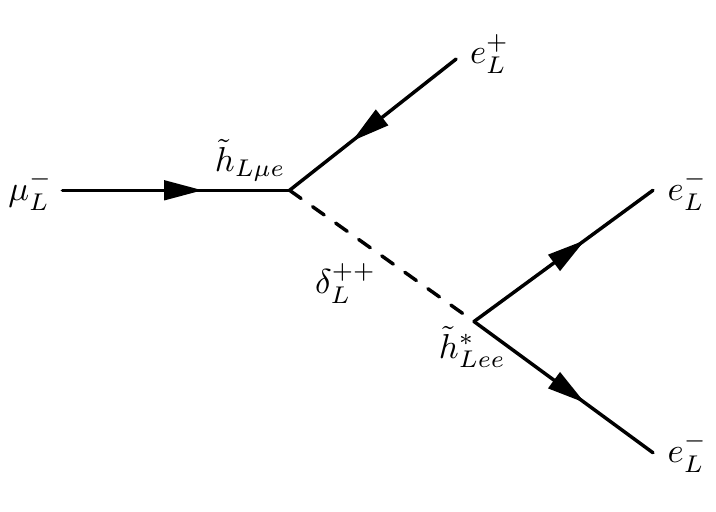}
		\label{fig:m3eTreeL}
	} \qquad
	\subfloat[]{
		\includegraphics[width = 0.3 \textwidth]{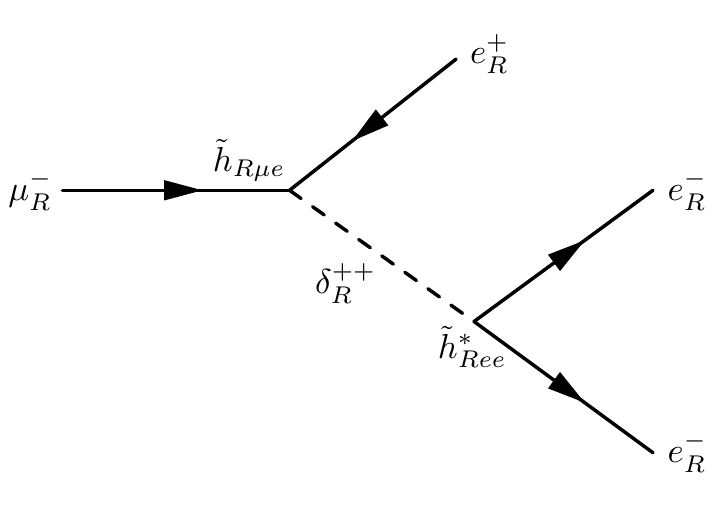}
		\label{fig:m3eTreeR}
	}
	\caption{Feynman diagrams of the tree-level processes of $\mu \to eee$.}
	\label{fig:m3eTreeFD}
\end{figure}

\begin{figure}[ht]
	\centering
	\subfloat[$B_{LL}^{\mu eee}$]{
		\includegraphics[width = 0.3 \textwidth]{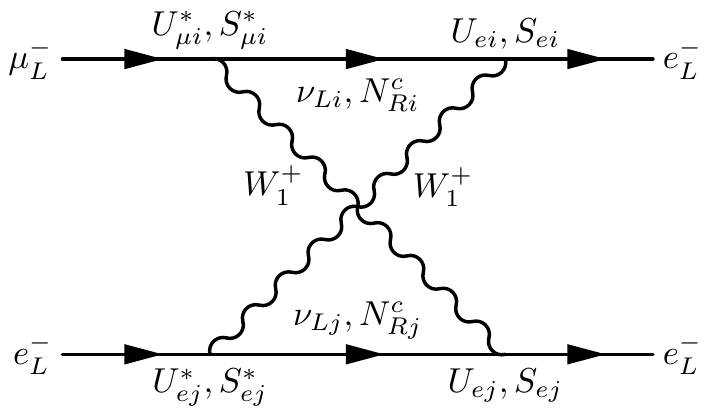}
		\label{fig:BeLL1}
	}
	\subfloat[$B_{LL}^{\mu eee}$]{
		\includegraphics[width = 0.3 \textwidth]{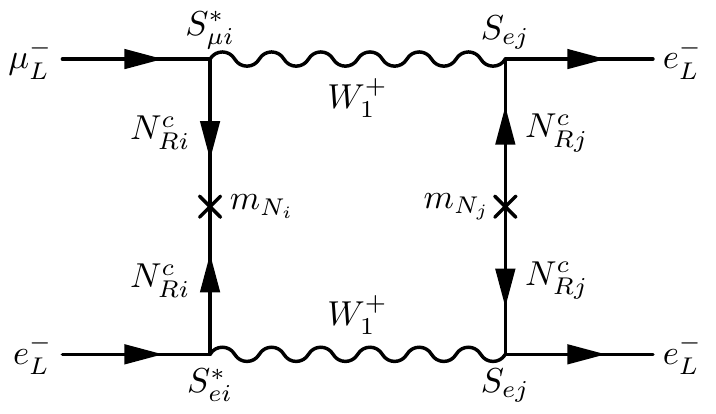}
		\label{fig:BeLL2}
	}
	\subfloat[$B_{RR}^{\mu eee}$]{
		\includegraphics[width = 0.3 \textwidth]{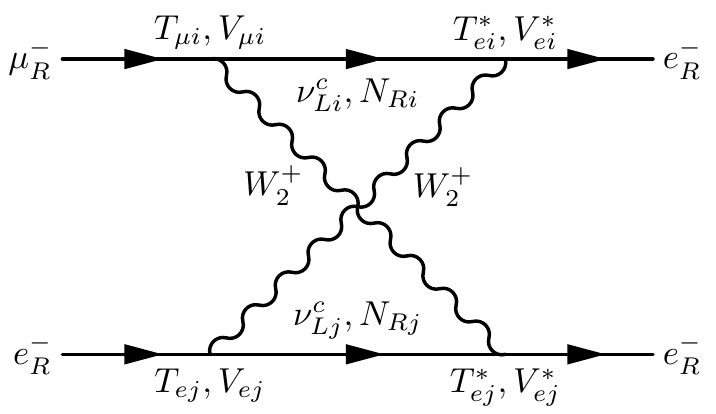}
		\label{fig:BeRR1}
	} \\
	\subfloat[$B_{RR}^{\mu eee}$]{
		\includegraphics[width = 0.3 \textwidth]{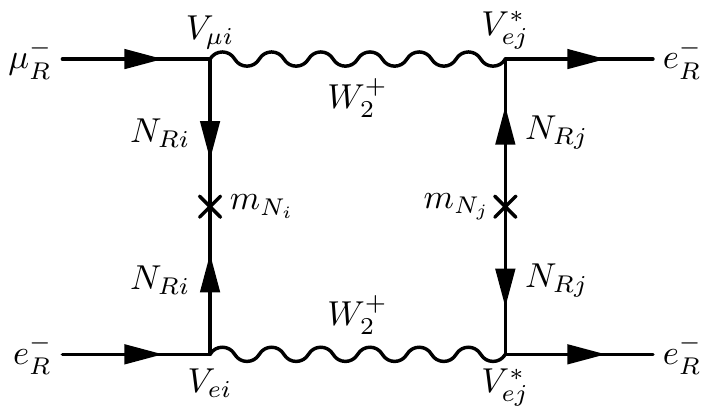}
		\label{fig:BeRR2}
	}
	\subfloat[$B_{LR}^{\mu eee}$]{
		\includegraphics[width = 0.3 \textwidth]{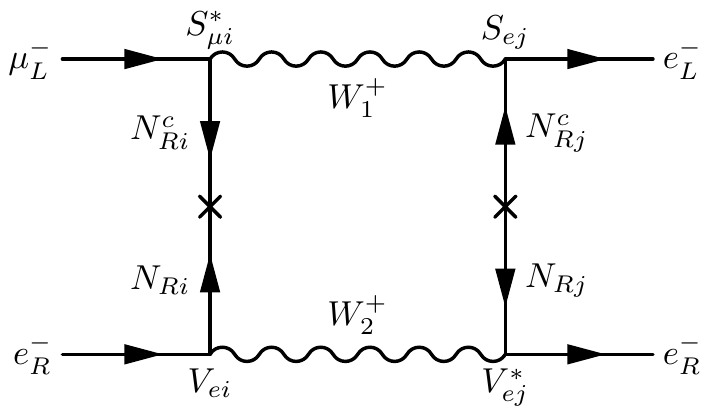}
		\label{fig:BeLR}
	}
	\subfloat[$B_{RL}^{\mu eee}$]{
		\includegraphics[width = 0.3 \textwidth]{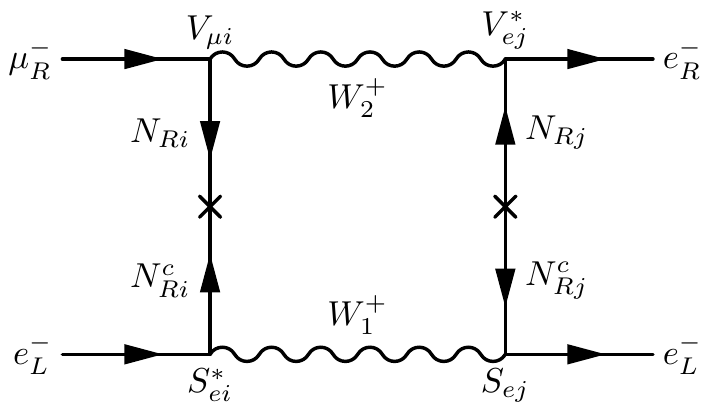}
		\label{fig:BeRL}
	}
	\caption{Feynman diagrams of $B^{\mu eee}$. Note that the arrows in neutrino propagators indicate the directions of the propagation of $\nu_i = \nu_{Li} + \nu_{Li}^c$ or $N_i = N_{Ri} + N_{Ri}^c$.}
	\label{fig:BeFD}
\end{figure}

The tree-level contribution to $\mu \to eee$ is
\begin{align}
	\text{BR}_{\mu \to eee}^\text{tree} = \frac{\alpha_W^4 m_\mu^5}{24576\pi^3 m_{W_L}^4 \Gamma_\mu} \frac{(4\pi)^2}{2 \alpha_W^2} \left( \big| \tilde{h}_{L \mu e} \tilde{h}_{Lee}^* \big|^2 \frac{m_{W_L}^4}{m_{\delta_L^{++}}^4} + \big| \tilde{h}_{R \mu e} \tilde{h}_{Ree}^* \big|^2 \frac{m_{W_L}^4}{m_{\delta_R^{++}}^4} \right).
\end{align}
The Feynman diagrams of the tree-level processes are given in figure \ref{fig:m3eTreeFD}.
The one-loop type-I seesaw contribution is given by \cite{LFVS, CLFVSUSY}
\begin{align}
	\text{BR}_{\mu \to eee}^\text{type-I} &= \frac{\alpha_W^4 m_\mu^5}{24576\pi^3 m_{W_L}^4 \Gamma_\mu}
		\bigg[ 2 \bigg\{ \bigg| \frac{1}{2} B_{LL}^{\mu eee} + F_L^{Z_1} - 2 s_W^2 \big( F_L^{Z_1} - F_L^\gamma \big) \bigg|^2
			+ \bigg| \frac{1}{2} B_{RR}^{\mu eee} - 2 s_W^2 \big( F_R^{Z_1} - F_R^\gamma \big) \bigg|^2 \bigg\} \nonumber \\
			&\qquad \qquad + \bigg| 2 s_W^2 \big( F_L^{Z_1} - F_L^\gamma \big) - B_{LR}^{\mu eee} \bigg|^2
			+ \bigg| 2 s_W^2 \big( F_R^{Z_1} - F_R^\gamma \big) - \big( F_R^{Z_1} + B_{RL}^{\mu eee} \big) \bigg|^2 \nonumber \\
			&\qquad \qquad + 8 s_W^2 \bigg\{ \text{Re}\bigg[ \big( 2 F_L^{Z_1} + B_{LL}^{\mu eee} + B_{LR}^{\mu eee} \big) G_R^{\gamma *} \bigg]
				+ \text{Re}\bigg[ \big( F_R^{Z_1} + B_{RR}^{\mu eee} + B_{RL}^{\mu eee} \big) G_L^{\gamma *} \bigg] \bigg\} \nonumber \\
			&\qquad \qquad - 48 s_W^4 \bigg\{ \text{Re}\bigg[ \big( F_L^{Z_1} - F_L^\gamma \big) G_R^{\gamma *} \bigg]
				+ \text{Re}\bigg[ \big( F_R^{Z_1} - F_R^\gamma \big) G_L^{\gamma *} \bigg] \bigg\} \nonumber \\
			&\qquad \qquad + 32 s_W^4 \big( |G_L^\gamma|^2 + |G_R^\gamma|^2  \big) \bigg\{ \ln{\bigg( \frac{m_\mu^2}{m_e^2} \bigg)} - \frac{11}{4} \bigg\} \bigg],
\end{align}
and the interference terms are
\begin{align}
	\text{BR}_{\mu \to eee}^\text{tree+type-I} &= \frac{\alpha_W^4 m_\mu^5}{24576\pi^3 m_{W_L}^4 \Gamma_\mu} \frac{2 (4\pi)}{\alpha_W} \times \nonumber \\
		&\qquad \bigg[ \frac{m_{W_L}^2}{m_{\delta_L^{++}}^2} \text{Re}\bigg[ \tilde{h}_{L \mu e}^* \tilde{h}_{Lee} \bigg\{ 2 s_W^2 F_L^\gamma + 4 s_W^2 G_R^\gamma + B_{LL}^{\mu eee} + F_L^{Z_1} (1 - 2 s_W^2) \bigg\} \bigg] \nonumber \\
			&\qquad \qquad + \frac{m_{W_L}^2}{m_{\delta_R^{++}}^2} \text{Re}\bigg[ \tilde{h}_{R \mu e}^* \tilde{h}_{Ree} \bigg\{ 2 s_W^2 F_R^\gamma + 4 s_W^2 G_L^\gamma + B_{RR}^{\mu eee} -2 s_W^2 F_R^{Z_1} \bigg\} \bigg] \bigg].
\end{align}
The form factors for the off-shell photon exchange are
\begin{align}
	F_L^\gamma &= \sum_{i = 1}^3 \left[ S_{\mu i}^* S_{ei} F_\gamma (x_i) - \tilde{h}_{L \mu i} \tilde{h}_{Lei}^* \bigg( \frac{2}{3} \frac{m_{W_L}^2}{m_{\delta_L^{++}}} \ln{\frac{m_\mu^2}{m_{\delta_L^{++}}}} + \frac{1}{18} \frac{m_{W_L}^2}{m_{H_1^+}} \bigg) \right], \\
	F_R^\gamma &= \sum_{i = 1}^3 \left[ V_{\mu i} V_{ei}^* \bigg( \xi^2 F_\gamma (x_i) + \frac{m_{W_L}^2}{m_{W_R}^2} F_\gamma (y_i) \bigg)
		- \tilde{h}_{R \mu i} \tilde{h}_{Rei}^* \frac{2}{3} \frac{m_{W_L}^2}{m_{\delta_R^{++}}} \ln{\frac{m_\mu^2}{m_{\delta_R^{++}}}} \right].
\end{align}
For the $Z_1$-exchange diagrams, the form factors are given by
\begin{align}
	F_L^{Z_1} &= \sum_{i,j = 1}^3 S_{\mu i}^* S_{ej} \bigg[ \delta_{ij} \big\{ F_Z (x_i) + 2 G_Z (0, x_i) \big\} \nonumber \\
		&\qquad \qquad + (S^\mathsf{T} S^*)_{ij} \big\{ G_Z (x_i, x_j) - G_Z (0, x_i) - G_Z (0, x_j) \big\} + (S^\dagger S)_{ij} H_Z (x_i, x_j) \bigg], \\
	F_R^{Z_1} &= \sum_{i = 1}^3 V_{\mu i} V_{ei}^* \bigg[ \frac{8 \zeta_3 c_W^2}{\sqrt{1 - 2 s_W^2}} \bigg\{ F_Z (y_i) + 2 G_Z (0, y_i) - \frac{y_i}{2} \bigg\} + 2 \left( \frac{\kappa_1 \kappa_2}{v_\text{EW} v_R} \right)^2 D_Z (y_i, x_i) \nonumber \\
		&\qquad \qquad + \left( \frac{\kappa_1 ^2 - \kappa_2^2}{\sqrt{2} v_\text{EW} v_R} \right)^2 D_Z (y_i, z_i) \bigg]
\end{align} 
where $z_i = (m_{N_i} / m_{H_2^+})^2$, $c_W \equiv \cos{\theta_W}$, and $\zeta_3$ is the $Z_1$-$Z_2$ mixing parameter given by equation \ref{eq:NGmixing}. The Feynman diagrams that contribute to $F_{L,R}^\gamma$ and $F_{L,R}^{Z_1}$ are presented in reference \cite{LFVnonSUSY}. The form factors of the box diagrams are written as
\begin{align}
	B_{LL}^{\mu eee} &= -2 \sum_{i = 1}^3 S_{\mu i}^* S_{ei} \big[ F_\text{Xbox} (0, x_i) - F_\text{Xbox} (0, 0) \big] \nonumber \\
		&\qquad + \sum_{i,j = 1}^3 S_{\mu i}^* S_{ej} \bigg[ -2 S_{ej}^* S_{ei} \big\{ F_\text{Xbox} (x_i, x_j) - F_\text{Xbox} (0, x_j) - F_\text{Xbox} (0, x_i) + F_\text{Xbox} (0, 0) \big\} \nonumber \\
			&\qquad \qquad + S_{ei}^* S_{ej} G_\text{box} (x_i, x_j, 1) \bigg], \displaybreak[0] \\
	B_{RR}^{\mu eee} &= -2 \frac{m_{W_L}^2}{m_{W_R}^2} \sum_{i,j = 1}^3 V_{\mu i} V_{ei}^* \big[ F_\text{Xbox} (0, y_i) - F_\text{Xbox} (0, 0) \big] \nonumber \\
		&\qquad + \frac{m_{W_L}^2}{m_{W_R}^2} \sum_{i,j = 1}^3 V_{\mu i} V_{ej}^* \bigg[ -2 V_{ej} V_{ei}^* \big\{ F_\text{Xbox} (y_i, y_j) - F_\text{Xbox} (0, y_j) - F_\text{Xbox} (0, y_i) + F_\text{Xbox} (0, 0) \big\} \nonumber \\
			&\qquad \qquad + V_{ei} V_{ej}^* G_\text{box} (y_i, y_j, 1) \bigg], \displaybreak[0] \\
	B_{LR}^{\mu eee} &= \frac{1}{2} \frac{m_{W_L}^2}{m_{W_R}^2} \sum_{i,j = 1}^3 S_{\mu i}^* S_{ej} V_{ei} V_{ej}^* G_\text{box} \bigg( x_i, x_j, \frac{m_{W_L}^2}{m_{W_R}^2} \bigg), \displaybreak[0] \\
	B_{RL}^{\mu eee} &= \frac{1}{2} \frac{m_{W_L}^2}{m_{W_R}^2} \sum_{i,j = 1}^3 V_{\mu i} V_{ej}^* S_{ei}^* S_{ej} G_\text{box} \bigg( x_i, x_j, \frac{m_{W_L}^2}{m_{W_R}^2} \bigg).
\end{align}
Here, the masses of light neutrinos and the momenta of external fields are assumed to be zero. The Feynman diagrams of the box diagrams are presented in figure \ref{fig:BeFD}.

\subsubsection{$\mu \to e$}
\begin{figure}[ht]
	\centering
	\subfloat[$B_{LL}^{\mu euu}$]{
		\includegraphics[width = 0.3 \textwidth]{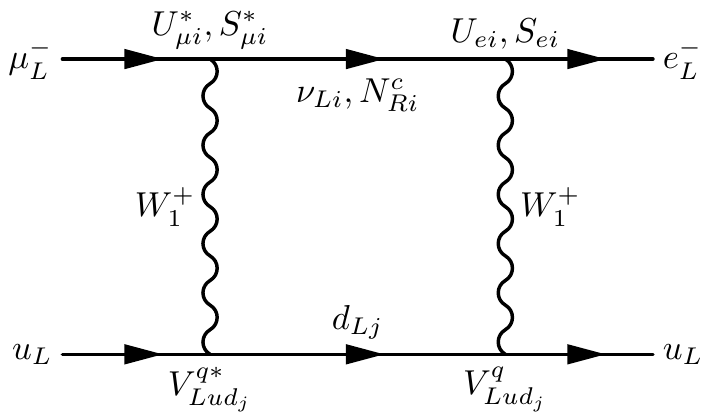}
		\label{fig:BuLL}
	}
	\subfloat[$B_{LL}^{\mu edd}$]{
		\includegraphics[width = 0.3 \textwidth]{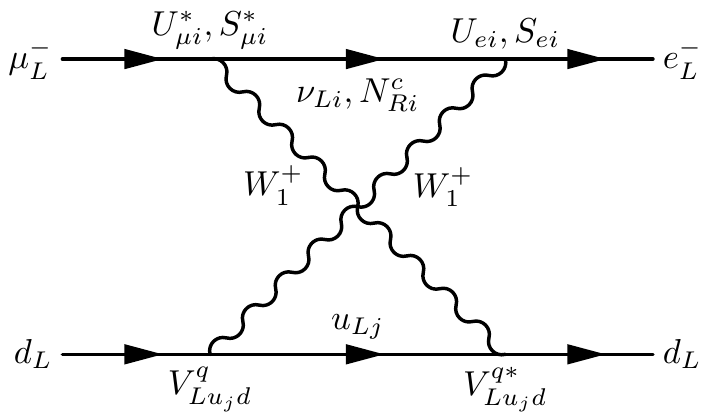}
		\label{fig:BdLL}
	} \\
	\subfloat[$B_{RR}^{\mu euu}$]{
		\includegraphics[width = 0.3 \textwidth]{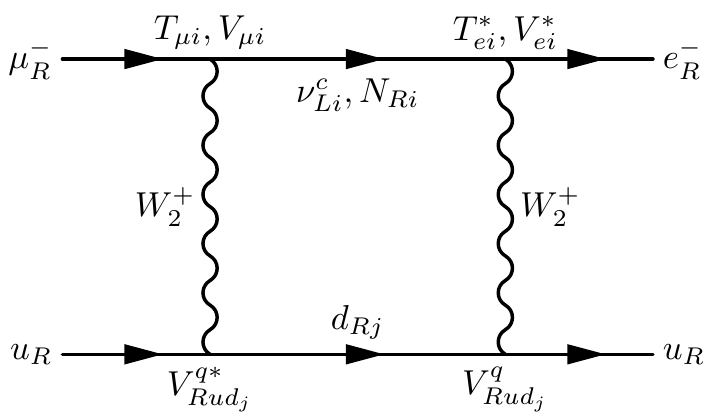}
		\label{fig:BuRR}
	}
	\subfloat[$B_{RR}^{\mu edd}$]{
		\includegraphics[width = 0.3 \textwidth]{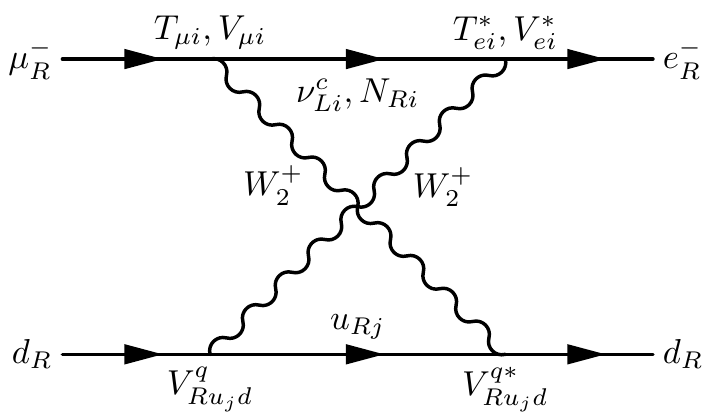}
		\label{fig:BdRR}
	}
	\caption{Feynman diagrams of $B^{\mu eqq}$.}
	\label{fig:BqFD}
\end{figure}

The $\mu \to e$ conversion rate is given by \cite{LFVnonSUSY, EWRL, mutoeTypeI, CLFVSUSY}
\begin{align}
	\text{R}_{\mu \to e}^{A(N,Z)} = \frac{\alpha_\text{em}^3 \alpha_W^4 m_\mu^5}{16\pi^2 m_{W_L}^4 \Gamma_\text{capt}} \frac{Z_\text{eff}^4}{Z} \big| F_p (-m_\mu^2) \big|^2 \big( \big| Q_L^W \big|^2 + \big| Q_R^W \big|^2 \big).
	\label{eq:Rmutoe}
\end{align}
Here, $A$, $N$, and $Z$ are the mass, neutron, and atomic numbers of a nucleus, respectively, and $Z_\text{eff}$ is the effective atomic number. The parameter $F_p$ is the nuclear form factor, $\Gamma_\text{capt}$ is the capture rate, and $\alpha_\text{em} \equiv e^2 / (4\pi)$. The values of $F_p$ and $\Gamma_\text{capt}$ of various nuclei are summarized in table \ref{tab:FCmutoe} \cite{mutoeTypeI}.
\begin{table}[ht]
	\center
	\begin{tabular}{|l||l|l|l|}
		\hline
		Nucleus $^A_Z \text{N}$ & $Z_\text{eff}$ & $|F_p (-m_\mu^2)|$ & $\Gamma_\text{capt}~(10^6~s^{-1})$ \\ \hline
		$^{27}_{13} \text{Al}$ & 11.5 & 0.64 & 0.7054 \\ \hline
		$^{48}_{22} \text{Ti}$ & 17.6 & 0.54 & 2.59 \\ \hline
		$^{197}_{79} \text{Au}$ & 33.5 & 0.16 & 13.07 \\ \hline
		$^{208}_{82} \text{Pb}$ & 34.0 & 0.15 & 13.45 \\ \hline
	\end{tabular}
	\caption{Form factors and capture rates of various nuclei associated with $\mu \to e$ conversion.}
	\label{tab:FCmutoe}
\end{table}
The form factors in equation \ref{eq:Rmutoe} are given by
\begin{align}
	Q_{L,R}^W = (2Z + N) \bigg[ W_{L,R}^u - \frac{2}{3} s_W^2 G_{R,L}^\gamma \bigg] + (Z + 2N) \bigg[ W_{L,R}^d + \frac{1}{3} s_W^2 G_{R,L}^\gamma \bigg]
\end{align}
and
\begin{align}
	W_{L,R}^u &= \frac{2}{3} s_W^2 F_{L,R}^\gamma + \bigg( -\frac{1}{4} + \frac{2}{3} s_W^2 \bigg) F_{L,R}^{Z_1} + \frac{1}{4} \bigg( B_{LL,RR}^{\mu euu} + B_{LR,RL}^{\mu euu} \bigg), \\
	W_{L,R}^d &= -\frac{1}{3} s_W^2 F_{L,R}^\gamma + \bigg( \frac{1}{4} - \frac{1}{3} s_W^2 \bigg) F_{L,R}^{Z_1} + \frac{1}{4} \bigg( B_{LL,RR}^{\mu edd} + B_{LR,RL}^{\mu edd} \bigg).
\end{align}
The box diagram form factors are
\begin{align}
	B_{LL}^{\mu euu} &= \sum_{i = 1}^3 S_{\mu i}^* S_{ei} [F_\text{box} (0, x_i) - F_\text{box} (0, 0)], \displaybreak[0] \\
	B_{LL}^{\mu edd} &= \sum_{i = 1}^3 S_{\mu i}^* S_{ei} \big[ F_\text{Xbox} (0, x_i) - F_\text{Xbox} (0, 0) \nonumber \\
		&\qquad \qquad + |V_{Ltd}^q|^2 \{F_\text{Xbox} (x_t, x_i) - F_\text{Xbox} (0, x_i) - F_\text{Xbox} (0, x_t) + F_\text{Xbox} (0, 0)\} \big], \displaybreak[0] \\
	B_{RR}^{\mu euu} &= \sum_{i = 1}^3 V_{\mu i} V_{ei}^* [F_\text{box} (0, x_i) - F_\text{box} (0, 0)], \displaybreak[0] \\
	B_{RR}^{\mu edd} &= \sum_{i = 1}^3 V_{\mu i} V_{ei}^* \big[ F_\text{Xbox} (0, x_i) - F_\text{Xbox} (0, 0) \nonumber \\
		&\qquad \qquad + |V_{Rtd}^q|^2 \{F_\text{Xbox} (x_t, x_i) - F_\text{Xbox} (0, x_i) - F_\text{Xbox} (0, x_t) + F_\text{Xbox} (0, 0)\} \big],
%
%
\end{align}
and $B_{LR}^{\mu eqq} = B_{RL}^{\mu eqq} = 0$ due to their chiral structures. Here, $x_t = m_t^2 / m_{W_L}^2$ and $y_t = m_t^2 / m_{W_R}^2$ where $m_t$ is the mass of a top quark, and the masses of all the other quarks as well as light neutrinos are assumed to be zero. The matrix $V_L^q$ is the Cabibbo-Kobayashi-Maskawa matrix, and $V_R^q$ is its RH counterpart. Note that $V_L^q \neq V_R^q$ for nonzero $\alpha$, although $V_{Ltd}^q = V_{Rtd}^q$ is assumed for the numerical analysis in this paper. The momenta of external fields are also assumed to be zero. The Feynman diagrams of the box diagrams are given in figure \ref{fig:BqFD}.

\subsubsection{Loop functions} \label{sec:loopf}
The loop functions of CLFV are
\begin{align}
	F_\gamma (x) &= \frac{7 x^3 - x^2 - 12 x}{12 (1 - x)^3} - \frac{x^4 - 10 x^3 + 12 x^2}{6 (1 - x)^4} \ln{x}, \displaybreak[0] \\
	G^\gamma_1 (x) &= -\frac{2 x^3 + 5 x^2 - x}{4 (1 - x)^3} - \frac{3 x^3}{2 (1 - x)^4} \ln{x}, \displaybreak[0] \\
	G^\gamma_2 (x) &= \frac{x^2 - 11 x + 4}{2 (1 - x)^2} - \frac{3 x^2}{(1 - x)^3} \ln{x}, \displaybreak[0] \\
	F_Z (x) &= -\frac{5 x}{2 (1 - x)} - \frac{5 x^2}{2 (1 - x)^2} \ln{x}, \displaybreak[0] \\
	G_Z (x, y) &= -\frac{1}{2 (1 - x)} \left[ \frac{x^2 (1 - y)}{1 - x} \ln{x} - \frac{y^2 (1 - x)}{1 - y} \ln{y} \right], \displaybreak[0] \\
	H_Z (x, y) &= \frac{\sqrt{x y}}{4 (x - y)} \left[ \frac{x (x - 4)}{1 - x} \ln{x} - \frac{y (y - 4)}{1 - y} \ln{y} \right], \displaybreak[0] \\
	D_Z (x, y) &= x \left( 2 - \ln{\frac{y}{x}} \right) + \frac{x (-8 + 9 x - x^2) - x^2 (8 - x) \ln{x}}{(1 - x)^2} + \frac{x y (1 - y + y \ln{y})}{(1 - y)^2} \nonumber \\
		&\qquad + \frac{2 x y (4 - x) \ln{x}}{(1 - x) (1 - y)} + \frac{2 x (x - 4 y) \ln{\frac{y}{x}}}{(1 - y) (x - y)}, \displaybreak[0] \\
	F_\text{box} (x, y) &= \left( 4 + \frac{x y}{4} \right) I_2 (x, y, 1) - 2 x y I_1 (x, y, 1), \displaybreak[0] \\
	F_\text{Xbox} (x, y) &= -\left( 1 + \frac{x y}{4} \right) I_2 (x, y, 1) - 2 x y I_1 (x, y, 1), \displaybreak[0] \\
	G_\text{box} (x, y, \eta) &= -\sqrt{x y} \left[ (4 + x y \eta) I_2 (x, y, \eta) - (1 + \eta) I_1 (x, y, \eta) \right]
\end{align}
where
\begin{align}
	I_1 (x, y, \eta) &= \left[ \frac{x \ln{x}}{(1 - x) (1 - \eta x) (x - y)} + (x \leftrightarrow y) \right] - \frac{\eta \ln{\eta}}{(1 - \eta) (1 - \eta x) (1 - \eta y)}, \\
	I_2 (x, y, \eta) &= \left[ \frac{x^2 \ln{x}}{(1 - x) (1 - \eta x) (x - y)} + (x \leftrightarrow y) \right] - \frac{\ln{\eta}}{(1 - \eta) (1 - \eta x) (1 - \eta y)}, \\
	I_i (x, y, 1) &\equiv \lim_{\eta \to 1} I_i (x, y, \eta).
\end{align}

\subsection{Neutrinoless double beta decay}
\begin{figure}[ht]
	\centering
	\subfloat[$\eta_\nu$]{
		\includegraphics[width = 0.3 \textwidth]{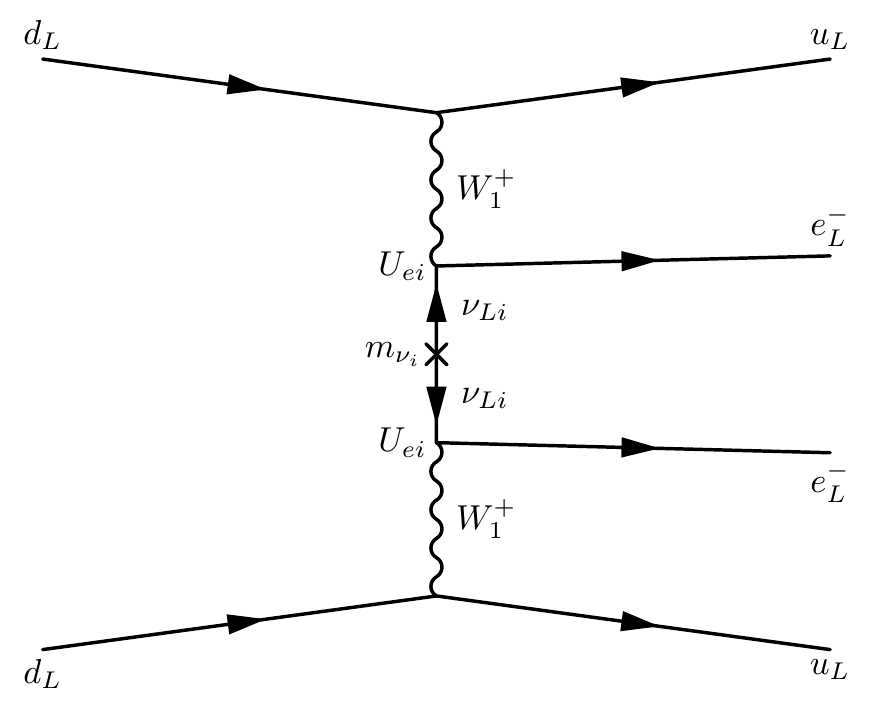}
		\label{fig:etnFD}
	}
	\subfloat[$\eta^L_{N_R}$]{
		\includegraphics[width = 0.3 \textwidth]{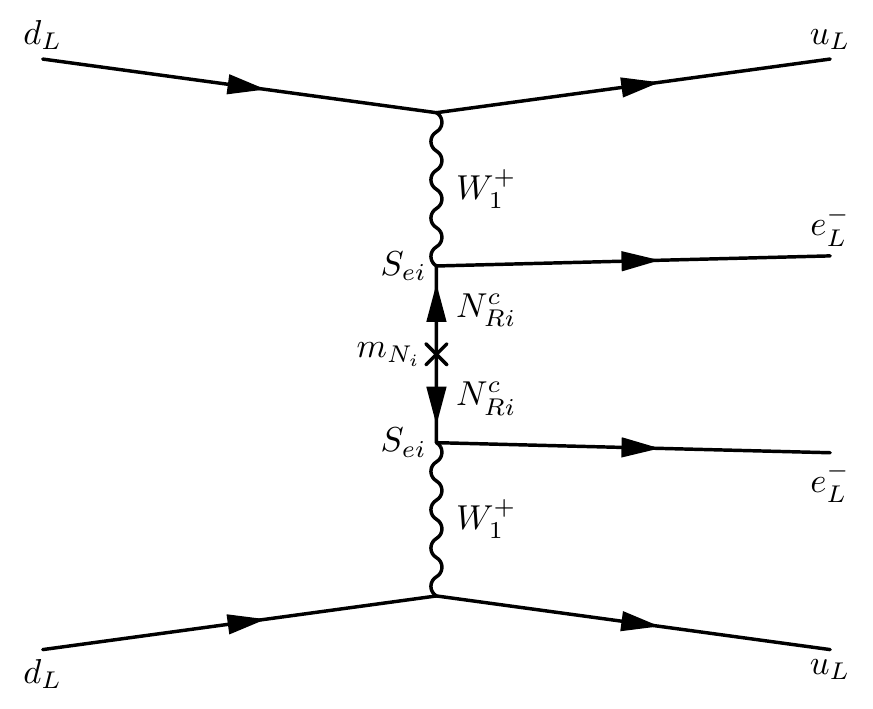}
		\label{fig:etLNRFD}
	}
	\subfloat[$\eta^R_{N_R}$]{
		\includegraphics[width = 0.3 \textwidth]{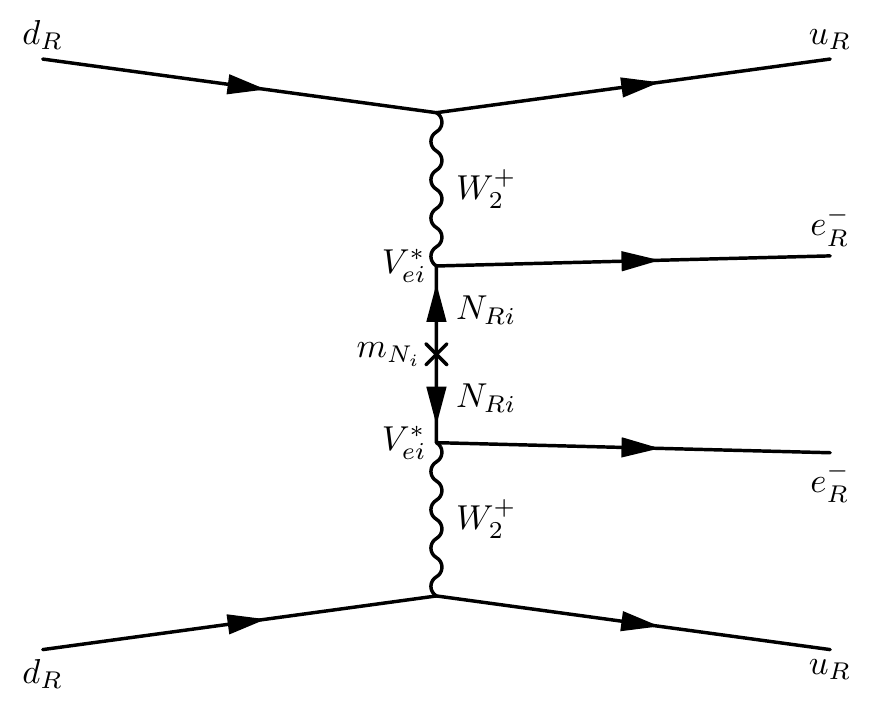}
		\label{fig:etRNRFD}
	} \\
	\subfloat[$\eta_{\delta_R}$]{
		\includegraphics[width = 0.3 \textwidth]{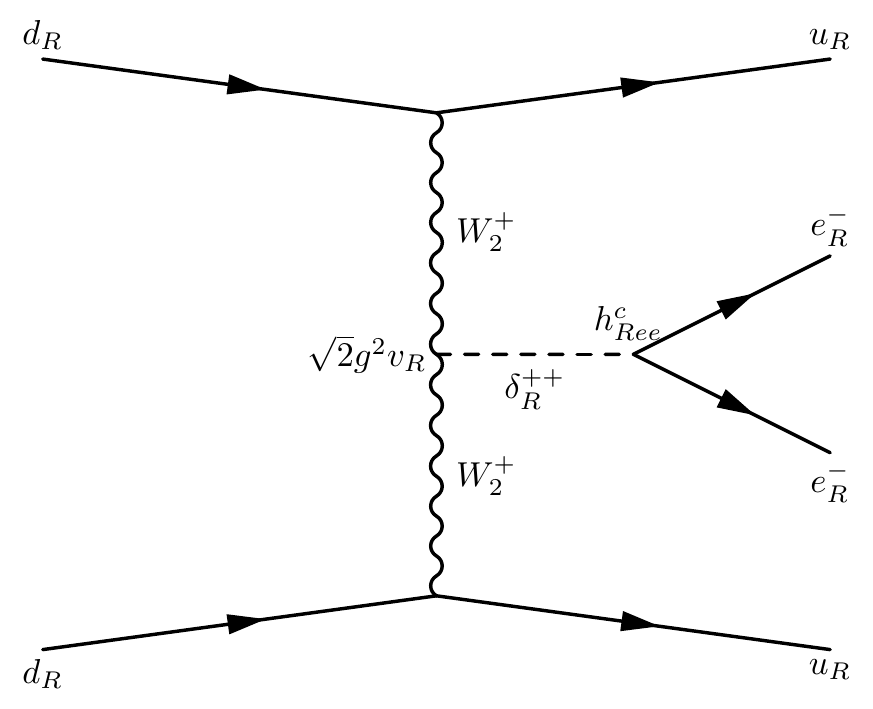}
		\label{fig:etdRFD}
	}
	\subfloat[$\eta_\lambda$]{
		\includegraphics[width = 0.3 \textwidth]{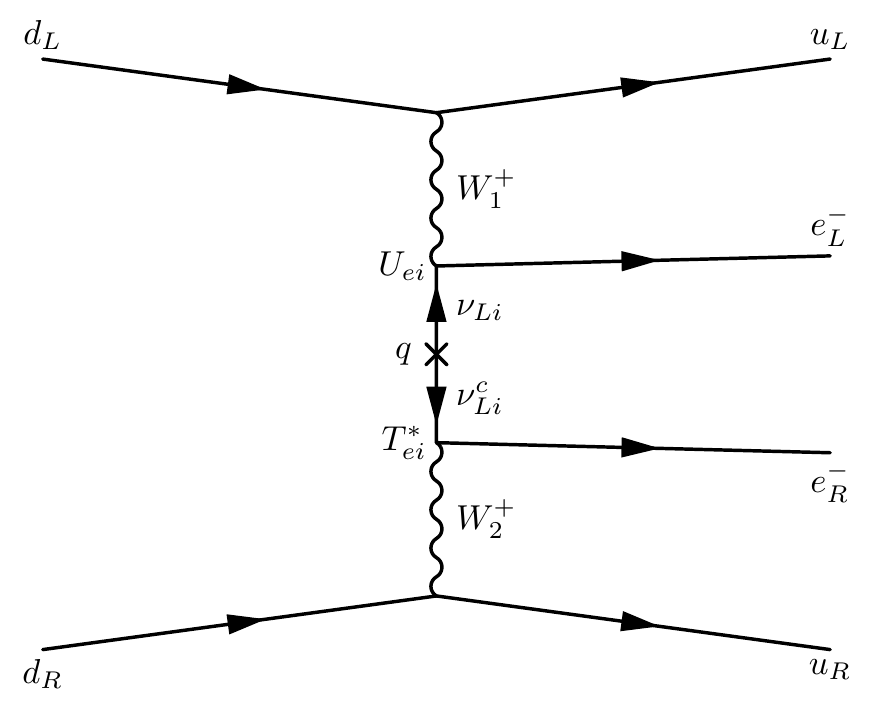}
		\label{fig:etlFD}
	}
	\subfloat[$\eta_\eta$]{
		\includegraphics[width = 0.3 \textwidth]{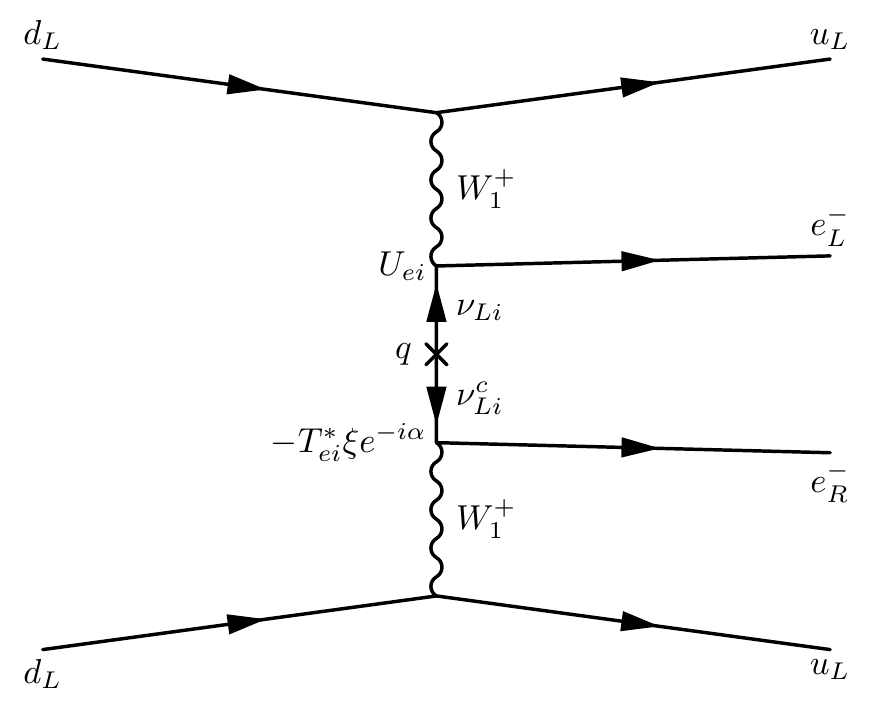}
		\label{fig:etetFD}
	}
	\caption{Feynman diagrams of $0 \nu \beta \beta$. Here, $W_L^+ \approx W_1^+ + \xi e^{-i \alpha} W_2^+$ and $W_R^+ \approx -\xi e^{i \alpha} W_1^+ + W_2^+$. The coupling $h_R^c \equiv V^{\ell \mathsf{T}}_R h V^\ell_R = M_R^c / (\sqrt{2} v_R)$ is the Yukawa coupling matrix in the charged lepton mass basis. The typical momentum transfer of the processes is $q \approx 100$ MeV.}
	\label{fig:0nbbFD}
\end{figure}

The dimensionless parameter associated with the $W_L$- and light neutrino exchange is
\begin{align}
	\eta_\nu = \frac{\sum_{i = 1}^3 (U_{ei})^2 m_{\nu_i}}{m_e}.
\end{align}
For the $W_L$- and heavy neutrino exchange, we have
\begin{align}
	\eta_{N_R}^L = m_p \sum_{i = 1}^3 \frac{(S_{ei})^2}{m_{N_i}}
\end{align}
where $m_p$ is the mass of a proton. For the $W_R$- and heavy neutrino exchange, the parameter is given by
\begin{align}
	\eta_{N_R}^R = m_p \bigg( \frac{m_{W_L}}{m_{W_R}} \bigg)^4 \sum_{i = 1}^3 \frac{(V_{ei}^*)^2}{m_{N_i}}.
\end{align}
For the $\delta_R^{++}$-exchange, we have
\begin{align}
	\eta_{\delta_R} = \frac{\sum_{i = 1}^3 (V_{ei})^2 m_{N_i}}{m_{\delta_R^{++}}^2 m_{W_R}^4} \frac{m_p}{G_F^2}.
\end{align}
For the $\lambda$-diagram with final state electrons of different helicities, the parameter is written as
\begin{align}
	\eta_\lambda = \bigg( \frac{m_{W_L}}{m_{W_R}} \bigg)^2 \sum_{i = 1}^3 U_{ei} T_{ei}^*.
\end{align}
For the $\eta$-diagram with $W_L$-$W_R$ mixing,
\begin{align}
	\eta_\eta = -\xi e^{-i \alpha} \sum_{i = 1}^3 U_{ei} T_{ei}^*.
\end{align}
The Feynman diagrams corresponding to those parameters are given in figure \ref{fig:0nbbFD}.
The phase space factors $G_{01}^{0 \nu}$ and matrix elements $\mathcal{M}^{0 \nu}$ for various processes that lead to $0 \nu \beta \beta$ are summarized in table \ref{tab:PM0nbb} \cite{LRSMLFV, PF0nbb, 0nbbIH, T0nbb, CPbb, ID0nbb, CPbbN, 0nbbQ, WNbb}. The inverse half-life is written as
\begin{align}
	[T_{1/2}^{0 \nu}]^{-1} &= G_{01}^{0 \nu} \left( |\mathcal{M}^{0 \nu}_\nu|^2 |\eta_\nu|^2 + |\mathcal{M}^{0 \nu}_N|^2 |\eta^L_{N_R}|^2 + |\mathcal{M}^{0 \nu}_N|^2 |\eta^R_{N_R} + \eta_{\delta_R}|^2 + |\mathcal{M}^{0 \nu}_\lambda|^2 |\eta_\lambda|^2 + |\mathcal{M}^{0 \nu}_\eta|^2 |\eta_\eta|^2 \right) \nonumber \\
		&\qquad + \text{interference terms}.
\end{align}
\begin{table}[ht]
	\center
	\begin{tabular}{|l||l|l|l|l|l|}
		\hline
		Isotope & $G_{01}^{0 \nu}~(10^{-14}~\text{yrs.}^{-1})$ & $\mathcal{M}^{0 \nu}_\nu$ & $\mathcal{M}^{0 \nu}_N$ & $\mathcal{M}^{0 \nu}_\lambda$ & $\mathcal{M}^{0 \nu}_\eta$ \\ \hline
		$^{76}$Ge & 0.686 & $2.58 - 6.64$ & $233 - 412$ & $1.75 - 3.76$ & $235 - 637$ \\ \hline
		$^{82}$Se & 2.95 & $2.42 - 5.92$ & $226 - 408$ & $2.54 - 3.69$ & $209 - 234$ \\ \hline
		$^{130}$Te & 4.13 & $2.43 - 5.04$ & $234 - 385$ & $2.85 - 3.67$ & $414 - 540$ \\ \hline
		$^{136}$Xe & 4.24 & $1.57 - 3.85$ & $164 - 172$ & $1.96 - 2.49$ & $370 - 419$ \\ \hline
	\end{tabular}
	\caption{Phase space factors and matrix elements associated with $0 \nu \beta \beta$.}
	\label{tab:PM0nbb}
\end{table}

\subsection{Electric dipole moments of charged leptons}
\begin{figure}[ht]
	\centering
	\subfloat[]{
		\includegraphics[width = 0.3 \textwidth]{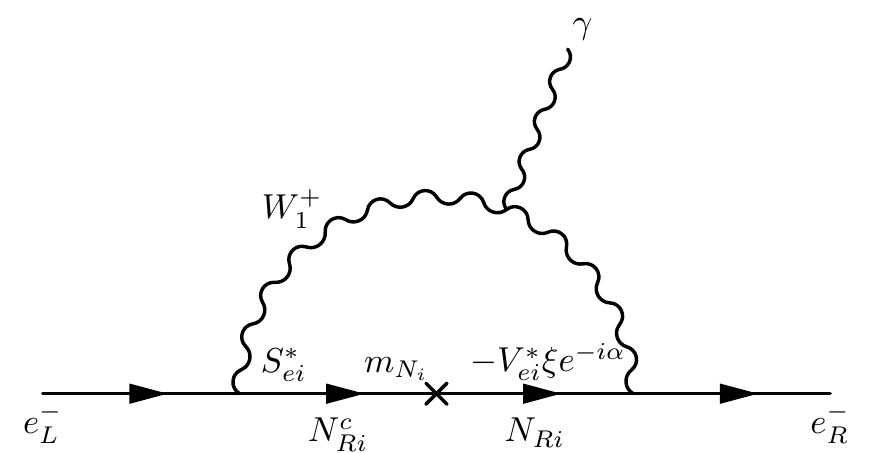}
		\label{fig:EDMe1}
	} \qquad
	\subfloat[]{
		\includegraphics[width = 0.3 \textwidth]{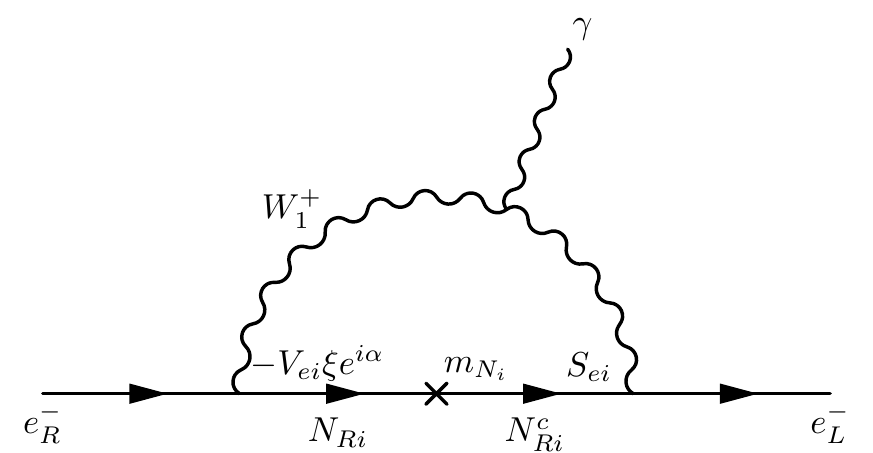}
		\label{fig:EDMe2}
	}
	\caption{Feynman diagrams contributing to the EDM of $e$.}
	\label{fig:EDMeFD}
\end{figure}

The EDM of the charged lepton $\ell_\alpha$ $(\alpha = e, \mu, \tau)$ is given by \cite{LRSMDM, EDMeLRSM}
\begin{align}
	d_\alpha = \frac{e \alpha_W}{8\pi m_{W_L}^2} \text{Im} \bigg[ \sum_{i = 1}^3 S_{\alpha i} V_{\alpha i} \xi e^{i \alpha} G_2^\gamma (x_i) m_{N_i} \bigg].
\end{align}
The Feynman diagrams that generate the EDM of an electron are given in figure \ref{fig:EDMeFD}.

\section{Benchmark model parameters and their predictions} \label{sec:BM}
The benchmark model parameters and their predictions are summarized in tables \ref{tab:BMin} and \ref{tab:BMout}. These parameters are chosen to obtain BR$_{\mu \to e \gamma}$, BR$_{\mu \to eee}$, R$_{\mu \to e}$, and $T_{1/2}^{0 \nu}$ large enough to be observable in near-future experiments.
\begin{table}[!htp]
	\center
	\small
	\begin{tabular}{|l|l||l|l|}
		\hline
		Parameter & Value & Parameter & Value \\ \hline
		$\log_{10}{(m_{\nu_3} / \text{eV})}$ & $-10.2$ & $\log_{10}{(\kappa_2 / \text{GeV})}$ & $-1.12$ \\ \hline
		$m_{W_R}$ & 3.60 TeV & $\alpha$ & $0.7843093682120977 \pi$ rad \\ \hline
		$\delta_D$ & $-0.700 \pi$ rad & $\log_{10}{(|A_{11}| / \text{GeV})}$ & $-8.20$ \\ \hline
		$\delta_{M1}$ & $-0.0640 \pi$ rad & $A_{11} / |A_{11}|$ & $1$ \\ \hline
		$\delta_{M2}$ & $0.850 \pi$ rad & $A_{22} / |A_{22}|$ & $-1$ \\ \hline
		$\theta_{L12}$ & $0.287 \pi$ rad & $A_{33} / |A_{33}|$ & $-1$ \\ \hline
		$\theta_{L13}$ & $0.387 \pi$ rad & $\theta_{A_{12}}$ & $-0.5970870460412485 \pi$ rad \\ \hline
		$\theta_{L23}$ & $0.546 \pi$ rad & $\theta_{A_{13}}$ & $0.26505775139215687 \pi$ rad \\ \hline
		$\delta_{L1}$ & $-0.488 \pi$ rad & $\theta_{A_{23}}$ & $-0.6679707059438431 \pi$ rad \\ \hline
		$\delta_{L2}$ & $-0.953 \pi$ rad & $\log_{10}{\alpha_3}$ & $0.520$ \\ \hline
		$\delta_{L3}$ & $-0.769 \pi$ rad & $\log_{10}{(\rho_3 - 2 \rho_1)}$ & $0.328$ \\ \hline
		$\delta_{L4}$ & $-5.30 \cdot 10^{-5} \pi$ rad & $\log_{10}{\rho_2}$ & $0.450$ \\ \hline
	\end{tabular}
	\caption{Benchmark parameters for large CLFV and $0 \nu \beta \beta$. The predictions from these parameters are given in table \ref{tab:BMout}.}
	\label{tab:BMin}
	\center
	\footnotesize
	\begin{tabular}{|l||l|}
		\hline
		Parameter & Value \\ \hline
		$m_{W_R}$ & 3.60 TeV \\ \hline
		$m_{\nu_1}$ & 0.0631 eV \\ \hline
		$m_{\nu_2}$ & 0.0637 eV \\ \hline
		$m_{\nu_3}$ & 0.0807 eV \\ \hline
		$m_{N_1}$ & 0.139 TeV \\ \hline
		$m_{N_2}$ & 0.280 TeV \\ \hline
		$m_{N_3}$ & 4.13 TeV \\ \hline
		$m_{H_1^+}$ & 8.08 TeV \\ \hline
		$m_{H_2^+}$ & 10.1 TeV \\ \hline
		$m_{\delta_L^{++}}$ & 8.09 TeV \\ \hline
		$m_{\delta_R^{++}}$ & 18.6 TeV \\ \hline
		$\kappa_1$ & 246 GeV \\ \hline
		$\kappa_2 e^{i \alpha}$ & $0.0759 e^{i 0.784 \pi}$ GeV \\ \hline
		$\alpha_3$ & $3.31$ \\ \hline
		$\rho_3 - 2 \rho_1$ & $2.13$ \\ \hline
		$\rho_2$ & $2.82$ \\ \hline
	\end{tabular}
	\begin{tabular}{|l||l|l|}
		\hline
		& Prediction & Near-future sensitivity \\ \hline
		BR$_{\mu \to e \gamma}$ & $5.98 \cdot 10^{-14}$ & $< 5.0 \cdot 10^{-14}$ (Upgraded MEG) \\ \hline
		BR$_{\tau \to \mu \gamma}$ & $1.94 \cdot 10^{-13}$ & \makecell{$\cdot$} \\ \hline
		BR$_{\tau \to e \gamma}$ & $4.85 \cdot 10^{-13}$ & \makecell{$\cdot$} \\ \hline
		BR$_{\mu \to eee}$ & $8.12 \cdot 10^{-14}$ & $< 1.0 \cdot 10^{-15}$ (PSI) \cite{PSI} \\ \hline
		R$_{\mu \to e}^\text{Al}$ & $2.17 \cdot 10^{-13}$ & $< 3.0 \cdot 10^{-17}$ (COMET) \\ \hline
		R$_{\mu \to e}^\text{Ti}$ & $4.13 \cdot 10^{-13}$ & $< 1.0 \cdot 10^{-18}$ (PRISM/PRIME) \\ \hline
		R$_{\mu \to e}^\text{Au}$ & $3.98 \cdot 10^{-13}$ & \makecell{$\cdot$} \\ \hline
		R$_{\mu \to e}^\text{Pb}$ & $3.83 \cdot 10^{-13}$ & \makecell{$\cdot$} \\ \hline
		$|\eta_\nu|$ & $1.21 \cdot 10^{-7}$ & $\lesssim 1.4 \cdot 10^{-7}$ (CUORE) \\ \hline
		$|\eta^L_{N_R}|$ & $4.97 \cdot 10^{-15}$ & \makecell{$\cdot$} \\ \hline
		$|\eta^R_{N_R}|$ & $4.77 \cdot 10^{-10}$ & \makecell{$\cdot$} \\ \hline
		$|\eta_{\delta_R}|$ & $4.24 \cdot 10^{-11}$ & \makecell{$\cdot$} \\ \hline
		$|\eta_\lambda|$ & $4.61 \cdot 10^{-10}$ & \makecell{$\cdot$} \\ \hline
		$|\eta_\eta|$ & $2.81 \cdot 10^{-13}$ & \makecell{$\cdot$} \\ \hline
		$T_{1/2}^{0 \nu} \big|_\text{Ge}$ & $2.12 \cdot 10^{26} - 1.31 \cdot 10^{27}$ yrs. & \makecell{$\cdot$} \\ \hline
		$T_{1/2}^{0 \nu} \big|_\text{Se}$ & $6.11 \cdot 10^{25} - 3.43 \cdot 10^{26}$ yrs. & \makecell{$\cdot$} \\ \hline
		$T_{1/2}^{0 \nu} \big|_\text{Te}$ & $5.91 \cdot 10^{25} - 2.41 \cdot 10^{26}$ yrs. & $> 2.1 \cdot 10^{26}$ yrs. (CUORE) \\ \hline
		$T_{1/2}^{0 \nu} \big|_\text{Xe}$ & $1.05 \cdot 10^{26} - 5.48 \cdot 10^{26}$ yrs. & \makecell{$\cdot$} \\ \hline
		$|d_e|$ & $|\minus 2.98 \cdot 10^{-31}|~e \cdot$cm & \makecell{$\cdot$} \\ \hline
		$|d_\mu|$ & $|1.99 \cdot 10^{-31}|~e \cdot$cm & \makecell{$\cdot$} \\ \hline
		$|d_\tau|$ & $|\minus 3.13 \cdot 10^{-31}|~e \cdot$cm & \makecell{$\cdot$} \\ \hline
	\end{tabular}
	\caption{Predictions from the benchmark model parameters of table \ref{tab:BMin}. Only near-future experiments that would detect the corresponding processes are presented here.}
	\label{tab:BMout}
\end{table}

The Yukawa coupling matrices $f$, $\tilde{f}$ in the symmetry basis calculated from these parameters are
\small
\begin{align}
	f &= \left( \begin{array}{ccc}
			-0.117629 & -0.0954074 - 0.303042 i & -0.287722 - 0.316317 i \\
			-0.0954074 + 0.303042 i & 0.858098 & -0.581546 - 0.997804 i \\
			-0.287722 + 0.316317 i & -0.581546 + 0.997804 i & 1.55438 \\
		\end{array} \right) \cdot 10^{-6}, \displaybreak[0] \\
	\tilde{f} &= \left( \begin{array}{ccc}
			9.02581  & 0.362808  - 3.15221 i & -0.217594 + 0.423914 i \\
			0.362808 + 3.15221 i & 1.53907 & 3.98014 \cdot 10^{-4} - 0.328771 i \\
			-0.217594 - 0.423914 i & 3.98014 \cdot 10^{-4} + 0.328771 i & 0.260124  \\
		\end{array} \right) \cdot 10^{-3}.
\end{align}
\normalsize
The charged lepton and Dirac neutrino mass matrices in the symmetry basis are
\small
\begin{align}
	M_\ell &= \frac{1}{\sqrt{2}} (f \kappa_2 e^{i \alpha} + \tilde{f} \kappa_1) \nonumber \\
		&= \left( \begin{array}{ccc}
			1.57002 - 3.95569 \cdot 10^{-9} i & 0.0631099 - 0.548321 i & -0.0378502 + 0.0737391 i \\
			0.0631098 + 0.548321 i & 0.267718 + 2.88565 \cdot 10^{-8} i & 6.92918 \cdot 10^{-5} - 0.0571891 i \\
			-0.0378501 - 0.0737391 i & 6.92247 \cdot 10^{-5} + 0.0571891 i & 0.0452481 + 5.22714 \cdot 10^{-8} i \\
		\end{array} \right)~\text{GeV}, \displaybreak[0] \\
	M_D &= \frac{1}{\sqrt{2}} (f \kappa_1 + \tilde{f} \kappa_2 e^{-i \alpha}) \nonumber \\
		&= \left( \begin{array}{ccc}
			-3.97641 - 3.03524 i & -1.37761 + 0.668135 i & 0.733973 + 0.446252 i \\
			0.742466 - 0.912148 i & 0.849485 - 0.517565 i & -1.12232 - 1.59841 i \\
			0.448861 - 0.299905 i & -0.901194 + 1.59814 i & 2.59511 - 0.0874759 i \\
		\end{array} \right) \cdot 10^{-4}~\text{GeV}.
\end{align}
\normalsize
The mixing matrices that diagonalize $M_\ell$ are
\small
\begin{align}
	V^\ell_L &= \left( \begin{array}{ccc}
			0.215620 + 3.59016 \cdot 10^{-5} i  & 0.272630 & 0.0353401 + 0.936980 i \\
			-0.174794 - 0.555520 i & 0.00850025 - 0.736518 i & -0.340224 + 0.0506041 i \\
			-0.527503 + 0.579736 i & 0.526439 - 0.325580 i & 0.0374439 - 0.0332209 i \\
		\end{array} \right), \\
	V^\ell_R &= \left( \begin{array}{ccc}
			0.215620 & 0.272630 & 0.0353401 + 0.936980 i \\
			-0.174886 - 0.555491 i & 0.00850025 - 0.736518 i & -0.340224 + 0.0506041 i \\
			-0.527407 + 0.579824 i & 0.526439 - 0.325580 i & 0.0374439 - 0.0332209 i \\
		\end{array} \right).
\end{align}
\normalsize
The neutrino mass matrices in the charged lepton mass basis are written as
\small
\begin{align}
	M_\nu^c &= U_\text{PMNS} M_\nu^\text{diag} U_\text{PMNS}^\mathsf{T} \nonumber \\
		&= \left( \begin{array}{ccc}
			6.14141 + 0.604007 i & -0.641188 + 1.37500 i & -0.414134 - 0.161926 i \\
			-0.641188 + 1.37500 i & 5.21993 + 3.90978 i & -0.721679 + 2.37952 i \\
			-0.414134 - 0.161926 i & -0.721679 + 2.37952 i & 5.35910 + 4.32684 i \\
		\end{array} \right) \cdot 10^{-11}~\text{GeV}, \displaybreak[0] \\
	M_D^c &= V^{\ell \dagger}_L M_D V^\ell_R \nonumber \\
		&= \left( \begin{array}{ccc}
			-0.887458 -0.00113569 i & -0.596983 - 1.80367 i & -0.364728 - 0.967911 i \\
			-0.596682 + 1.80377 i & 2.44772 - 0.204264 i & 0.650485 - 0.676299 i \\
			-0.364567 + 0.967972 i & 0.650486 + 0.676299 i & -3.86700 - 3.43503 i \\
		\end{array} \right) \cdot 10^{-4}~\text{GeV}, \displaybreak[0] \\
	M_R^c &= -M_D^{c \mathsf{T}} (M_\nu^c)^{-1} M_D^c \nonumber \\
		&= \left( \begin{array}{ccc}
			327.179 - 124.513 i & -141.421 - 201.931 i & 36.0396 + 816.162 i \\
			-141.421 - 201.931 i & 56.2978 + 60.4971 i & 517.744 - 74.6682 i \\
			36.0396 + 816.162 i & 517.744 - 74.6682 i & -2486.91- 2973.37 i \\
		\end{array} \right)~\text{GeV}.
\end{align}
\normalsize
The neutrino mixing matrices are given by
\small
\begin{align}
	U &= U_\text{PMNS}
	= \left( \begin{array}{ccc}
			0.824240 & 0.535780 + 0.109200 i & 0.131084 - 0.0667906 i \\
			-0.365548 + 0.0658493 i & 0.632967 + 0.173591 i & -0.585126 - 0.298136 i \\
			0.420911 + 0.0741679 i & -0.516908 + 0.0551401 i & -0.659043 - 0.335799 i \\
		\end{array} \right), \displaybreak[0] \\
	S &= \left( \begin{array}{ccc}
			-0.492113 - 0.340868 i & 0.999284 + 0.0561499 i & 0.239615 + 0.0281506 i \\
			-0.0475962 + 0.503081 i & -0.231028 - 1.26661 i & -0.00795814 - 0.320325 i \\
			0.232020 - 0.00648341 i & -0.401571 + 0.125068 i & -0.175188 + 0.136668 i \\
		\end{array} \right) \cdot 10^{-6}, \displaybreak[0] \\
	T &= \left( \begin{array}{ccc}
			-6.53107 - 6.47350 i & -8.46370 + 5.72968 i & -1.16360 - 8.20634 i \\
			2.04202 - 6.05309 i & -4.69170 - 5.30774 i & 3.49735 - 2.06263 i \\
			-1.83711 - 0.641098 i & 0.0608069 + 1.46932 i & 0.103607 - 0.502026 i \\
		\end{array} \right) \cdot 10^{-7}, \displaybreak[0] \\
	V &= \left( \begin{array}{ccc}
			-0.183724 + 0.375972 i & 0.879386 + 0.0740900 i & 0.195953 - 0.0876577 i \\
			-0.881006 + 0.210057 i & -0.242230 - 0.320460 i & -0.0720947 - 0.114618 i \\
			-0.0677616 + 0.00212502 i & 0.177470 - 0.168300 i & -0.408123 + 0.876937 i \\
		\end{array} \right).
\end{align}
\normalsize
The Yukawa coupling matrix $h$ in the symmetry basis is
\small
\begin{align}
	h &= \frac{1}{\sqrt{2} v_R} V^{\ell *}_R M_R^c V^{\ell \dagger}_R \nonumber \\
		&= \left( \begin{array}{ccc}
			0.206578 + 0.223735 i & 0.120506 - 0.0241230 i & -0.0469350 - 0.0641918 i \\
			0.120506 - 0.0241230 i & 0.00351664 - 0.0376782 i & -0.0257606 + 0.00173595 i \\
			-0.046935 - 0.0641918 i &  -0.0257606 + 0.00173595 i & -0.00335158 + 0.0385022 i \\
		\end{array} \right),
\end{align}
\normalsize
and the normalized Yukawa couplings $\tilde{h}_L$,  $\tilde{h}_R$ in the charged lepton mass basis are
\small
\begin{align}
	\tilde{h}_L &= \frac{2}{g} V^{\ell \mathsf{T}}_L h V^\ell_L \nonumber \\
		&= \left( \begin{array}{ccc}
			0.0908945 - 0.0345568 i & -0.0392741 - 0.0560986 i & 0.00997325 + 0.226713 i \\
			-0.0392741 - 0.0560986 i & 0.0156383 + 0.0168047 i & 0.143818 - 0.0207412 i \\
			0.00997325 + 0.226713 i & 0.143818 - 0.0207412 i & -0.690808 - 0.825936 i \\
		\end{array} \right), \\
	\tilde{h}_R &= \frac{2}{g} V^{\ell \mathsf{T}}_R h V^\ell_R \nonumber \\
		&= \left( \begin{array}{ccc}
			0.0908830 - 0.0345871 i & -0.0392835 - 0.0560921 i & 0.0100110 + 0.226712 i \\
			-0.0392835 - 0.0560921 i & 0.0156383 + 0.0168047 i & 0.143818 - 0.0207412 i \\
			0.0100110 + 0.226712 i & 0.143818 - 0.0207412 i & -0.690808 - 0.825936 i \\
		\end{array} \right).
\end{align}
\normalsize
Note that $\tilde{h}_L \approx \tilde{h}_R$ since we are considering the cases of $V^\ell_L \approx V^\ell_R$ for the TeV-scale phenomenology.


\end{document}